\documentclass[aps,rmp,twocolumn,showpacs,groupedaddress,preprintnumbers,amsmath,amssymb]{revtex4-2}

\usepackage[usenames,dvipsnames]{xcolor}

\usepackage[normalem]{ulem}
 \newcommand{\stkout}[1]{\ifmmode\text{\sout{\ensuremath{#1}}}\else\sout{#1}\fi}

\usepackage{float}
\usepackage{graphicx}
\usepackage{dcolumn}
\usepackage{bm}
\usepackage{multirow}
\usepackage{amsmath,amssymb}

\begin{document}
\title{Synchrotron Radiation Techniques and their Application to Actinide Materials}

 \author{R.~Caciuffo}
    \affiliation{European Commission\char44\ Joint Research Centre (JRC)\char44\ Postfach 2340\char44\ DE-76125 Karlsruhe\char44\ Germany}
      \affiliation{Istituto Nazionale di Fisica Nucleare\char44\ Via Dodecaneso 33\char44\ IT-16146 Genova\char44\ Italy}
 
 \author{G.~H.~Lander}
   \affiliation{European Commission\char44\ Joint Research Centre (JRC)\char44\ Postfach 2340\char44\ DE-76125 Karlsruhe\char44\ Germany}
 \affiliation{Interface Analysis Centre\char44\ School of Physics\char44\ University of Bristol\char44\ Tyndall Avenue\char44\
    Bristol\char44\ BS8 1TL\char44\ United Kingdom}

    \author{G.~\surname{van der Laan}}
   \affiliation{Diamond Light Source\char44\ Harwell Science and Innovation Campus\char44\ Didcot OX11 0DE\char44\ United Kingdom}

\date{\today}

\begin{abstract}
Research on actinide materials, both basic and applied, has been greatly advanced by the general techniques available from high-intensity photon beams from x-ray synchrotron sources. The most important single reason is that such x-ray sources can work with minute (e.g., microgram) samples, and at this level the radioactive hazards of actinides are much reduced. We start by discussing the form and encapsulation procedures used for different techniques, then discuss the basic theory for interpreting the results. By reviewing a selection of x-ray diffraction (XRD), resonant elastic x-ray scattering (REXS), x-ray magnetic circular dichroism (XMCD), resonant and non-resonant inelastic scattering (RIXS, NIXS), dispersive inelastic x-ray scattering (IXS), and conventional and resonant photoemission experiments, we demonstrate the potential of synchrotron radiation techniques in studying lattice and electronic structure, hybridization effects, multipolar order, and lattice dynamics in actinide materials.
\end{abstract}


\maketitle

\tableofcontents

\section{Introduction}
Actinides are the heaviest chemical elements available on a macroscopic scale. The complexity of their electronic structure often produces exotic physical properties, such as heavy-Fermi-liquid ground states \cite{coleman07} and unconventional superconductivity \cite{sarrao02,aoki07}. The richness of actinide physics has multiple origins \cite{moore09,caciuffo14}. The strong Coulomb repulsion between electrons in the open 5$f$ shell favors electron localization and the formation of large magnetic moments. This is contrasted by the effects of the hybridization between 5$f$ and conduction or neighboring-atom electronic states, promoting an opposite tendency towards itinerancy. This competition between localization and itinerancy results in frail, narrow-band 5$f$ states that can be driven by small perturbations towards one behavior or the other. The complexity of actinides is largely due to this instability. Moreover, when quantum fluctuations become large enough, magnetism disappears and new kind of order may develop, unveiling new physics beyond the "standard" Landau–Fermi liquid theory \cite{loehneysen07}.

Relativistic effects are a second source of complexity in actinide materials. Whereas the mass increase due to core electrons moving at almost 70\% of speed light affects the orbit size and the screening of the nuclear charge, strong spin–orbit coupling and the presence of unquenched orbital degrees of freedom give rise to a rich variety of phenomena involving dipole and higher-order electromagnetic multipole interactions. These interactions influence the dynamics of the system and may also drive exotic phase transitions with \textit{hidden} (non-dipolar) order parameters that have inspired many different theoretical models \cite{santini09,mydosh20}.

X-ray synchrotron radiation (SR) techniques provide powerful tools to unravel the complexity of actinide materials. These element- and shell-specific techniques probe spatial and temporal fluctuations of structural and electronic degrees of freedom, allowing one to observe hidden order parameters and to characterize elementary excitations with high sensitivity and resolution \cite{caciuffo21}. Contrary to neutron scattering, SR experiments only require samples on the microgram scale. This is important for actinides, as large quantities  considerably raise the radioactive inventory, thus breaching safety limits imposed at a general user facility, and large single crystals are also rare to obtain. 

\begin{figure}
   \centering
\includegraphics[width=1.0\columnwidth]{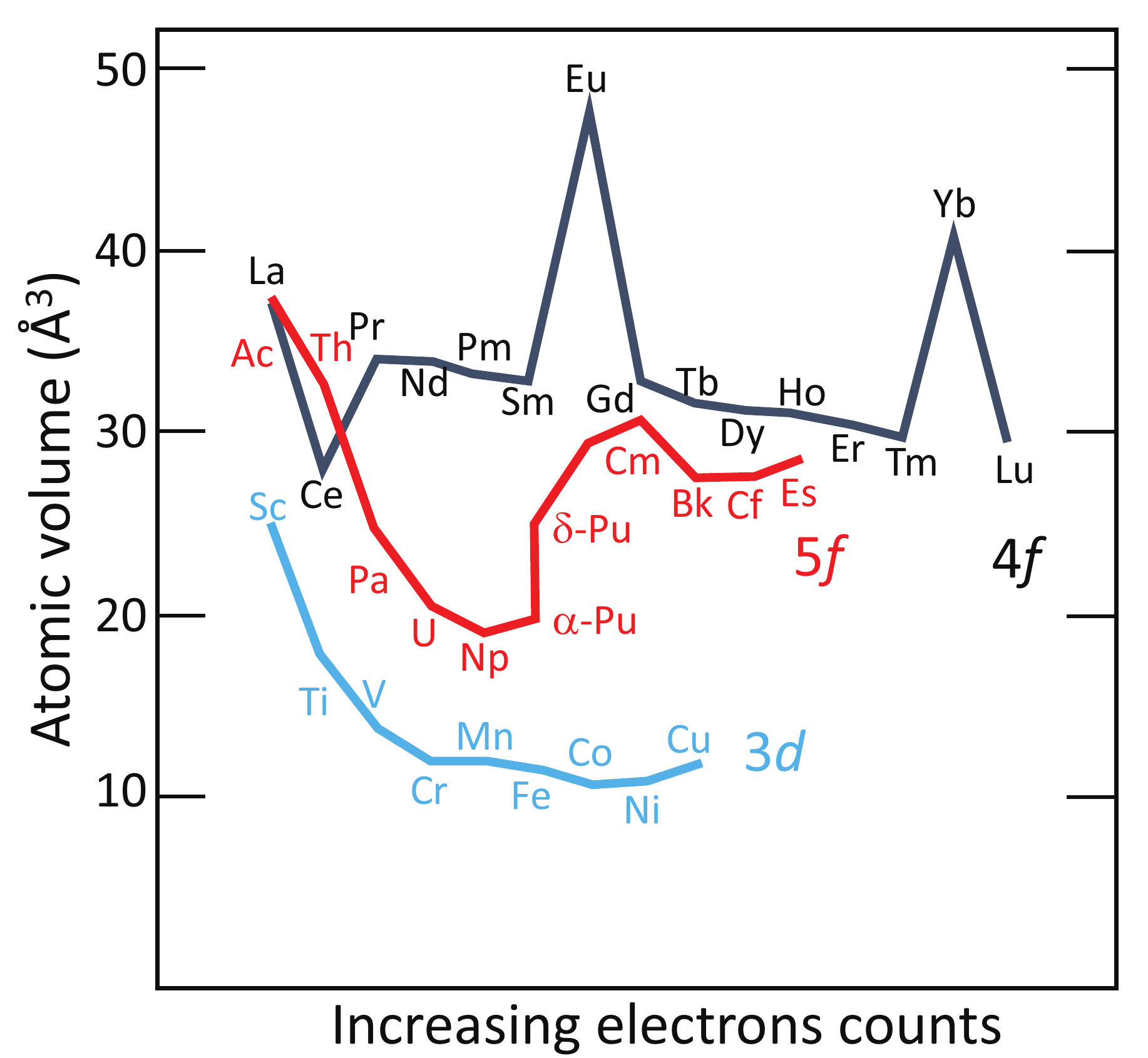}
  \caption{Atomic volume of the transition (3\textit{d}), rare-earth (4\textit{f}), and actinide (5\textit{f}) elements as a function of electron count. Note the unusual shape of the curve for the actinide series. In the case of Eu and Yb the large expansion is due to these elements being stable in the divalent state. All other rare-earth elements have the trivalent ground state.  Reproduced from \cite{caciuffo21}.
}
\label{atomicvol}
 \end{figure}

A central feature of the research on actinides can be illustrated by Fig.~\ref{atomicvol} and the atomic volume as a function of electron count across the 3\textit{d}, 4\textit{f}, and 5\textit{f} series of elements. For 3\textit{d} elements, the additional 3\textit{d} electrons result in a \textit{contraction} of the atomic volume as each additional electron adds to the cohesion of the element. In the 4\textit{f} (rare-earth) series, apart from the two divalent elements Eu and Yb, the atomic volume remains practically constant across the series. This is because the 4\textit{f} electrons are spatially located close to the nucleus and  not involved in the bonding. However, for the 5\textit{f} (actinide) series both behaviors are observed; an initial drop in the volume up to
$\alpha$-Pu, suggesting the 5\textit{f} states are contributing to the bonding and are therefore \textit{itinerant} for the light actinides, and then a strong expansion for other phases of Pu and through to heavier actinide elements, hence suggesting a \textit{localization} of the 5\textit{f} states from Am onwards.

Actinide elements, unlike most metals, crystallize in open-packed, low-symmetry structures.
The behavior of elemental plutonium, exhibiting six ambient pressure allotropes, is exemplary \cite{zachariasen55,zachariasen57,zachariasen63a,zachariasen63b,clark19}. A key factor in the metallic actinides systems is the position of the 5$f$ bands with respect to the conduction electrons, and in ionic systems we need to know the excited states and their energy differences, especially if covalency effects are induced by mixing the 5$f$ states with those of $p$ (or $d$) character from the neighboring anions. As we shall see in this review, there are many spectroscopic methods that can address these questions using x-ray techniques, including photoemission. Angle-resolved photoemission has been particularly valuable in this respect, as the results can be compared directly to theoretical predictions. SR techniques, providing a growing arsenal of tools for structural characterizations at atomic and mesoscopic level, as well as for electronic structure investigations, are now recognized as a major source of information on the properties of nuclear fuel, the stability of nuclear waste forms, and the behavior of actinide materials in the environment. Although these topics are of great industrial and societal importance, we will not cover them exhaustively in this review, which is mainly focussed on the physics.

\section{Samples and beamlines}\label{secbeam}
\begin{figure}
   \centering
\includegraphics[width=1.0\columnwidth]{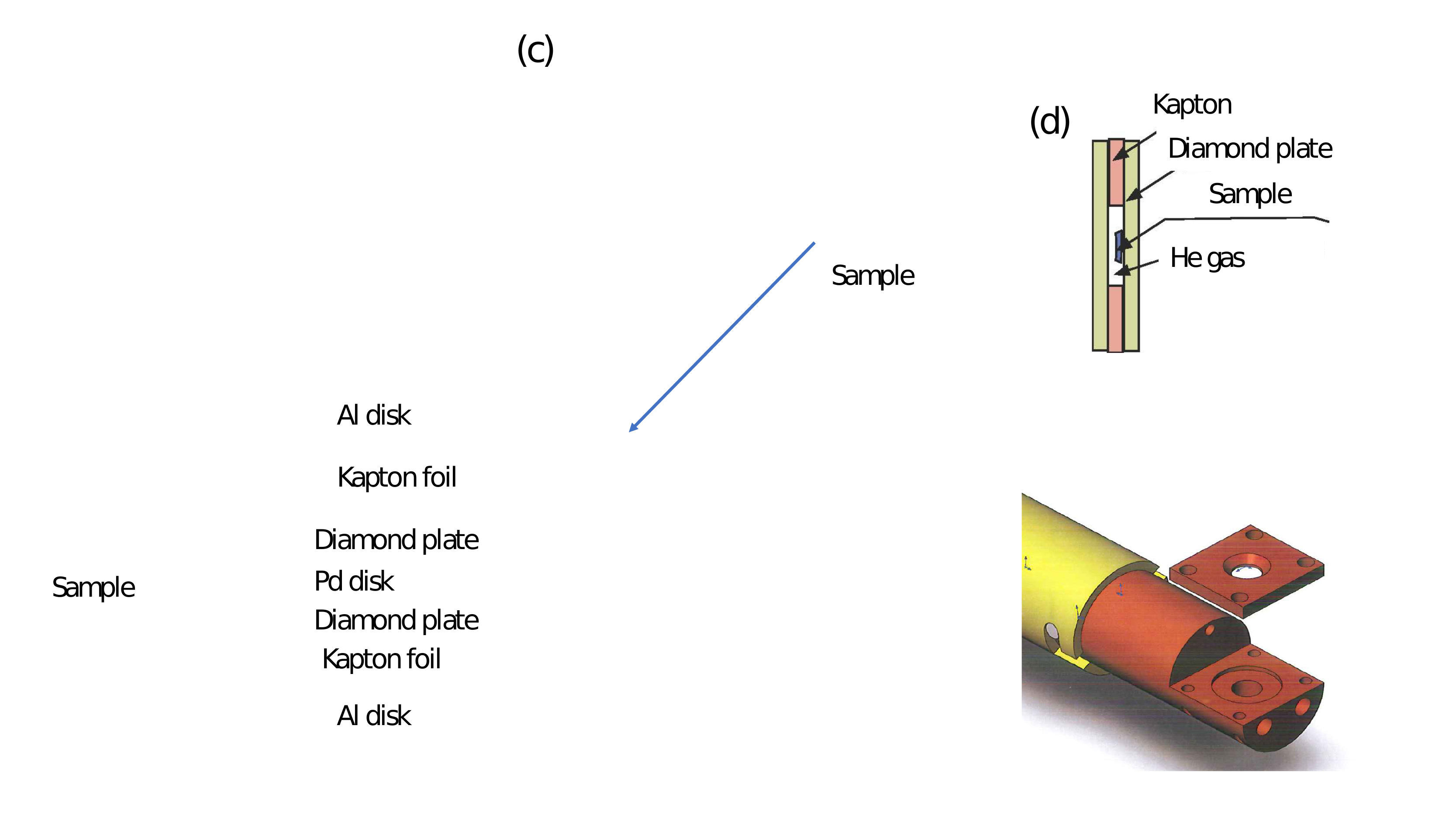}
  \caption{Hermetic sample holders used for investigating actinide materials with different synchrotron-radiation-based techniques. (a) Low-temperature powder x-ray diffraction \cite{hill13}; (b) inelastic x-ray scattering in transmission geometry \cite{walters15}; (c) x-ray magnetic circular dichroism \cite{lander19} and non-resonant inelastic x-ray scattering  \cite{sundermann20}; (d) inelastic x-ray scattering in reflection geometry \cite{maldonado16}. Panels (c) and (d) shows also a picture of the $^{248}$Cm (0.6$\times$0.8$\times$0.1 mm$^{3}$; m $\approx$ 650 $\mu$g) 
and NpO$_2$ (0.78$\times$0.56$\times$0.25 mm$^{3}$; m $\approx$ 1.2 mg) samples used for the studies described by \cite{lander19,maldonado16}.
}
\label{sampleholder}
 \end{figure}

The safety concerns raised by the radiotoxicity of the samples demand the implementation of strict rules and procedures to exclude any potential health risks and contamination of the beamline. Handling of radioisotopes should be undertaken in a properly regulated radiological facility and hermetic sample holders with multiple level of containment should be used. Some examples are shown in Fig.~\ref{sampleholder}. Panel (a) shows a photograph of the sample holder developed at the Joint Research Centre, Karlsruhe, for low-temperature x-ray diffraction (XRD) experiments on powder transuranium samples. It consists of a 1-mm diameter polyimide capillary, containing the powder material mixed with a low-viscosity epoxy resin, inserted into a drilled-out plexiglass rod, which is in turn enveloped within a 4-mm diameter polyimide tube \cite{hill13}. Panel (b) shows a zoomed-in view of a sample holder used for inelastic x-ray scattering (IXS) measurements in transmission geometry, with the sample sandwiched inside two single-crystal diamond plates, protected by Kapton foils and an aluminum case \cite{walters15}.  Figure~\ref{sampleholder}(c) shows a sample holder used for x-ray magnetic circular dichroism (XMCD) and non-resonant inelastic x-ray scattering (NIXS) measurements. The sample sitting in the middle of the aluminum support is covered by a Be window glued with epoxy resin and protected by a 6-$\mu$m thickness Kapton foil. The external cover of the capsule is kept in place by screws made from a diamagnetic material \cite{lander19}. Figure~\ref{sampleholder}(d) shows the drawings of a sample holder used for IXS experiments in reflection geometry. The sample is kept between two 0.5 mm thick single-crystal diamond slabs separated by a hollow Kapton foil \cite{maldonado16}.

The presence of windows covering the sample can degrade the signal-to-noise quality due to the absorption of incoming and outgoing photons or because a signal originating from the windows can overlap and swamp the one coming from the sample. Their material and thickness must therefore be chosen with care. Moreover, especially if the windows are not transparent, hitting a very small sample with a x-ray beam with dimensions of microns can be challenging. It is then useful to glue the sample onto a single crystal substrate and to monitor the intensity of a Bragg peak from the substrate during a $x$-$y$ position scan, looking for the dip of minimum intensity expected at the sample position. Care should be taken if using any epoxy glue in a confined volume with $\alpha$-emitting samples, as the glue may radiolyze producing reactive products such as HF \cite{mannix99a}. 

 \begin{figure}
   \centering
\includegraphics[width=1.0\columnwidth]{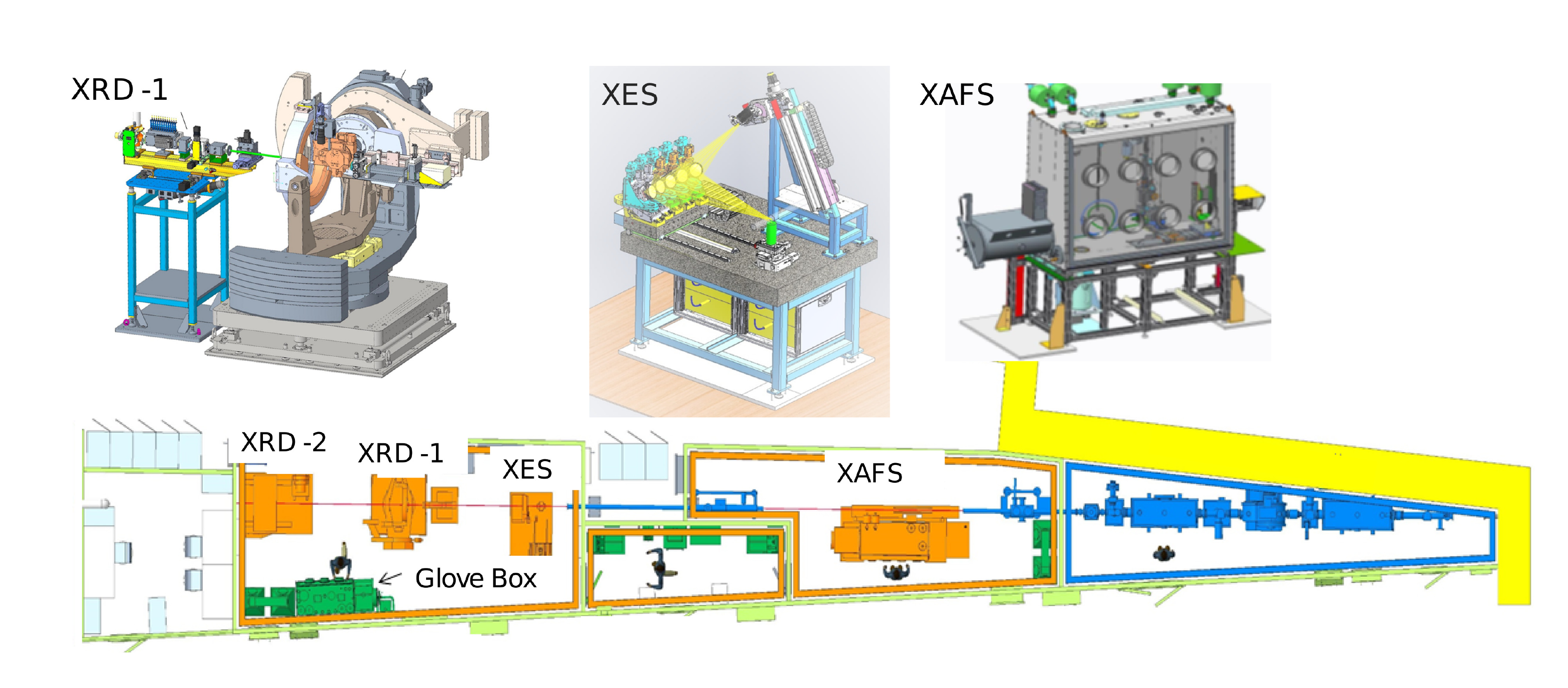}
  \caption{Schematic layout of the ROBL-II beamline at the European Synchrotron Radiation Facility (ESRF), Grenoble. The insets show the powder x-ray diffraction, x-ray emission spectroscopy, and the x-ray absorption fine-structure stations. Adapted from \cite{scheinost21}.
}
\label{ROBL}
 \end{figure}

Pioneering work on actinides stimulated the construction of \textit{dedicated} beamlines: BL27B
\cite{konishi96} at the Photon Factory in Tsukuba and BL23SU \cite{saitoh01,saitoh12} at SPring-8, in Sayo, Hy\={o}go Prefecture, Japan; BL 11-2 \cite{bargar02} at the Stanford Synchrotron Radiation Laboratory (SSRL), Menlo Park, California; MARS \cite{solari09} at SOLEIL, Gif-sur-Yvette, France; CAT-ACT \cite{dardenne09,rothe12,zimina17,schacherl22a} at the Karlsruhe Research Accelerator (KARA, former ANKA) in Karlsruhe, Germany; MicroXAS \cite{borca09} at the Swiss Light Source (SLS), Villigen, Switzerland; and ROBL-II \cite{kvashnina16,scheinost21} at the European Synchrotron Radiation Facility (ESRF), Grenoble, France. Generally, dedicated beamlines have several experimental stations. For instance, the ROBL-II beamline, located at a bending magnet  of the ESRF (Fig.~\ref{ROBL}), has four stations dedicated to (i) fluorescence and transmission detection for x-ray absorption fine-structure (XAFS) spectroscopy, including (conventional) x-ray absorption near-edge structure (XANES) and extended x-ray absorption fine-structure (EXAFS) spectroscopies, (ii) high-energy-resolution fluorescence-detection x-ray absorption near-edge spectroscopy (HERFD-XANES), x-ray emission spectroscopy (XES) and resonant inelastic x-ray scattering (RIXS), (iii) powder x-ray diffraction (PXRD), surface-sensitive crystal truncation rod (CTR) and resonant anomalous x-ray reflectivity (RAXR) measurements, and (iv) single crystal x-ray diffraction (SCXRD) and in-situ/in-operando PXRD.

The construction of dedicated  actinide beamlines at many synchrotrons, as discussed above, has greatly increased the number of experiments that have been performed over the last twenty years on actinide materials. In most cases the capability of these special actinide beamlines is sufficient, but they are usually designed to perform a variety of different tasks, so they are often not at the cutting edge of synchrotron research. It is thus important in this context that the synchrotron staff can define conditions at which actinides can be used at any beamline. This has been our experience at the ESRF, where the allowed quantities of each radioactive isotope are well defined, and we have developed, together with the staff, special sample holders, several shown in Fig.~\ref{sampleholder}. Diamond Light Source (Harwell, UK) has a dedicated radioactive materials laboratory to prepare hot samples for measurements at the beamlines.

Examples of experiments on transuranium samples that have been performed on general beamlines at the ESRF  are high-pressure diffraction \cite{lindbaum01,heathman2005}, resonant elastic x-ray scattering (REXS) \cite{paixao02}, XMCD \cite{halevy12,wilhelm13,magnani15,magnani17,lander19}, IXS  \cite{wong03,maldonado16}, NIXS \cite{sundermann20}, and RIXS \cite{heathman10}.
Of course, the amount of the allowed radioisotopes will then be less than at a dedicated actinide beamline. For example, at a general beamline at the ESRF the radioactivity should not exceed 10 $\mu$Ci (370 kBq), which gives an amount of $^{239}$Pu of 160 $\mu$g. On the other hand, at the ROBL beamline experiments can be performed on samples with a maximum activity of 185 MBq, a factor 500 higher than on non-dedicated beamlines. 

Some synchrotrons state that they will not allow any radioactive material on the beamlines. However, our experience has been that one has to convince the management of the need for the special techniques available, and work with them constructively to satisfy the relevant safety conditions.

\section{Introduction to the  theory}
\label{sec:theory}

 By providing a theoretical foundation of the most explored x-ray spectroscopies we are able to point out the conceptually common aspects and classify the different processes (elastic and inelastic, resonant and non-resonant, first- and second-order transitions, coherent and incoherent).

The various kinds of x-ray spectroscopies practiced at synchrotrons around the globe have much in common: clearly they are all based on photon absorption or scattering, as opposed to, say, electron or neutron scattering which do entail different mechanisms. Subject to Fermi's golden rule, photon processes are either direct transitions from initial to final state, or they are second-order coherent transitions involving an intermediate state. In a few exceptional cases they are higher-order processes like the ones in non-linear optics. 
Here we provide a basic introduction that serves to put the different spectroscopic techniques into perspective and place them on a common platform. 
One can make a distinction between the various electric- and magnetic-multipole transitions, such as dipole- and quadrupole transitions, each of which can be excited with circularly or linearly polarized x-rays, amounting to different selection rules.
Furthermore, one can distinguish between non-resonant and resonant transitions, which can be either elastic or inelastic. 
 In a resonant transition a core electron is excited into a discrete state above the Fermi level, giving a strong intensity enhancement as well as a sensitivity to the local environment. Bragg scattering is an example of elastic scattering, which has a non-resonant term (Thomson scattering) and a resonant one known as anisotropy of the tensor of susceptibility (ATS) or resonant elastic x-ray scattering (REXS). Inelastic scattering can also be divided up into non-resonant inelastic x-ray scattering (NIXS) and resonant inelastic x-ray scattering (RIXS).
 
 The following theoretical description is based on quantum mechanics and provides a step-by-step approach to x-ray spectroscopies. After presenting the electronic Hamiltonian that empowers us to obtain the initial and final states, the interaction Hamiltonian is presented. By Fermi's golden rule, the transition contains  first-order (direct) and second-order components, which can be further classified according to the number of times that the vector potential occurs.

\subsection{Interaction of radiation with electronic matter}\label{secInteraction}

\subsubsection{The Hamiltonian}\label{secHamiltonian}

The Hamiltonian in the nonrelativistic limit with relativistic corrections for photon-matter interaction can be separated into an electronic, radiation, and interaction part,
\begin{equation}
H   = H_{\mathrm{el}} + H_{\mathrm{rad}} + H_{\mathrm{int}} .
\end{equation}
For $n$ electrons moving about a point nucleus of charge of an atom, the Hamiltonian for the electronic part  can be written in the central field approximation as
\begin{equation}
H_{\mathrm{el}} =  \sum_{n=1}^{N}  \left[ \frac{\mathbf{p}_n^2}{2m} + V(\mathbf{r}_n) + \frac  {e \hbar}{2m^2c^2} \, \mathbf{s}_n \cdot ( \nabla V(\mathbf{r}_n) \times \mathbf{p}_n) \right] ,
\label{eq:H-el}
\end{equation}
where  $e$ is the elementary charge, $m$ is the electron rest mass, and $c$ is the speed of light. 
The first term is the kinetic energy, which contains the momentum operator ${\mathbf{p}}_n$ = $-i \hbar \nabla_n$.
The second term is the potential energy, $V$, which depends on the position vector
 ${\mathbf{r}}_n$  of the $n$-th electron. This potential energy collects terms such as the electrostatic (Coulomb) interactions and external (magnetic and crystal) fields.  The last term is the spin-orbit interaction, in which $\mathbf{s}_n$ is the spin vector operator, which finds its origin in the Dirac equation. 

The Hamiltonian for the radiation field is
\begin{equation}
H_{\mathrm{rad}} =  \sum_{\mathbf{k} \bm{\varepsilon}} \hbar \omega_\mathbf{k}
 \left[ a^\dagger(\mathbf{k} \bm{\varepsilon}) a(\mathbf{k}\bm{\varepsilon}) + \frac{1}{2} \right]  ,
\end{equation}
where $\hbar \omega_\mathbf{k}$ is the energy of a photon with wavevector $\mathbf{k}$ and polarization vector $\bm{\varepsilon}$. The $a^\dagger$ and $a$ are photon creation and annihilation operators, respectively.

An electromagnetic field consists of two vector fields, an electric field $ \mathbf {E} (\mathbf {r} )$ and a magnetic field $ \mathbf {B} (\mathbf {r} )$. Both are time-dependent vector fields that in vacuum depend on the vector potential field $ \mathbf {A} (\mathbf {r} )$ and the scalar field $ \phi (\mathbf {r} )$,
\begin{align}
\mathbf {B} (\mathbf {r}  ) & =  \nabla \times \mathbf {A} (\mathbf {r}  ) , \\
\mathbf {E} (\mathbf {r}  ) & = - \nabla \phi (\mathbf {r}  ) - \frac{\partial \mathbf {A} (\mathbf {r} )}{c \partial t} .
\end{align}
Choosing the Coulomb gauge, for which $\nabla \cdot \mathbf{A} = 0$, makes $\mathbf{A}$ into a transverse field. The Fourier expansion of the vector potential is then
\begin{equation}
\mathbf{A}(\mathbf{r}) = \sum_{\mathbf{k} \varepsilon} \sqrt{ \frac{\hbar}{2 \omega_{\mathbf{k}} \mathrm{\textsf{V}} \epsilon_0}}
\left[ a^\dagger(\mathbf{k} \bm{\varepsilon}) \bm{\varepsilon}^\ast  e^{-i \mathbf{k} \cdot \mathbf{r}}
+ a (\mathbf{k}\bm{\varepsilon} ) \bm{\varepsilon}  e^{i \mathbf{k} \cdot \mathbf{r}}  \right] ,
\label{eq:A-rad}
\end{equation}
where \textsf{V} is the normalization volume, $\epsilon_0$ is the vacuum permittivity. 
The time evolution is obtained by replacing $(\mathbf{k} \cdot \mathbf{r})$ by ($\mathbf{k} \cdot \mathbf{r}-\omega_{\mathbf{k}} t$).

For x-rays, the inverse of $\mathbf{k}$ is large compared to the core orbitals involved  in the transition,
and for ($\mathbf{k} \cdot \mathbf{r} \ll 1$) Eq.~(\ref{eq:A-rad}) simplifies to the dipole approximation
$ \mathbf{A}(\mathbf{r}) \sim \sum_{\mathbf{k} \varepsilon} 
[ a^\dagger(\mathbf{k}\bm{\varepsilon}) \bm{\varepsilon}^\ast + a (\mathbf{k}\bm{\varepsilon} ) \bm{\varepsilon}  ] $.

The  interaction Hamiltonian that  involves terms with the vector potential $\mathbf{A}$ is \cite{blume85,blume88,altarelli13,schulke07}
\begin{align}
H_{\mathrm{int}} = &   \sum_{n=1}^N   \left[
\frac{e^2}{2mc^2}  \mathbf{A}^2 ( \mathbf{r}_n) 
 -  \frac{e}{mc} \mathbf{A}( \mathbf{r}_n) \cdot  \mathbf{p}_n \right.  \nonumber \\
  & \makebox[0.2in]{} -   \frac{e \hbar}{mc} \,  \mathbf{s}_n  \cdot ( \nabla \times \mathbf{A} (\mathbf{r}_n ) ) \nonumber \\ 
  & \makebox[0.2in]{} + \left.  \frac{e \hbar}{2 m^2 c^3} \, \mathbf{s}_n \cdot
   [ \frac{\partial \mathbf{A}(\mathbf{r}_n)}{\partial t}  \times ( \mathbf{p}_n - \frac{e}{c} \mathbf{A}(\mathbf{r}_n))]
 \right]  \nonumber \\
  &  \equiv  H'_1 + H'_2 + H'_3 + H'_4  .
 \label{eq:H-int}
\end{align}
For brevity, we will often omit  the summation over $n$ in the following.

To shed some light on Eq.~(\ref{eq:H-int}), the generalized momentum  in the presence of an electromagnetic field is
 $\mathbf{\Pi}$ = ${\mathbf{p}}-\frac{e}{c} {\mathbf{A}}$. 
 Expanding the non-relativistic kinetic energy operator
$\frac{1}{2m} ({\mathbf{p}}-\frac{e}{c} {\mathbf{A}})^2$
gives the terms $H'_1$ and $H'_2$ in $H_{\mathrm{int}}$, while the term
$\mathbf{p}^2/2m$ is captured in $H_\mathrm{el}$.
The $H'_1$ interaction gives non resonant photon scattering, $H'_2$ interacts with the charge, and gives an electric-multipole transition.

The term $H'_3$ with $\mathbf{s}  \cdot (\nabla \times {\mathbf{A}})$ gives an  interaction of the spin
$\mathbf{s}$  with the magnetic field $\mathbf{B} = \nabla \times \mathbf{A}$ of the radiation, and leads to magnetic  transitions. 
Compared to charge scattering, the magnetic scattering is smaller by a factor $(\hbar \omega / mc^2)^2$ (of the order of 10$^{-4}$ at the $M$ edges of uranium). 

 The origin of the relativistic term $H'_4$ is the spin-orbit interaction of the electron spin and the radiation field, not to be confused with the atomic spin-orbit interaction given in Eq. (\ref{eq:H-el}). 
 

\subsubsection{Kramers-Heisenberg formula}

\begin{figure}
	\includegraphics[width = 0.45\textwidth]{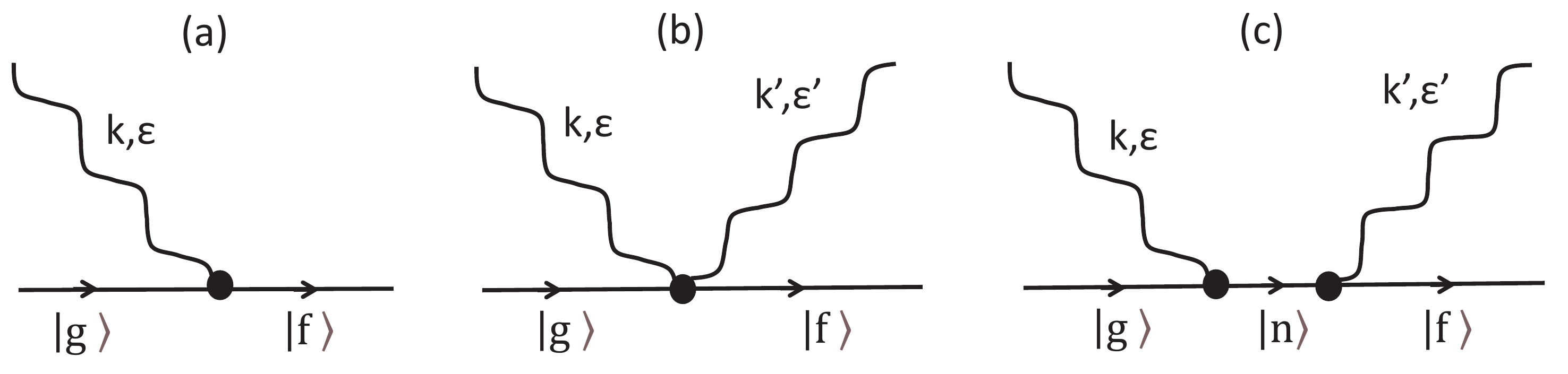}
	\caption{ Feynman diagrams representing the  scattering amplitude.
	(a) First-order diagram for photon absorption with interaction $H'_2$ or $H'_3$;
	(b) first-order diagram for photon scattering with interaction $H'_1$ or $H'_4$;
	and
	(c) second-order diagram for photon scattering (see text). Not shown is another second order Feynman diagram with emission followed by absorption, that contributes to the non-resonant scattering.
}
	\label{fig:Diagrams}
 \end{figure}          

Armed with the interaction Hamiltonian, we are  ready to write down the transition probability.
Applying Fermi's golden rule in first- and second-order perturbation, we arrive at the Kramers-Heisenberg (KH) formula for the number of transitions per unit time \cite{sakurai67}
\begin{align}
w & =  \frac{2 \pi}{\hbar}  \sum_f \left| 
 \langle f  | H_{\mathrm{int}} | g \rangle  
+ \sum_n 
  \frac{ \langle f | H_{\mathrm{int}} | n \rangle \langle n | H_{\mathrm{int}} | g \rangle}{E_g - E_n + \hbar \omega + i \Gamma_n/2}  \right|^2
  \nonumber \\
 & \makebox[0.2in]{}  \times  \delta(E_g - E_f  + \hbar \omega - \hbar \omega^\prime)  ,
\label{eq:Fermis}
\end{align}
\noindent
where the delta function indicates the energy conservation for initial state $| g \rangle$ and final state $| f \rangle$. 
Also implicitly included in the initial state is a photon that is annihilated by 
the vector potential $\mathbf{A}$ given in Eq.~(\ref{eq:A-rad}).

Equation (\ref{eq:Fermis}) contains a direct and an indirect transition; the latter contains an intermediate (virtual) state $| n \rangle$.
The modulus square over these two transitions terms allows for 
interference between different 
$|n \rangle$ intermediate states leading to the same  final state $|f \rangle$.

A formula equivalent  to Eq.~(\ref{eq:Fermis}) was first obtained by H.~A.~Kramers and W.~Heisenberg in 1925 using the correspondence principle \cite{kramers25}. 
A classical example is provided by the Rayleigh scattering, which describes the elastic scattering of light with wavelengths much longer than the particle size,
where $| f \rangle $ = $| g  \rangle $, $\hbar \omega =  \hbar \omega'$.
The interaction  $ \bm{\varepsilon} \cdot \mathbf{p}$  (dipole approximation) with the  incoming and outgoing photon  results  in a scattering cross section that varies as the inverse fourth power of the wavelength (Rayleigh's law). This theory explains why the sky is blue and the sunset is red.

In the non-resonant case, where the incident photon energy is far away from the atomic binding energy, we can ignore the second-order perturbation term, and the scattering is due to  $\mathbf{A}^2$, which in the dipole approximation becomes 
 $\bm{\varepsilon}^{\prime \ast} \cdot \bm{\varepsilon}$, which is independent of the propagation direction. This gives the  Thomson scattering, which is insensitive to the nature of the electron bonding. The Thomson cross section for light scattering by free (unbound) electrons is given by 
\begin{equation}
\frac{d \sigma}{d \Omega} = r^2_0 (\bm{\mathbf{\varepsilon}}^{\prime \ast} \cdot \bm{\mathbf{\varepsilon}} )^2 ,
\label{eq: Thomson}
\end{equation}
where $r_0 \equiv e^2/m c^2  \equiv   2.818 \times 10^{-15}~\mathrm{m}$ is the classical electron radius or Thomson scattering length,
$e$ is the elementary charge, $m$ is the electron mass, and $c$ is the speed of light.
The  cross-section is independent of $\omega$, but has a strong polarization dependence as a function of the scattering angle. For linearly polarized x-rays the intensity vanishes at a  scattering angle of $90^\circ$, which can be used to suppress the elastic scattering when measuring other types of scattering. We have more to say about Thomson scattering in Sec.~{\ref{sec:Thomson}.

The KH formula can also be applied to inelastic scattering of light, with $\omega \ne \omega'$ and $| f \rangle \ne | g \rangle$, where in the optical region  it is called the Raman effect.

Here, our  aim is to describe different types of x-ray spectroscopies, and
the KH formula facilitates  just this by plugging in the various interactions. 
 As seen from Eq.~(\ref{eq:A-rad}), the vector potential $\mathbf{A}$ is linear in the creation and annihilation operators of the photon.
Thus $\mathbf{A} |g \rangle$ annihilates a photon in the initial state, whereas $\langle f | \mathbf{A}$ creates a photon in the final state.

SR studies imply the presence of incoming photons, either  without or with emission of a secondary photon, corresponding to absorption and scattering, respectively. Emission-only processes, such as radiative de-excitation and bremsstrahlung, fall beyond the frame of these studies.
We will also neglect the effects of electron recoil ($\hbar \omega \ll m_e c^2$) that occurs for Compton scattering.

We will first consider the first-order perturbation, shown in Fig.~\ref{fig:Diagrams}(a), which corresponds to photon absorption. 
The vector potential $\mathbf{A}$  is linear for the interaction terms $H'_2$ and $H'_3$.
The interaction $H'_2 \sim \mathbf{p} \cdot \mathbf{A}$ gives electric and magnetic multipole transitions (see Sec.~\ref{sec:mat-el}), in which a photon is absorbed under excitation of an electron. This process encompasses both absorption (XAS) and photoemission (PES) where an electron is excited into a discrete and continuum state, respectively.
 The interaction term $H'_3 \sim \mathbf{s}  \cdot (\nabla \times {\mathbf{A}}) = {\mathbf{s}} \cdot   {\mathbf{B}}$ gives the pure magnetic absorption, which is much smaller.

The first-order perturbation is quadratic in $\mathbf{A}$  for the interaction terms $H'_1$ or $H'_4$, which describes photon scattering (photon-in photon-out). This is shown  by the so-called ``seagull'' diagram in Fig.~\ref{fig:Diagrams}(b).
$ H'_1 \sim \mathbf{A}^2$ gives a non-resonant  process since it does not contain the momentum operator $\mathbf{p}$. Besides the already mentioned elastic process of Thomson scattering, there is also an inelastic process (NIXS),  see Sec.~\ref{sec:NIXS}.

Next, we consider the second-order perturbation term in the KH formula, as shown in Fig.~\ref{fig:Diagrams}(c).
 This contains a scattering process involving two photons, i.e., twice the interaction $ \mathbf{A} \cdot \mathbf{p}$, and an intermediate state $| n \rangle$, such as in REXS and RIXS.
Alternatively, there can also be second-order interaction processes that are first-order in $\mathbf{A}$, such as resonant photoemission spectroscopy (RPES), in which the photon absorption (interaction $ \mathbf{A} \cdot \mathbf{p}$) is followed by a Coulomb decay  (see Sec.~\ref{sec:rpes}).
All the above processes will be treated in more detail below.

Finally, non-linear  optical effects, such as the third-order process of second-harmonic generation (SHG), are also possible, although they remain largely unexplored in the x-ray region due to the relatively low photon flux compared to  lasers in the optical region.
The recent advent of free-electron lasers (FELs) in  energy ranges from extreme ultraviolet to x rays will allow us to explore these effects involving core-level resonances. 
In SHG an incident photon excites an electron of the atom, which is promoted to an empty state, and a second photon excites it to the next level. The state then de-excites to the equilibrium ground state under emission of a photon that, due to energy conservation, has twice the energy and frequency of the original photons.
As in the case of natural circular dichroism, the process  of SHG for circular dichroism is  parity-odd and time-even, which allows to measure the helicity of a material from the piezoelectric crystal class \cite{vanderlaan21}, such as the noncentrosymmetric uranium(IV)  fluoride U$_3$F$_{12}$(H$_2$O).
 

\subsubsection{Green's function approach}

In the KH formula, the denominator containing the intermediate states can  be written in terms of the Green's
function, which is also referred to as the intermediate-state propagator, which describes the
system in the presence of a core hole  
\begin{equation}
G(z) = \frac{1}{z - H} = \sum_n \frac{ | n \rangle \langle n |}{z - E_n} ,
\label{eq:Green}
\end{equation}
where $ | n \rangle$ forms a complete basis set and $z = E_g + \hbar \omega + i \Gamma/2$.
It reduces  the scattering amplitude to the compact expression
\begin{equation}
F_{gf} = \langle f | H_{\mathrm{int}}^{ \dagger} G(z) H_{\mathrm{int}} | g \rangle  .
\end{equation}
This formulation is frequently encountered in both RPES and RIXS.

\subsubsection{Initial- and final-state wave functions}

The atomic wave functions can be obtained from the electronic Hamiltonian given in Eq.~(\ref{eq:H-el}), in which $V(\mathbf{r})$ may contain external fields such as  magnetic, electric, and crystal field.

The electronic wave functions  $\psi_i$ for the initial states are obtained by finding the eigenvalues  $E_i $ of the Hamiltonian using $H_{\mathrm{el}} \psi_i = E_i \psi_i $.
By including the core hole potential in  $H_{\mathrm{el}}$, it can also be used to obtain the final states $|f \rangle$ = $c^{}_{m' \sigma'}\ell^{\dagger}_{m\sigma} | g \rangle$, where $c^{}_{m' \sigma'}$ is the annihilation operator of a core electron $c$ with quantum numbers $m'$ and $\sigma'$, 
 and $\ell^{\dagger}_{m\sigma}$ is the creation operator of an electron $\ell$ in a (partly) empty shell with quantum numbers $m$ and $\sigma$. 
 
 The wavefunction can be separated into an angular and radial part. The angular part depends on the angular
quantum numbers of the basis states of the configuration and are independent of the radial wave functions. 
General analytical  methods for calculating angular coefficients have been
computerized by \cite{cowan81}. The basis wave functions are an anti-symmetrized product of
one-electron functions, so-called Slater determinants. In spherical symmetry these wave functions are eigenfunctions
of the total angular momentum $J$ and its components
$M_J$. 

Upon application of a magnetic field $\textbf{B}$ along the quantization direction $\hat{z}$ the $(2J+1)$-fold degenerate ground state splits up into its $M_J$ sublevels with energy $\mathrm{g} \mu_{\mathrm{B}}B M_J$ of which only the level $M_J = -J$ is populated at $T$ = 0 K, where
$\mu_{\mathrm{B}} = e \hbar/(2m)$ is the Bohr magneton and $\mathrm{g}$ is the Land\'e $\mathrm{g}$ factor.
 At finite temperatures the levels are populated by the  $Z^{-1} \exp (-\mathrm{g} \mu_{\mathrm{B}}B M_J/ k_{\mathrm{B}} T)$ probability factor, where $Z$ is the partition function and $k_{\mathrm{B}}$ is the Boltzmann factor. 

If the spin-orbit interaction is much smaller than the electrostatic interactions, the states are characterized by quantum numbers $\alpha LS$, where $\alpha$ is a suitable quantity for distinguishing between terms having the same values of the orbital and spin angular momenta $L$ and $S$.

Two different kind of basis sets, i.e., $LS$- and $jj$-coupled wave functions, are commonly used. 
The electrostatic interaction is diagonal
in $LS$ coupling, whereas the spin-orbit interaction is diagonal in $jj$ coupling. In the $LS$-coupling
scheme, the various one-electron orbital momenta $\ell$ are coupled together successively to give a total orbital momentum,
and the various one-electron spin momenta $s$
are coupled to give a total spin,
\begin{equation}
\{[(( \ell_a s_a) L_a S_a, \ell_b s_b) L_b S_b, \cdots , \ell_n s_n] L_n S_n \} J_n ,
\end{equation}
with triangulation rules such as $L_b = |L_a - \ell_b|, \cdots, L_a + \ell_b$, or in short $\Delta(L_b, L_a, \ell)$.

In the other scheme of $jj$ coupling, each $\ell$ and $s$ are
coupled to give a total angular momentum $j$, and the
various $j$ are then coupled to give successive values of $J$,
\begin{equation}
\{ [ ( \ell_a s_a j_a) J_a, ( \ell_b s_b j_b)]J_b, \cdots, ( \ell_n s_n j_n) \} J_n .
\end{equation}

In practice, neither electrostatic nor spin-orbit interaction can be neglected. In this intermediate coupling case,
any multielectronic wavefunction can be written as a linear combination $\psi = \sum_{LSJ} c_{LSJ}  \psi_{LSJ}$ = $\sum_{jj'J} c_{jj'J}  \psi_{jj'J}$, where $c_{LSJ}$ and $c_{jj'J}$ are wave function coefficients \cite{vanderlaan96}.

\subsubsection{Specific x-ray spectroscopies}
In the remainder of this theory section, we will give the fundamentals in terms of transition probabilities for the photon-matter interaction of the various specific x-ray techniques. In order to create some structure in this, we will use the term expansion in the KH formula, thereby gradually building toward more complexity. We start with describing processes involving first-order absorption [Fig.~\ref{fig:Diagrams}(a)], such as XAS, XMCD, and PES. Thereafter, we move on to the first-order scattering [Fig.~\ref{fig:Diagrams}(b)], such as Thomson scattering and NIXS, and finally the second-order scattering [Fig.~\ref{fig:Diagrams}(c)], such as REXS and RIXS, which contain an intermediate state and allow for resonant enhancement. However, we will not keep the same order in the experimental part. There we start with techniques that reader are most familiar with, such as XRD, and gradually going to more ‘exotic’ techniques.

\subsection{X-ray absorption spectroscopy (XAS)}

\subsubsection{Transition matrix element}
\label{sec:mat-el}

XAS represents an atomic transition from a core level into an unoccupied, discrete state below the continuum. Transitions to states at higher energies are treated in Sec. \ref{XAFS}.
As the excited electron remains in the solid and is not directly detectable, the XAS is measured either by the x-ray transmission probability or by the decay of the created core holes, which gives fluorescence, Auger, and secondary electron yield \cite{vanderlaanCCR14}.

The XAS process arises from the $\mathbf{A} \cdot \mathbf{p}$ term in $H_{\mathrm{int}}$
with its transition probability given by the first-order term in Fermi's golden rule
\begin{equation}
T_{gf} = \frac{2 \pi}{\hbar} | \langle f | \mathbf{A} \cdot \mathbf{p}  |g \rangle |^2 \, \delta(E_g - E_f + \hbar \omega) .
\end{equation}

The matrix element $\langle f | {\mathbf{A}} \cdot {\mathbf{p}} |g \rangle$ can also feature as the  initial step in  second-order processes, which makes it rewarding to examine  its properties.
Converting it to the length form by using the commutator between the position vector $\mathbf{r}$ of an electron and the Hamiltonian of the unperturbed atom, $H$, 
\begin{equation}
{\mathbf{p}} \equiv -i \hbar \nabla = \frac{m}{i\hbar} \, \left[ {\mathbf{r}}, H \right] ,
\end{equation}
\noindent
and using the annihilation part of the vector potential, ${\mathbf{A}} ({\mathbf{k}},{\mathbf{r}}) = {\mathbf{\bm{\varepsilon}}} \, e^{i {\mathbf{k}} \cdot {\mathbf{r}}}$, gives
\begin{equation}
\langle f |  {\mathbf{A}} \cdot {\mathbf{p}} | g \rangle
= \frac{m}{i\hbar} (E_f -E_g)   \langle f |  ( {\mathbf{\bm{\varepsilon}}} \cdot {\mathbf{r}} ) e^{i {\mathbf{k}} \cdot {\mathbf{r}}} | g \rangle .
\end{equation}

Next, we  show how to factorize the scalar operator $\mathbf{A} \cdot \mathbf{p} \sim ( \bm{\varepsilon} \cdot \mathbf{r} ) e^{i \mathbf{k} \cdot \mathbf{r}}$ in the matrix element. The transition operator can be factorized into a part that depends only on the  geometry of the experiment (which is commonly called the \textit{photon part}, or sometimes the geometrical or angular factor) and another part that depends only on the materials properties (which is commonly called the \textit{matter part} or sometimes the dynamic or physical factor). For simplicity we assume spherical symmetry and neglect crystal field interaction. This gives a proper way to assign the different  multipole moments of the electromagnetic radiation field.
Furthermore, it offers a first glance on the rather powerful method of moment recoupling, which is usually done using spherical tensor algebra \cite{vanderlaan06a}.

Cartesian and spherical components of a dipole $\mathbf{r} = (x, y, z)$ are related by
$r_{\pm1} = \mp(x \pm iy)/ \sqrt{2}$, $r_0 = z$.
Cartesian tensors are conceptually more simple and closely related to the natural coordinate system of the crystal, but become unwieldy for higher multipole moments.
Spherical tensors are irreducible and follow general rules of angular momentum algebra \cite{edmonds57}.
In both formulations, the aim is to separate the cross section into a matter part and a geometric part. While the resonant amplitude of the matter part requires a detailed knowledge of the electronic wavefunction, the geometric part  can be factored out. This also allows us to handle the polarization and angular part of the matrix elements.

Spherical tensors can be coupled together using the tensor coupling theorem
\begin{equation}
 [  C^{(z')}  , C^{(z'')}  ] ^{(z)}_\zeta \equiv \sum_{\zeta ' \zeta''} C^{(z')}_{\zeta ' }  C^{(z'')}_{ \zeta''}  
C^{z \zeta}_{z' \zeta ', z'' \zeta''},
\label{eq:tensorcoupling}
\end{equation}
where
$C^{(z)}_{\zeta}  = \sqrt{4 \pi/(2z+1)} \, Y^{(z)}_{\zeta}$ is the renormalized spherical harmonic and
$ C^{z \zeta}_{z' \zeta ', z'' \zeta''}  = \langle z' \zeta', z'' \zeta'' | z \zeta \rangle$
is the Clebsch-Gordan coefficient.

For $({\mathbf{k}} \cdot {\mathbf{r}}) \ll 1$, the exponential factor in the vector potential $\mathbf{A}$ can be expanded as
\begin{equation}
 e^{i {\mathbf{k}} \cdot {\mathbf{r}}}
 = \sum_{L=0}^{\infty} \frac{ (i \mathbf{k} \cdot \mathbf{r})^L }{L!}
=  1+ i {\mathbf{k}} \cdot {\mathbf{r}} - \frac{1}{2} ({\mathbf{k}} \cdot {\mathbf{r}})^2+\cdots  ,
\label{e-ikr}
\end{equation}
and using the tensor coupling theorem of Eq.~(\ref{eq:tensorcoupling}) we can  decompose the transition operator into multipole moments of rank $z$, which  couple  to a scalar product.
\begin{align}
( \bm{\varepsilon}  \cdot \mathbf{p}  ) e^{i {\mathbf{k}} \cdot {\mathbf{r}}} & \propto  i^L
\sum_{L=0}^{\infty}\sum_{z=L}^{L+1} \sum_{\zeta=-z}^{+z} \left[ [ \bm{\varepsilon}, \mathbf{k}^{(L)}  ]^{(z)}_{-\zeta} [ \mathbf{p}, \mathbf{r}^{(L)} ]^{(z)}_\zeta \right]^{(0)}_0 \nonumber \\
& =    \bm{\varepsilon} \cdot \mathbf{p} + i  \sum_{z=1}^{2} \sum_{\zeta=-z}^{+z} 
\left[ [ \bm{\varepsilon}, \mathbf{k}  ]^{(z)}_{-\zeta} [ \mathbf{p}, \mathbf{r} ]^{(z)}_\zeta \right]^{(0)}_0
+ \cdots .
\label{eq:p-exp}
\end{align}   
\noindent

\noindent
Thus, this gives a sum over products of tensors of rank $z$ for a geometric factor, containing  $\bm{\varepsilon}$ and $\mathbf{k}$, and a dynamic factor, containing $\mathbf{p}$  and $\mathbf{r}$.
 On the last line of Eq.~(\ref{eq:p-exp}) we only retain the terms  $L =0$  and $L =1$ that are discussed next.

The $L=0$ term is $\bm{\varepsilon} \cdot \mathbf{p} \propto \omega \bm{\varepsilon} \cdot \mathbf{r}$, and corresponds to the electric dipole (E1) term. Both the geometrical factor $\bm{\varepsilon}$ and dynamic factor $\mathbf{r}$ are time-even and parity-odd so that the matrix element
$\langle f | \mathbf{r} | g \rangle$ is non-zero only if the initial and final states have opposite parity.

The $L=1$ term  factorizes into a geometric factor $[ \bm{\varepsilon}, \mathbf{k} ]^{(z)}$ and
dynamic factor $[ \mathbf{p}, \mathbf{r} ]^{(z)}$, which are parity even. For $z=0$, the geometric factor vanishes due to transversality, $\bm{\varepsilon} \cdot \mathbf{k} =0$.
The $z=1$ term gives the magnetic dipole contribution (M1) with a geometric factor 
$[ \bm{\varepsilon}, \mathbf{k} ]^{(1)} \propto (\mathbf{k} \times \bm{\varepsilon})$ and dynamic factor 
$[ \mathbf{p}, \mathbf{r} ]^{(1)} \propto (\mathbf{p} \times \mathbf{r}) \propto$ orbital momentum operator  $\mathbf{L}$, which is time-odd.
The $z=2$ term gives the electric quadrupole term (E2) with dynamic factor
$[ \mathbf{p}, \mathbf{r} ]^{(2)}$   $\propto$ $\mathbf{L}^{(2)}$ $\propto$ the charge quadrupole, which is time-even.

Thus the leading terms in the transition amplitude can be written as
\begin{eqnarray}
&& - \frac{m}{\hbar} (E_f-E_g)
\left[ \langle f |  {\mathbf{\bm{\varepsilon}}} \cdot {\mathbf{r}} | g \rangle
+ \frac{ i}{2} \langle f | ( {\mathbf{\bm{\varepsilon}}} \cdot {\mathbf{r}})( {\mathbf{k}} \cdot {\mathbf{r}}) |  g \rangle \right]
\nonumber \\
&&  \makebox[0.4in]{} -  \langle f | ({\mathbf{k}} \times  {\mathbf{\bm{\varepsilon}}} ) \cdot ( {\mathbf{L}} + \mathrm{g} {\mathbf{S}}) | g \rangle + \cdots ,
\end{eqnarray}

\noindent
where the first, second, and third term correspond to the electric dipole (E1), quadrupole (E2), and magnetic-dipole (M1) transition matrix elements, respectively. 
The E1 transition is  parity odd, whereas M1 and E2 are parity even.
The E2 and M1 transition probabilities are $(\alpha Z_{\mathrm{eff}})^2$ times smaller than that of E1, where $\alpha$ is  the fine-structure constant 
$e^2/ \hbar c$ $\approx$ 1/137.

Expressions for the geometric and dynamical part of electric- and magnetic-multipole transitions of arbitrary rank can be found in \cite{vanderlaan06a}. In essence,
 the $L$ term with parity $(-)^{L+1}$ separates into a magnetic $2^L$-pole with dynamic factor $[\mathbf{p}, \mathbf{r}^{(L)}]^{(L)}$ and an electric $2^{(L+1)}$-pole with dynamic factor $[\mathbf{p}, \mathbf{r}^{(L)}]^{(L+1)}$. 
 The exception is $L=0$, since there is no magnetic monopole.

The magnetic dipole operator does not incorporate the radial variable $\mathbf{r}$, so that its matrix
elements vanish if the radial part of the initial and final states are orthogonal. 
In the $LS$ coupling scheme, the one-electron magnetic dipole
transition rules are $| \Delta j | \le 1$, $ \Delta \ell  = 0$ , $ \Delta s  = 0$, and $ \Delta n  = 0$ (identical principal quantum
numbers). This means that magnetic dipole absorption is relevant only at low energy (typically
in the microwave and optical range). Its observation at higher energy implicates an
appreciable configuration interaction between the initial and final states, due to a
departure from a pure $LS$ coupling. However, such a configuration interaction
is negligible in core-level spectroscopy, because of the large energy difference between
the initial and final states. Nevertheless, E1M1 absorption in UV spectroscopy is a standard tool for characterizing chiral molecules.

The electric multipole transition operator with $z=L+1$ is proportional to
\begin{equation}
\sum_z \left[ [   C^{(1)} ({\mathbf{\bm{\varepsilon}}}) , C^{(z-1)} ({\mathbf{\bm{k}}})  ]^{(z)}_{-\zeta}  ,
C^{(z)}_\zeta({\mathbf{\bm{r}}}) \right] ^{(0)}_0 ,
\label{eq:coupling}
\end{equation}
where the factor $C^{(z)}_\zeta({\mathbf{\bm{r}}})$ of the matter part gives the spectra $I^{z}_\zeta  (\omega)$.
For instance, for the electric-dipole transition ($z=1$), where the  ${\mathbf{\bm{k}}}$ dependence is absent, the transition operator is
\begin{equation}
{\bm{\varepsilon}} \cdot {\mathbf{r}}=\sum_{j=x,y,z} \varepsilon_j r_j
= r \sum_{\zeta =-1,0,1} C^{(1)\ast}_{\zeta}({\mathbf{\bm{\varepsilon}}})  C^{(1)}_\zeta ({\mathbf{\bm{r}}}),
\end{equation}

\noindent where $C_{\zeta}^{(1)\ast} = C_{-\zeta}^{(1)}$.

The electric $2^z$-pole matrix elements with components $\zeta$ for a transition operator $T^{(z)}_\zeta$ from a core state $|\alpha' c \gamma \rangle$ to an empty valence state $| \alpha \ell m \rangle$ are, according to the Wigner-Eckart theorem, proportional to the $3j$ symbol times the reduced matrix element

\begin{align}
I^{(z)}_\zeta (\omega) \propto \langle \alpha' c \gamma  | T^{(z)}_\zeta |  \alpha \ell m \rangle  & =  (-1)^{c-\gamma}
\left( \begin{array}{ccc}
c & z & \ell \\
-\gamma & \zeta & m \end{array} \right)  \nonumber \\
& \makebox[0.1in]{} \times  \langle \alpha' c \parallel T^{(z)} \parallel \alpha \ell \rangle ,
\end{align}
where the $3j$ symbol (or Wigner coefficient), denoted by the parentheses, is related to the Clebsch-Gordan coefficient as

\begin{equation}
\langle c \gamma z \zeta | \ell m \rangle = (-1)^{c + \zeta + m } \sqrt{2 \ell +1} \left( \begin{array}{ccc}
c & z & \ell \\
-\gamma & \zeta & m \end{array} \right).
\end{equation}

The $3j$-symbol  is zero unless the following conditions are satisfied: $c + z + \ell$ is an integer (integer perimeter rule) or an even integer if $ \gamma = \zeta =m = 0$ (parity rule); $|c -  \ell | \le z \le |c +  \ell  |$ [triangular inequality: $\Delta (c z \ell)$]; and $-\gamma + \zeta+ m = 0$ (rotational invariance).
We immediately recognize these physically relevant conditions as the selection rules for optical transitions. The reduced matrix element 
$\langle \alpha' c \parallel T^{(z)}  \parallel \alpha \ell\rangle$
is responsible for the parity rule, $c+ z + \ell $ = even.

Similar selection rules apply for the matrix elements in the multipole transition $\alpha JM \to \alpha' J'M'$ for  $jj$ coupled states
\begin{align}
\langle \alpha' J' M' | T^{(z)}_\zeta | \alpha J M \rangle  & =  (-1)^{J'-M'}
\left( \begin{array}{ccc}
J' & z & J \\
-M' & \zeta & M \end{array} \right)  \nonumber \\
& \makebox[0.1in]{} \times  \langle \alpha' J'  \parallel T^{(z)}   \parallel \alpha J \rangle ,
\end{align}
where the conservation of angular momentum is contained in the $3j$ symbol as $\Delta (J'zJ)$ and $M' = M+ \zeta$.

\subsubsection{Sum rules}

The polarization dependence of XAS is one of its greatest assets, making it sensitive to charge anisotropy and spin and orbital magnetism
\cite{thole92,carra93,vanderlaan98}.
 Starting from the electric-dipole spectra $I_\zeta  \equiv I^{(1)}_\zeta$ for left-circular ($\zeta=-1$), right-circular ($\zeta=+1$), and linear polarization along the beam direction ($\zeta=0$), we can make new linear combinations of the so-called fundamental spectra $I^x$, where $x$ is the angular momentum transferred from the photon to the atom.
This gives the isotropic spectrum, $I^{0} = (I_{-1} + I_{0} + I_{+1})$, the x-ray magnetic circular dichroism (XMCD) [difference between left- and right-circular polarization: $I^{1} = (I_{-1} - I_{+1})$, and the x-ray linear dichroism (XLD): $I^{2} = (I_{-1} - 2 I_{0} + I_{+1})$.

 The spectra $I^0$, $I^1$, and $I^2$ provide  sum rules that relate the energy-integrated intensities $\rho^x = \int I^x dE$ over the spin-orbit separated regions of the core-level spectrum with the expectation values of the ground-state moments.
For a deep core level, the large spin-orbit interaction splits the spectrum into two separate manifolds with good quantum numbers $j_{\pm} = \ell \pm s$. For  example, the core $3d \to 5f$ transition gives rise to a multiplet spectrum  $3d^{10}5f^n \to 3d^9 5f^{n+1}$, which splits  into  $3d_{5/2}$ and $3d_{3/2}$ manifolds
($M_5$ and $M_4$ edges, for uranium at $\sim$3552 and 3728 eV, respectively). The sum rules for this excitation are the same as for the $4d \to 5f$ ($N_{5,4}$ edges, for U at $\sim$736 and 778 eV). 
However, in the case of the $5d \to 5f$ ($O_{5,4}$ edges, for U at $\sim$94 and 103 eV) the core spin-orbit interaction is not large enough compared to the  $5d$-$5f$ electrostatic interaction to fully separate both edges, leading to  $jj$ mixing between the edges, making the spin sum rule inaccurate.

 \subsubsection{Isotropic spectrum}

In this and the following subsection we present the expressions of the XAS and XMCD sum rules for the $d^{10}f^n \to d^{9}f^{n+1}$ electric-dipole transitions at the $M_{4.5}$ and $N_{4.5}$ absorption edges.

 For the isotropic XAS spectrum the integral over the two absorption edges $\rho^0_{j+}$ and $\rho^0_{j-}$ is proportional to the number of holes $n_h = (14-n)$ in the $5f$ shell, which can be used  to normalize  the dichroic sum rules.

The weighted intensity difference between the two edges is proportional to the expectation value of the angular part of the spin-orbit operator, $\bm{\ell} \cdot \textbf{s}$ \cite{thole88}
\begin{equation}
\frac{\rho^0_{j+} - \frac{3}{2} \rho^0_{j-}}{\rho^0_{j+} + \rho^0_{j-}} 
=  \frac{2}{3} \frac{\langle \bm{\ell} \cdot \textbf{s} \rangle}{\langle n_h \rangle}  ,
\label{eq:spin-orbit-rule}
\end{equation}
where 
\begin{align}
\langle \bm{\ell} \cdot \textbf{s} \rangle  & =   -2   \langle n_h^{5/2}\rangle + \frac{3}{2} \langle n_h^{7/2}\rangle ,  \nonumber  \\
\langle n_h \rangle & =  \langle n_h^{5/2}\rangle +\langle n_h^{7/2}\rangle ,
\label{def-LS}
\end{align}
with $\langle n_h^{5/2}\rangle $ and $ \langle n_h^{7/2}\rangle$ the number of $j = 5/2$ and $j=7/2$  holes in the $f$ shell.

Alternatively, we can  relate the spin-orbit  expectation value to the branching ratio
\begin{align}
B \equiv   \frac {\rho^0_{j+}} { \rho^0_{j-}+\rho^0_{j+} }
 & =  \frac{3}{5} + \frac{4}{15} \frac{\langle \bm{\ell} \cdot \textbf{s} \rangle}{\langle n_h \rangle} \nonumber  \\
 & =  \frac{3}{5} + \frac{-8 \langle n_h^{5/2} \rangle + 6 \langle n_h^{7/2} \rangle}{15 \langle n_h \rangle} ,
\label{eq:BR}
\end{align}
where the fraction 3/5 is the statistical ratio for $B$ in the absence of spin-orbit interaction. 
 
\subsubsection{X-ray magnetic circular dichroism (XMCD)}

For the XMCD the  integral over both edges is proportional to the expectation value of the orbital moment $ L_z $. For the $d \to f$ transition we have 
\begin{equation}
\frac{\rho^1_{j+} + \rho^1_{j-}}{\rho^0_{j+} + \rho^0_{j-}}  =  \frac{1}{3} \frac{\langle L_z \rangle}{n_h}   .
\label{eq:orbital-rule}
\end{equation}

The  weighted difference over the core spin-orbit-split intensities is proportional to the ground state expectation value of the effective spin moment 
$ S_{z, {\mathrm{eff}}} $, 
which  comprises  the spin moment  $  S_z  $ and the magnetic dipole term 
$ \mathbf{T} =  \sum_i \mathbf{s}_i - 3  \mathbf{\bm{r}}_i ( \mathbf{\bm{r}}_i \cdot s_i)$,
\begin{equation}
\frac{\rho^1_{j+} - \frac{3}{2} \rho^1_{j-}}{\rho^0_{j+} + \rho^0_{j-}} 
=  \frac{2}{3} \frac{ \langle S_{z, {\mathrm{eff}}} \rangle  }{n_h}  
=   \frac{2}{3}  \frac{ [ \langle S_z \rangle +  3 \langle T_z \rangle ] }{n_h}   .
\label{eq:spin-rule}
\end{equation}
While the spin moment $\mathbf{S}$ is isotropic, $\mathbf{T}$ gives the anisotropy of the spin moment due to the coupling with the charge quadrupole moment.

The number of $f$-holes, $n_h = (14-n)$ is often known in the case of the more localized rare earths \cite{thole85}, but this is usually not so for the actinides \cite{moore09}. The number of holes cancels out in the orbital-to-spin moment ratio, 

\begin{equation}
 \frac{\langle L_z \rangle} {\langle S_{z, {\mathrm{eff}}} \rangle  } = 
 2  \,  \frac {\rho^1_{j+} + \rho^1_{j-}}{\rho^1_{j+} - \frac{3}{2} \rho^1_{j-}  } \, .
\label{eq:orbital2spin-rule}
\end{equation}

Further, there are sum rules  for the x-ray magnetic linear dichroism, relating the integrated intensities to the charge anisotropy and the anisotropic part of the spin-orbit interaction of the $f$ electrons \cite{vanderlaan99}.

\subsubsection{X-ray absorption fine structure (XAFS)} \label{XAFS}

So far we looked at transitions into the localized 5$f$ states. We will now  consider transitions to states at higher energies.
In the solid, the core-level absorption spectrum can be separated  in transitions to bound (discrete)  final states and  continuum states below and above the ionization potential due to excitations of the photoelectron in the vacuum. While the naming of these regions is very much a matter of taste, it makes sense to classify them  according to the applicable theoretical models. In the case of atomic-like transitions,  the onset of the edge shows intense multiplet structure that we  call XAS. The region with excitations to continuum states can be  divided up into  the x-ray absorption near-edge structure (XANES) and extended x-ray absorption fine structure (EXAFS), at respective higher energies above the edge.
The XANES region gives information about the electronic structure of the binding electrons, and is often calculated using
multiple scattering theory.
In the EXAFS region,  the core electron is excited into a continuum state that still feels the potential of the neighboring atoms. The  interference of the outgoing and backscattered electron waves by the neighboring atoms
depends on the wave-nature of the photoelectron and
is particularly sensitive to the radial distances of the various shells of neighboring atoms around the absorbing one. 
It is convenient to think of XAFS in terms of the photoelectron wavenumber,
$k =\sqrt {2m(E-E_0)}/\hbar $, where $E_0$ is the threshold  absorption energy \cite{lytle99}.

The modulation of the x-ray absorption coefficient  $\mu(E) \sim | \langle f |  H | g \rangle |^2$  at energies near and above an x-ray absorption edge is captured by 
the EXAFS equation  \cite{stohr92}
\begin{align}
\chi (E) &  \equiv \frac{\mu (E) - \mu_0 (E)} {\Delta \mu_0 (E)} \nonumber \\
& = \sum_j \frac{N_j f_j(k) e^{-2 k^2 \sigma_j^2} }{k R_j^2} \sin [2k R_j + \delta_j(k)]  ,
\end{align}
where $\mu_0(E)$ is a smooth background function representing the absorption of the isolated atom, and $\Delta \mu_0(E)$  is the measured jump in the absorption $\mu(E)$ at the threshold energy $E_0$.

The sum is over ``shells'' of similar neighboring atoms with coordination number $N$, distance $R$, and  mean-square disorder $\sigma^2$.
The scattering amplitude $f(k)$ and phase shift $ \delta(k)$ depend on the atomic
number $Z$ of the scattering atom, which can be used to determine the species of the neighboring atoms.

\subsection{X-ray photoemission spectroscopy (PES)}

\subsubsection{Core-level photoemission}

Photoelectron emission spectroscopy is well-established as a popular method to study the electronic structure of materials. Due to the small electron elastic escape depth, PES is rather surface sensitive, so that good vacuum conditions are needed to conduct measurements of a prepared surface. The photoelectron inelastic mean free path varies as a function of kinetic energy with a minimum around 40 eV. Bulk sensitivity is acquired at high photon energy, but at a cost of reduced cross-section and often energy resolution. One distinguishes between XPS and UPS, when using soft x-rays and ultraviolet radiation, respectively. XPS is primarily performed with Al or Mg $K\alpha$ radiation from a lab x-ray source or monochromatized radiation from the synchrotron; UPS is mainly performed using He I or He II gas discharge lamp in the laboratory. A benefit of the different photon energies, especially using tunable SR, is that it gives different relative cross-sections of the transitions involved, thereby providing a way to distinguish between them. PES is also performed in angle-resolved fashion using  single-crystal samples, or at resonance using x-ray energies that coincide with a core-valence excitation \cite{terry02}.

In the PES process, a photon $h\nu$ is absorbed under emission of an electron with kinetic energy, $E_{\mathrm{kin}}$. Energy conservation requires that 
$E_{\mathrm{kin}} =h\nu +E_N - E_{N-1}$,
 where $E_N$ and $E_{N-1}$ are the energies of the $N$-electron initial state and the $(N-1)$-electron final state. 
The energy $E_B = h\nu- E_{\mathrm{kin}} = E_{N-1} - E_N$ is usually called the electron binding energy, however, it would be better to call this the electron removal energy. Only when the electrons do not feel each other, PES gives the one-electron density-of-states (DOS). However, in correlated materials, such as many of the actinide metals, PES should be regarded as a probe of the many-electron state.

In contrast to XAS, where the excited electron goes in unoccupied valence state, in PES it goes into a continuum state, and reaches a detector that analyses its kinetic energy. The ionization has major implications for the screening of the photo-excited hole. Thus, in the actinide atom, the valence and core PES can be represented by the transitions 
$5f^n \to 5f^{n-1} \epsilon$
and $5f^n \to \underline{c} 5f^n \epsilon$, respectively, where 
 $\epsilon$ is a continuum state far above the Fermi level and $\underline{c}$ denotes a core hole.
In the so-called sudden approximation, one assumes that the excited photoelectron has no interaction with the state left behind, so that in the calculation the photoelectron state can be decoupled from the atomic state. The PE spectrum, as a function of binding energy  is expressed as
 \begin{equation}
 I(E_B) = \sum_{n m m' \sigma \sigma'} | \langle f_n | \epsilon_{m' \sigma'}^\dagger c_{m \sigma} | g \rangle |^2 \,\delta_{\sigma,\sigma^\prime} 
 \delta(E_B + E_g - E_{f_n})  ,
 \label{eq:pes}
\end{equation}
where   the ground and final state $|g \rangle$  and  $|f_n \rangle$  have energy $E_g$ and  ${E_f}_n$, respectively. $c_{m \sigma}$ is the annihilation operator of an electron $c$ with quantum numbers $m$ and $\sigma$; and $\epsilon^{\dagger}_{m'\sigma'}$ is the creation operator of a continuum electron $\epsilon$ with quantum numbers $m'$ and $\sigma'$.
The  $5f$ PES of the actinide atoms displays intense multiplet structure, which has been calculated in intermediate coupling by \cite{gerken83}.

In contrast to XAS, core level PES is well suited to determine the hybridization between the valence electrons
\cite{vanderLaan86}.  In XAS the core electron is excited into an unoccupied $f$ state, effectively screening the core hole. The potential energy change is $U_{ff}-Q_{cf}$, where $U_{ff}$ is the Coulomb interaction between two $f$ electrons and $Q_{cf}$ that between core and $f$ electron.  If
$U_{ff}$ $\approx$ $Q_{cf}$, as is roughly the case, the change in hybridization (mixing) between initial and final state is small. In PES the core electron is excited into the continuum, leaving the core hole unscreened. The potential energy change is $-Q_{cf}$. Hence the change in hybridization in the final state is large, which can be used to determine the hybridization in the initial state as is demonstrated by the
example of Pu $4f$ core-level PES in Sec. \ref{clpes}.

\subsubsection{Angle-resolved PES (ARPES)}

At low photon energies, typically used in angle-resolved photoemission spectroscopy (ARPES), the photon momentum can be neglected. 
Taking advantage of total energy and momentum conservation laws, one can relate the kinetic energy and momentum of the photoelectron to the binding energy $E_B$ and crystal momentum $\hbar \mathbf{k}$ inside the solid.

The photoemission is measured as a function of kinetic energy
$E_{\mathrm{kin}} = \hbar \omega - E_B - \phi $ and angle $\vartheta$ relative to the surface normal,  where $\phi $ is the work function of the material.
The Bloch wave vector $\mathbf{k}$ can be linked to the measured electron momentum $ | \mathbf {p} | = \sqrt{ 2 m_e E_{\mathrm{kin}}}$, 
so that
\begin{align} 
\mathbf{p}_{\parallel} & = \hbar \mathbf{k}_{\parallel} =   \sqrt{2 m_e E_{\mathrm{kin}} } \, \sin{\vartheta} , \nonumber \\
\mathbf{p}_{\perp} & = \hbar \mathbf{k}_{\perp} =\sqrt{2 m_e (E_{\mathrm{kin}} \cos^2 \vartheta + V_0) } ,
\end{align}
where $\hbar \mathbf{k}_{\parallel}$ and $\hbar \mathbf{k}_{\perp}$ are the components parallel and normal to the surface, respectively, and the inner potential $V_0$ is an a-priori unknown parameter.
Upon going to larger angles $\vartheta$, one actually probes electrons with $\mathbf{k}$  in higher-order Brillouin
zones. By subtracting the corresponding reciprocal lattice vector $\mathbf{G}$, one obtains the reduced electron crystal
momentum in the first Brillouin zone.

\subsubsection{Resonant PES (RPES)}
\label{sec:rpes}

In the $t$-matrix approach the transition probability for RPES is determined by an operator 
\cite{vanderlaan1999-RPES}
\begin{equation}
T = D + VGT = D + VGD + VGVGT = \sum_{n=0}^{\infty} (VG)^n D,
\end{equation}
which is first order in the dipole operator $D$ and infinite order in the Coulomb decay operator $V$. The Green's function is
$G = 1/(H- h \nu - E_g - i \eta )$, where $H$ is the unperturbed Hamiltonian,  and $\eta$ is a small positive number, cf.~Eq.~(\ref{eq:Green}). For $\zeta$-polarized light, the probability for the
angle-integrated photoemission in resonance with a core level can be expressed as
\begin{align}
I_\zeta  & (h \nu, E_B) = | \langle f | T_\zeta | g \rangle |^2 \, \delta (E_g -E_f + h \nu) \nonumber \\
&= \sum_f \left| \langle f | D_\zeta | g \rangle + \sum_n
\frac{ \langle f | V | n \rangle \langle n| D_\zeta | g \rangle}{ E_n - E_g - h \nu -i \Gamma_n/2} + \cdots \right|^2 \nonumber \\
&  \ \ \ \ \times \delta (E_g -E_f + h \nu) ,
\label{eq:rpes}
\end{align}
with decay full-width
\begin{equation}
\Gamma_n = 2 \pi | \langle f | V | n \rangle |^2  .
\end{equation}
For instance, for the $4d \to 5f$ resonant photoemission, $| g \rangle$, $| n \rangle$, and $| f \rangle$ are the eigenstates in the transition
$5f^N + h \nu \to 4d^9 5f^{N+1} \to 5f^{N-1}+\epsilon$.
Equation (\ref{eq:rpes}) may serve to illustrate the different ways to reach the final state, i.e., direct photoemission
from the initial to the final state, photoexcitation into an intermediate
state followed by decay into the final state, and the coherent superposition of these processes.

\subsection{Nonresonant scattering}

At incident photon energies far away from resonant excitation, the double differential cross section (DDCS) 
for x-ray scattering is obtained from the KH formula in first-order in  ${\mathbf{A}}^2$ (Fig.~\ref{fig:Diagrams}(b)), 
with vector potential   ${\mathbf{A}} ({\mathbf{k}},{\mathbf{r}}) = {\mathbf{\bm{\varepsilon}}} \, e^{i {\mathbf{k}} \cdot {\mathbf{r}}}$,
as
 \begin{align}
 \frac{d  \sigma}{ d \Omega d \omega} & = r^2_0 
 \frac{ \omega_{\mathbf{k}'} } {\omega_{\mathbf{k}} }
 ( \bm{\varepsilon}^{ \prime \ast} \cdot  \bm{\varepsilon})^2 
 \sum_f \left|   \langle f | \sum_j  e^{ i (\mathbf{k}' -\mathbf{k}) \cdot {\mathbf{r}}_j  } | g \rangle   \right|^2 
 \nonumber \\
 & \makebox[0.2in]{}  \times \delta(E_g - E_f + \hbar \omega - \hbar \omega') .
\label{eq:KH-nonres}
\end{align}
Defining  the photon momentum transfer (or scattering vector) $\mathbf{Q} \equiv \mathbf{k}' - \mathbf{k}$,
leads to a factor $e^{ i \mathbf{Q}  \cdot {\mathbf{r}}_j  }$ in Eq.~(\ref{eq:KH-nonres}).

\subsubsection{Thomson scattering}
\label{sec:Thomson}

The earlier introduced Thomson scattering is the elastic scattering of an electromagnetic wave from charged particles due to $ \mathbf{A}  ^2$. 
  While this falls more in the domain of conventional x-ray diffraction using cathode ray tubes, it is briefly discussed here for completeness.

As Thomson scattering is elastic, $\omega_{\mathbf{k}'} = \omega_{\mathbf{k}}$ and $| f \rangle  = | g \rangle$, which leads us to Eq.~(\ref{eq: Thomson}).
  The scattering cross section in a crystal can be written as
\begin{equation}
\frac{d  \sigma}{ d \Omega} = r_0^2 ( \bm{\varepsilon}^{ \prime \ast} \cdot  \bm{\varepsilon})^2 \left| F(\mathbf{Q}) \right|^2
= r_0^2 ( \bm{\varepsilon}^{ \prime \ast} \cdot  \bm{\varepsilon})^2  \left| \sum_j e^{i \mathbf{Q} \cdot \mathbf{r}_j} f_j(\mathbf{Q}) \right|^2, 
\label{eq:cross-section}
\end{equation}  
\noindent
where $f_j(\mathbf{Q})$ is the atomic form factor. Neglecting energy dispersion corrections, $f_j(\mathbf{Q})$ is a real number given by the Fourier transform of the charge distribution in the $j$-th atom; the sum is over the $n$ atoms of the unit cell basis, and
$\mathbf{Q}$ is equal to a reciprocal lattice vector $\mathbf{G}$.
 
The unit cell structure factor $F( \mathbf{Q}) = \sum_j f_j e^{i \mathbf{Q} \cdot \mathbf{r}_j}$ takes into account the phase factor that arise from the path difference of the x rays between the atoms.
$F( \mathbf{Q})$ is generally complex although it is purely real in centro-symmetric crystals with a suitable choice of origin (again, if dispersion corrections are negligible, see Sec. \ref{REXS}). Away from resonance, the charge scattering becomes an integral over the continuous electron density
$\rho(\mathbf{r})$ as
$ F(\mathbf{Q}) = \int d \mathbf{r} \, e^{i \mathbf{Q} \cdot \mathbf{r} } \rho(\mathbf{r})$.
 
Since the scattering cross-section in Eq.~(\ref{eq:cross-section}) is related to the absolute square of the Fourier transform of the electron density at momentum $\mathbf{Q}$,  the phase information is lost in the
measurement. This is known as the phase problem.

\subsubsection{Nonresonant inelastic x-ray scattering (NIXS)}
\label{sec:NIXS}

NIXS is  a photon-in photon-out process, which is a direct transition not involving an intermediate state.
Compared to XAS, the main benefit of  NIXS is that it gives us access to  higher multipole transitions; the main disadvantage is, as for all non-resonant techniques, the low cross section and the required narrow bandwidth optics that makes measurements challenging
\cite{schulke07,caciuffo10,vanderlaan12}.

Reshuffling the DDCS  in Eq.~(\ref{eq:KH-nonres}), the Thomson differential scattering cross section $(d \sigma/ d \Omega)_{\mathrm{Th}} $  can be factored out so that the technique measures a target excitation structure known as the {\it{dynamical structure factor}},  $S(\mathbf{Q},\omega)$, which is independent of the specific experimental geometry,  
\begin{equation}
\label{eq:DDCS}
\frac{d^2\sigma}{d \Omega d \omega}   =   \left( \frac{d \sigma}{d \Omega }  \right)_{\!\mathrm{Th}}  S(\mathbf{Q},\omega), 
\end{equation}
with
\begin{equation}
 S(\mathbf{Q},\omega)   =   \sum_f \left|   \langle f | \sum_j  e^{ i {\mathbf{Q}} \cdot {\mathbf{r}}_j  } | g \rangle   \right|^2 
 \delta(E_g - E_f + \hbar \omega)   ,
\label{eq:S}
\end{equation}
and 
\begin{equation}
\left( \frac{d \sigma}{d \Omega} \right)_{\mathrm{Th}} = r^2_0 \, \frac{   \omega_{\mathbf{k}^\prime}}{\omega_{\mathbf{k}}}
( {\bm{\varepsilon}}^{\prime \ast} \cdot {\bm{\varepsilon}})^2,
\end{equation}

\noindent
where  $\hbar  \omega = \hbar  \omega_{\mathbf{k}} - \hbar  \omega_{\mathbf{k}^\prime}$ is the energy transfer,  
 ${\mathbf{Q}}  = {\mathbf{k}} - {\mathbf{k}}' $ is the scattering vector (the momentum transfer in the excitation process), 
and  $d \Omega$ is the solid angle element of the scattered photons.


The dynamical structure factor $S(\mathbf{Q},\omega)$ in Eq.~(\ref{eq:S}) can be related to a density-density correlation function in space and time of the system \cite{vanhove54}.
  \begin{equation}
  S(\mathbf{Q},\omega) = \int d \mathbf{r} d t \, e^{i (\mathbf{Q} \cdot \mathbf{r} - \omega t)}
  \int d \mathbf{r}' \langle g | \rho (\mathbf{r}' ,0) \rho(\mathbf{r} + \mathbf{r}' , t) | g \rangle.
   \end{equation}
 This means that the inelastic analog of the Thomson scattering permits to study the spectrum of charge density fluctuations, such as phonons, electronic excitations and plasmons.

The transition operator  $ e^{ i {\mathbf{Q}} \cdot {\mathbf{r}} } $ in Eq.~(\ref{eq:S}) can be expanded as a sum over scalar products of spherical multipole tensors with rank $k$ and  components $\kappa = -k, \cdots, k$ as
 \begin{align}
 e^{i {\mathbf{Q}} \cdot {\mathbf{r}} }  
 = \sum_{k=0}^{\infty} \sum_{\kappa = -k}^{k} i^k (2k+1) 
  j_k(Q r)  C^{(k)*}_{\kappa} ( {\mathbf{\bm{Q}}} )  C^{(k)}_{\kappa} ( {\mathbf{\bm{r}}} ),
\label{eiqr}
\end{align}
\noindent
where  $j_k(Qr)$  are spherical Bessel functions of rank  $k$ and  $C^{(k)}_{\kappa} ( {\mathbf{\bm{r}}} )$ are renormalized spherical harmonics.

Assuming spherical symmetry, the interference terms ($k \neq k'$, where $k'$ is the rank of the conjugated multipole) vanish. Taking the radial-matrix elements constant over the spectral region of interest, $S(\mathbf{Q},\omega)$ (Eq. \ref{eq:S})
can be separated into an angular and a radial part:

\begin{align}
S(\mathbf{Q},\omega)   =   \sum_{k=0}^{\infty} I^k(\omega) \left|   \langle f | j_{k}(Qr) | g \rangle   \right|^2,
\label{eq:S1}
\end{align}

The angular part of the isotropic $2^k$-pole spectrum is
\begin{align}
 I^k(\omega) =  \sum_{f,\kappa} 
 |\langle f |  C^{(k)}_{\kappa} ( {\mathbf{\bm{r}}} )  | g \rangle   |^2   
 \delta(E_g - E_f + \hbar \omega).
\label{eq:Iomega}
\end{align}

Multipole moments $k$ for the $\ell \to \ell'$  transition are restricted by the triangle condition,   $| \ell-\ell ' | \leq k \leq \ell + \ell '$, and parity rule,  $\ell+\ell ' +k$ = even. Thus for $d \to f $ transitions,  $k=1$ (dipole), $k=3$ (octupole), and $k=5$ (triakontadipole) transitions are allowed. 

At low momentum transfer ($Q \to 0$) the radial matrix elements
$\left| \langle f | j_{k}(Qr) | g \rangle \right|$
are negligible for $k >$ 1 and only dipole transitions are important, similar to the case of soft XAS. On the other hand, for large enough $Q$ values (larger than $\sim$9 \AA$^{-1}$ at the uranium $O_{4,5}$ edges) $\left| \langle f | j_{0}(Qr) | g \rangle \right|$ vanishes and the spectra are dominated by the $k=3$ and $k=5$ contributions.

In the dipole approximation the NIXS spectrum for  $f^n + \hbar \omega_{\mathbf{k}}  \to d^9 f^{n+1} + \hbar \omega_{\mathbf{k}  ^\prime}$  is the same as  the  XAS spectrum for $f^n + \hbar \omega  \to d^9 f^{n+1}$.   
In the dipole approximation, $\hat{\bf{Q}} = {\mathbf{Q}}/Q$ plays in NIXS the same role as
$\bm{\varepsilon}$ in XAS. 
However, it should be noted that ${\hat{\bm{Q}}}$ is a polar vector allowing only hermitian matrix elements,  whereas  
 $\bm{\varepsilon}$ is an axial vector, allowing anti-hermitian matrix elements and hence enabling XMCD, making the latter a supreme tool for magnetism \cite{vanderlaanCCR14}.
 
Contrary to XAS, it is easy  in NIXS  to reach the range $Qa \ge 1$, where $a$ is the atomic radius, enabling to reach dipole-forbidden transitions with high probability, by increasing the scattering angle $\theta$ and thereby shifting $\mathbf{Q} = 2({\mathbf{k}}' - {\mathbf{k}}) \sin \theta$. 

The reason that XAS and NIXS give the same spectra, despite  different transition-matrix elements, is provided by the concept of the fundamental spectra, where the angular part of the transition probability is separated from the radial and geometrical part.   A requirement for this is that the radial part remains constant over the spectral range, which is usually fulfilled.

\subsubsection{Generalized spin-orbit sum rule}

Interestingly, the spin-orbit sum rule for XAS in Eq.~(\ref{eq:BR}) can be expanded to the case of NIXS. 
According to this sum rule, the initial-state spin-orbit interaction per hole is linearly related to the core-level branching ratio. 

For the $2^k$-multipole transition $d \to f$ we obtain the branching ratio of the core spin-orbit split levels as 
\cite{vanderlaan12b}
\begin{align}
B^{k} & = \frac{3}{5} + \frac{18-k(k+1)}{60} \frac{\langle \bm{\ell} \cdot \textbf{s} \rangle}{\langle n_h \rangle} \nonumber \\
& = \frac{3}{5} + [18-k(k+1)]  \frac{-4 \langle n_h^{5/2} \rangle + 3 \langle n_h^{7/2} \rangle}{120 \langle n_h \rangle},
\label{eq:BRk}
\end{align}
which generalizes the spin-orbit sum rule in Eq.~(\ref{eq:BR}).
Thus the various multipole moments $k$ have strongly different values of $B^{k}$.
For a $f_{7/2}$ hole we have $B^1=1$ since the excitation $d_{3/2} \to f_{7/2}$ is forbidden for dipole transitions. However, this transition becomes allowed for higher multipole moments, for which  $B^3= 3/4$ and $B^5= 3/10$.
For a $f_{5/2}$ hole we find $B^1=1/15$, $B^3= 2/5$, and $B^5 = 1$.
Equation (\ref{eq:BRk}) is very useful for the actinide $M_{4,5}$ and  $N_{4,5}$ edges, which unfortunately have a very low NIXS cross sections. Although for the  $O_{4,5}$ edges the $jj$ mixing is too large to separate the individual $5d_{3/2}$ and $5d_{5/2}$ edges, the spin-orbit sum rule is still very useful, since it gives a large intensity transfer to higher energy as $\langle \bm{\ell} \cdot \textbf{s} \rangle$ reduces.
For the ground state with lowest spin-orbit energy, higher $k$ spectra have a lower branching ratio.
Hence, in the low energy region, the overall intensity of the $k=3$ spectrum is higher than that of the $k=5$ spectrum, whereas this is opposite in the high energy region.

\subsubsection{Non-resonant magnetic scattering}

Non-resonant magnetic scattering  provides a direct method of distinguishing spin and orbital
magnetic moments for long-range magnetic structures  \cite{blume85,blume88}.
Pioneering experimental studies have been published by \cite{debergevin81}.

A pure spin contribution to the magnetic scattering is obtained by
the term $H'_4 \sim \partial \mathbf{A} /  \partial t \cdot (e/c) \mathbf{A}$ in the interaction Hamiltonian. 
On the other hand, an orbital contribution mainly arises  from the term $\mathbf{A} \cdot \mathbf{p}$ at the second-order.
The scattering amplitude can be written as 
\begin{equation}
f^{\mathrm{mag}}  = -i \frac{\hbar \omega_{\mathbf{k}} }{m c^2}    
  \langle g | \sum_j e^{i \mathbf{Q} \cdot \mathbf{r}_j}
 \left[ \frac{ i \mathbf{Q} \times \mathbf{p}_j}{\hbar k^2} \cdot \mathbf{P}_L - \mathbf{s}_j \cdot \mathbf{P}_S \right] | g \rangle  ,
\label{eq:fmag}
\end{equation}
with the polarization factors related to the orbital and spin moment
\begin{align}
\mathbf{P}_L & = ( \bm{\varepsilon}^{\prime \ast} \times \bm{\varepsilon})  , \\
\mathbf{P}_S & = ( \bm{\varepsilon}^{\prime \ast} \times \bm{\varepsilon} ) + (\bm{\mathbf{k}}' \times \bm{\varepsilon}^{\prime \ast}) (\bm{\mathbf{k}}' \cdot  \bm{\varepsilon} )
 -  (\bm{\mathbf{k}} \times \bm{\varepsilon})(\bm{\mathbf{k}} \cdot \bm{\varepsilon}^{\prime \ast})  \nonumber \\
& \makebox[0.2in]{} -  (\bm{\mathbf{k}}' \times \bm{\varepsilon}^{\prime \ast}) \times (\bm{\mathbf{k}} \times \bm{\varepsilon}) .
\end{align}
\noindent
The cross-section of the magnetic scattering  is  $\hbar^2 \omega_k^2 / m^2c^4$ times smaller as that of the charge scattering, which amounts to a factor $4 \times 10^{-4}$  at 10 keV incident photon energy.
However, the interference term between the two amplitudes can be used to extract the magnetic cross section.

We define the microscopic electric current density operator
\begin{align}
\mathbf{j} (\mathbf{r}) = \frac{-e}{2m} \sum_j [ \mathbf{p}_j \delta (\mathbf{r} -\mathbf{r}_j) 
+   \delta (\mathbf{r} -\mathbf{r}_j) \mathbf{p}_j ] \equiv c [ \nabla \times \mathbf{M}_L (\mathbf{r} )] ,
\end{align}
so that $\mathbf{j} (\mathbf{Q}) = - i c \mathbf{Q} \times  \mathbf{M}_L (\mathbf{Q})$, where $\mathbf{M}_L (\mathbf{Q})$ is the Fourier transform of the orbital magnetization.
 The Fourier transform of the spin magnetization is
 \begin{align}
 \mathbf{M}_S (\mathbf{Q}) = \frac{e \hbar}{mc} \sum_j e^{i \mathbf{Q} \cdot \mathbf{r}_j} \mathbf{s}_j .
\end{align}

\noindent
Then, Eq.~(\ref{eq:fmag}) can be written as
\begin{align}
f^{\mathrm{mag}} & = -i \frac{ \omega_{\mathbf{k}} }{e c} [  4 \sin^2 \theta  \,
  \langle g | \bm{\mathbf{Q}} \times [ \mathbf{M}_L (\mathbf{Q}) \times \bm{\mathbf{Q}} ] \cdot \mathbf{P}_L | g \rangle
 \nonumber \\
 & \makebox[0.2in]{} + \langle g | \mathbf{M}_S (\mathbf{Q}) \cdot \mathbf{P}_S  | g \rangle ].
\label{eq:fmag2}
\end{align}
The imaginary prefactor in Eq.~(\ref{eq:fmag2})   means that after taking the square modulus there is no interference between the Thomson and the magnetic scattering, unless the structure factors 
$\sum_j \langle g | e^{i \mathbf{Q} \cdot \mathbf{r}_j} | g \rangle$ are complex, in which case the crystallographic structure is noncentrosymmetric, leading to interference terms. 
Non-resonant magnetic scattering has been used to separate spin and orbital moments in UAs \cite{langridge97}.

 \begin{figure}
   \centering
   \includegraphics[width=1.0\columnwidth]{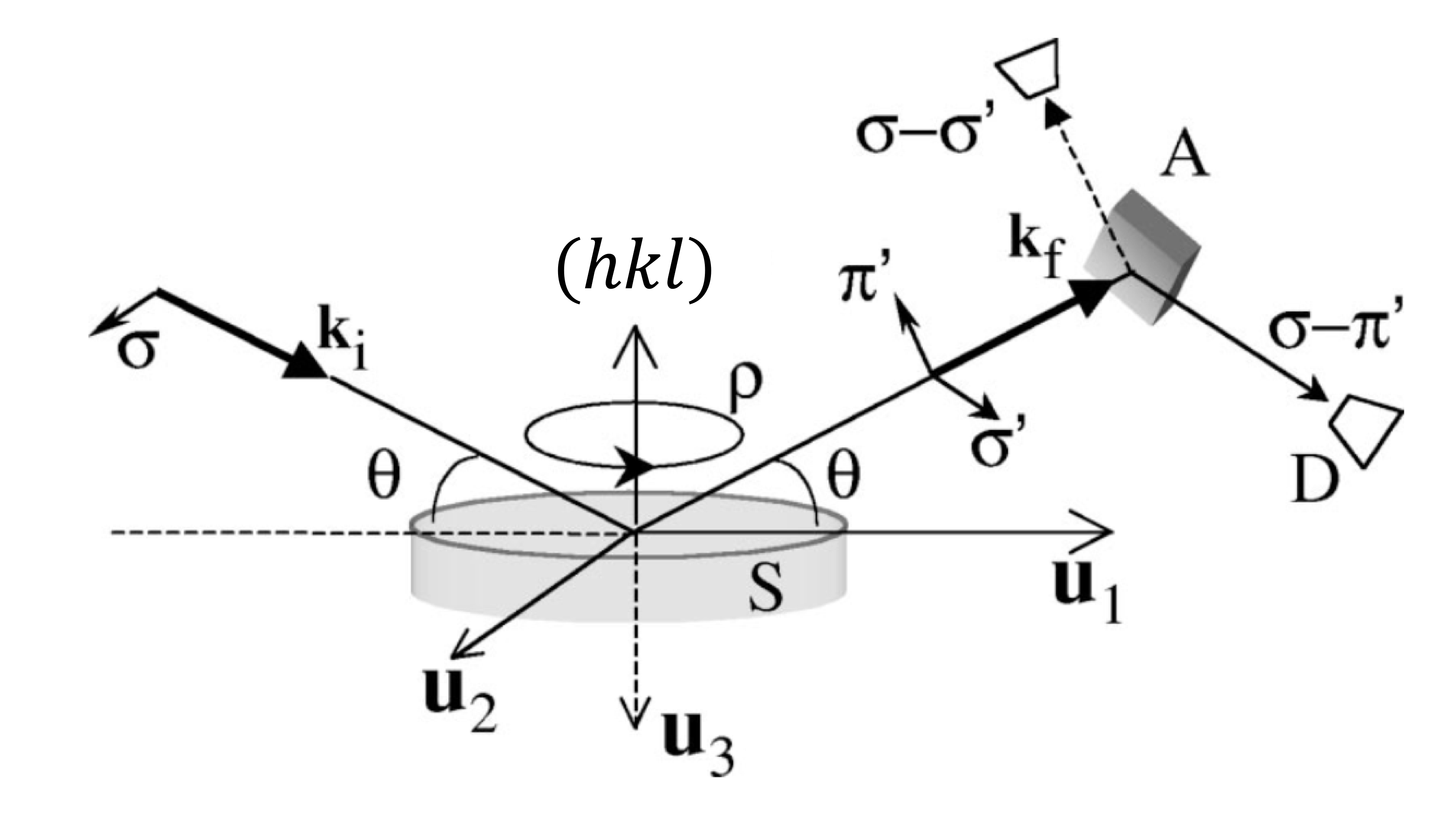}
  \caption{Schematic representation of the experimental geometry for non-resonant magnetic diffraction measurements. The incident photon beam is polarized perpendicularly to the scattering plane ($\sigma$ polarization, from the initial of the German word senkrecht). S is the sample, D is the detector, and A is the analyzer crystal allowing one to select the components of the scattered beam with polarization parallel ($\pi'$) or perpendicular ($\sigma'$) to the scattering plane.  Integrated intensities of the Bragg peaks are measured as a function of the photon energy for different values of the azimuthal angle ($\rho$), defining the crystal orientation about the scattering vector (\textit{hkl}). The unit vectors $\textbf{u}_{i}$ define the reference frame. Reproduced from \cite{caciuffo02}.} 
\label{RXESgeometry}
\end{figure}

With reference to the geometry described in Fig.~\ref{RXESgeometry}, the scattering amplitude $f^\mathrm{mag}$ can be written as \cite{blume88}
\begin{align}
& f^\mathrm{mag}  = { } \binom{f_{\sigma  \sigma'}}{f_{\sigma \pi'}}  \\\nonumber
& = \binom{\sin 2\theta \, M_{S2}(\mathbf{Q})}{\sin \theta \, \sin 2\theta \, [M_{S1}(\mathbf{Q}) +M_{L1}(\mathbf{Q})] + 2 \sin^{3} \theta \, M_{S3}(\mathbf{Q})},
\label{nonresonmag}
\end{align}

\noindent
where $\theta$ is the Bragg angle, $\mathbf{Q}$ is the scattering vector, $M_{Si}(\mathbf{Q})$ and $M_{Li}(\mathbf{Q})$ ($i$ = 1, 2, 3) are the cartesian components of $\textbf{M}_{S}(\textbf{Q})$ and $\textbf{M}_{L}(\textbf{Q})$ along the unit vectors $\textbf{u}_{i}$ shown in Fig.~\ref{RXESgeometry}). 

With an appropriate choice of the scattering geometry, measuring the integrated intensity of magnetic Bragg peaks as a function of the azimuthal angle $\rho$ allows one to determine the ratio $M_{L}(\mathbf{Q})/M_{S}(\mathbf{Q})$ and to obtain information on the relative orientation of spin and orbital magnetic moments.

\subsection{Resonant scattering}

 \subsubsection{Resonant elastic x-ray scattering (REXS)} \label{REXS}

The resonance process in REXS  gives a cross section enhancement of several orders in magnitude together with a strong dependence on the polarization of the incident and scattered beams. The latter  requires  a formal description of the atomic scattering amplitude as a tensor, instead of a scalar quantity, with  consequences for the angular dependence of the diffracted beam \cite{templeton80}.
It gives a sensitivity of the resonant scattering to the charge and magnetic order, as well as to the orbital order.
 
 The atomic form factor becomes an energy-dependent complex scalar quantity \cite{als-nielsen11}
 \begin{equation}
f(\mathbf{Q},\omega) = f_0(\mathbf{Q}) + f'(\omega) + if''(\omega).
\end{equation}
The non-resonant Thomson term, $f_0(\mathbf{Q})$, is the Fourier transform of the electronic charge distribution in the atom, modeled as  a cloud of free electrons with spatial extent comparable to the x-ray wavelength. Therefore, $f_0$ is dependent on $\mathbf{Q}$ but not on the photon energy
$\hbar \omega$. On the other hand, the resonant \textit{dispersion correction term} $f'(\omega) + f''(\omega)$ (also called \textit{anomalous scattering correction})
describes the effect of the electrons being bound in atoms. It is therefore dependent on $\omega$,  with resonances at energies corresponding to the atomic absorption edges. It is, however, almost $\mathbf{Q}$-independent when the electrons involved in the resonance excitation are in the $K$, $L$, or $M$ shells, whose spatial extent is much smaller than the x-ray wavelength.
We recall that the real and imaginary parts of the dispersion correction, $f'(\omega)$ and $f''(\omega)$, 
are related by the Kramers-Kronig transformation. 

According to  second-order perturbation theory with $H'_2$, the anomalous scattering
factor in the vicinity of the absorption edge is given by
\begin{equation}
f_{\mathrm{res}} = \frac{1}{m} \sum_n 
\frac{\langle g | (\bm{\varepsilon}^{\prime \ast} \cdot \mathbf{p}) e^{-i \mathbf{k}' \cdot \mathbf{r}} | n \rangle \langle n | (\bm{\varepsilon} \cdot \mathbf{p}) e^{i \mathbf{k} \cdot \mathbf{r}} | g \rangle}
{E_g -E_n + \hbar \omega_\mathbf{k} - i \Gamma_n/2} .
\end{equation}
%
where $\Gamma_n$ is the full width of the emission decay. Values of $\Gamma_n$ for several uranium edges are given in \cite{raboud99} and reported in Table \ref{attlengths}.

Expanding $e^{i \mathbf{k} \cdot \mathbf{r}}$  as in Eq.~(\ref{e-ikr}) leads to electric-dipole, dipole-quadrupole, and quadrupole transitions, respectively \cite{hill96}.
 

 %
 The resonant E1E1 scattering amplitude  in cartesian form is proportional to
\begin{equation}
F_{E1E1} =  (\mathbf{r} \cdot \bm{\varepsilon}')^\ast (\mathbf{r} \cdot \bm{\varepsilon}) = r_i^\ast r_j  {{\varepsilon}}_i ^{\prime \ast}{\varepsilon}_j 
=T_{ij} X_{ij} ,
\label{eq:FE1E1}
\end{equation}
with the second-rank tensors $T_{ij} = r_i^\ast r_j$ and $X_{ij} = {\varepsilon}_i^{\prime \ast} {\varepsilon}_j$, which represent the geometric and dynamic tensors for the  matter and  x-ray probe, respectively. 
Since the amplitude $F$ is a scalar, the tensors 
%
\begin{align}
\mathbf{T} = \left( \begin{array}{ccc}
r_1^\ast r_1  & r_1^\ast r_2    & r_1^\ast r_3  \\
r_2^\ast r_1   & r_2^\ast r_2   & r_2^\ast r_3  \\
r_3^\ast r_1  & r_3^\ast r_2   & r_3^\ast r_3 \end{array} \right);
%
\mathbf{X} = \left( \begin{array}{ccc}
\varepsilon_1^{\prime \ast} \varepsilon_1  & \varepsilon_1^{\prime \ast} \varepsilon_2   & \varepsilon_1^{\prime \ast} \varepsilon_3  \\
\varepsilon_2^{\prime \ast} \varepsilon_1  & \varepsilon_2^{\prime \ast} \varepsilon_2   & \varepsilon_2^{\prime \ast} \varepsilon_3  \\
\varepsilon_3^{\prime \ast} \varepsilon_1 &
\varepsilon_3^{\prime \ast} \varepsilon_2  & \varepsilon_3^{\prime \ast} \varepsilon_3
\end{array} \right)
\label{eq:TXmatrix}
\end{align}
must have the same symmetry properties, e.g., chiral and magnetic materials can only be probed with circular dichroism. 
The anisotropy of the tensor $\mathbf{T}$ allows one to observe transitions that are otherwise forbidden in Thomson scattering \cite{templeton80}. 

Since the tensor components are physical observables, the tensor must remain the same upon symmetry transformation in the point group of the atom. It implies that in cylindrical (i.e., SO(2)) symmetry we have
\begin{align}
\mathbf{T} = \left( \begin{array}{ccc}
F^{(0)} -\frac{1}{3}F^{(2)}  & -F^{(1)}  & 0 \\
F^{(1)} & F^{(0)} -\frac{1}{3}F^{(2)} & 0 \\
0 & 0 & F^{(0)} +\frac{2}{3}F^{(2)} \end{array} \right) ,
\label{eq:Tcylinder}
\end{align}
which is invariant against rotation about the $z$-axis. 
This tensor decomposes into a scalar $F^{(0)}$, a vector  $F^{(1)}$, and a traceless symmetric tensor  $F^{(2)}$, also known as deviator. 
The antisymmetric vector term $F^{(1)}$ is purely imaginary, which means it is time-odd, i.e., magnetic. Note that the  SO(2) model is valid when the crystal field is negligible compared to magnetic interactions. In the most general case, all nine elements of $\mathbf{T}$ are independent.

If the magnetic unit vector is along $\bm{\mathbf{z}}$, 
by substituting Eq.~(\ref{eq:Tcylinder}) into (\ref{eq:FE1E1})  we obtain the well-known expression for the resonant magnetic scattering amplitude \cite{hannon88,vanderlaan08}
\begin{align}
F_{E1E1} & = (\bm{\varepsilon} ^{\prime \ast} \cdot \bm{\varepsilon} ) F^{(0)} 
- i \bm{\mathbf{z}} \cdot  (\bm{\varepsilon} ^{\prime \ast}\times \bm{\varepsilon} ) F^{(1)} \nonumber \\
& +   (\bm{\varepsilon}^{\prime \ast} \cdot \bm{\mathbf{z}})  (\bm{\varepsilon} \cdot \bm{\mathbf{z}})  F^{(2)} .
\label{eq:hannon}
\end{align}

Since absorption is a special case of scattering it is possible to link the two together. The relation between the total  absorption coefficient $\sigma(\hbar \omega)$ (including all elastic and inelastic processes) and the imaginary  part of the forward scattering amplitude  is given by the optical theorem
\begin{equation}
\sigma (\hbar \omega) = \frac{4 \pi  n_0}{k} \, \mathrm{Im} \, F( \mathbf{k}^\prime = \mathbf{k}, \varepsilon ^\prime = \varepsilon, \hbar \omega) ,
\end{equation}
where  $n_0$ is the atomic density.
Applying the optical theorem to Eq.~(\ref{eq:hannon}) with $\bm{\mathbf{m}}$ along $\bm{\mathbf{z}}$ gives
\begin{equation}
\sigma =F^{(0)\prime \prime} 
- i \, \bm{\mathbf{m}} \cdot   (\bm{\varepsilon}^{\prime \ast} \times \bm{\varepsilon} ) F^{(1)\prime \prime} 
+ | \bm{\varepsilon} \cdot \bm{\mathbf{m}}|^2  F^{(2)\prime \prime} ,
\end{equation}
where the double prime indicates the imaginary part of the scattering amplitude.
$ F^{(0) \prime \prime}$, $ F^{(1) \prime \prime}$, and $ F^{(2) \prime \prime}$ correspond to the isotropic spectrum, XMCD, and XMLD, respectively.
Note that the XMCD, which is the only term linear in $\bm{\mathbf{m}}$, vanishes if  $\bm{\varepsilon}$ is real, and thus requires circular polarization.

  \subsubsection{Resonant inelastic x-ray scattering (RIXS)}
 In the RIXS process there is a single photon in the initial state with momentum $\hbar \mathbf{k}$, energy $\hbar \omega_\mathbf{k}$ and polarization $\varepsilon$
 which is scattered to $\hbar \mathbf{k}'$, $\hbar \omega_\mathbf{k}'$,  $\varepsilon'$ in the final state
\cite{kotani01,schulke07,ament11}. 
The differential cross section is
\begin{align}
\frac{d \sigma}{d \Omega} = 
 \frac{2 \pi}{\hbar}& \sum_f \left|  F_{gf}(\mathbf{k}, \mathbf{k}', \varepsilon, \varepsilon, \omega_{\mathbf{k}}, \omega_{\mathbf{k}'})
  \right|^2
  \nonumber \\
  & \times  \delta(E_g - E_f  + \hbar \omega_{\mathbf{k}} - \hbar \omega_{\mathbf{k}^\prime}) ,
\label{eq:RIXS}
\end{align}
where the resonant scattering amplitude is given by the second-order term in the KH formula
 with $H'_2 \sim \mathbf{A} \cdot \mathbf{p}$ as
\begin{align}
   F_{gf}(\mathbf{k}, \mathbf{k}', \varepsilon, \varepsilon', \omega_{\mathbf{k}}, \omega_{\mathbf{k}'}) & =
\sum_n 
  \frac{ \langle f | H'_2 | n \rangle \langle n | H'_2 | g \rangle}{E_g - E_n + \hbar \omega_{\mathbf{k}} + i \Gamma_n/2}
  \nonumber \\
&   = \langle f | H_2^{\prime \dagger} G(z) H'_2 | g \rangle  ,
\end{align}  
with  the Green's function $G(z)$ defined in Eq.~(\ref{eq:Green}).
  
RIXS should be considered as a coherent coupling of the (virtual) absorption with the re-emission process as indicated by the two matrix elements.
One can separate the full propagator $G$ into the unperturbed propagator $G_0 = (z - H_0)^{-1}$ and a term that contains the core-hole Hamiltonian $H_C$, using the identity $G = G_0 + G_0H_CG$. 
This  separates the  scattering amplitude into two parts that define the \textit{direct} and \textit{indirect} RIXS \cite{ament11}
\begin{align}
 F^{\mathrm{direct}}_{gf} & = \langle f | H_2^{\prime \dagger} G_0 H'_2 | g \rangle  , \nonumber \\
 F^{\mathrm{indirect}}_{gf} & = \langle f | H_2^{\prime \dagger} G_0 H_C G H'_2 | g \rangle .
\end{align}

 The physical picture that arises for direct RIXS is that an incoming photon promotes a core electron to an
empty valence state and subsequently an electron from a different state in the valence band
decays, annihilating the core hole, see Fig. \ref{RIXS}.
Indirect RIXS is more complicated. Here, the incoming photon promotes a core-electron to an itinerant state far above the electronic chemical potential. Subsequently, the electron in this same state decays again, filling the core-hole. Scattering of the x-rays occurs via the core-hole potential that is present in the intermediate state. It shakes up the electronic system, creating excitations to which the x-ray photon loses energy and momentum \cite{ament11}. However, it must be noticed that the two processes, direct and indirect RIXS, have the same initial and final states, differing only in the involved intermediate state(s), like in a two-slit experiment. Being impossible to say through which intermediate states the system went, the two processes interfere unless one of the two intermediate channels is forbidden by selection rules. Usually the direct and indirect RIXS processes have large cross sections for different sets of final states, therefore it is customary to analyze experimental results by considering only one of the two classes of process. For instance, at the $L_{2,3}$ edges of transition metals and at the $M_{4,5}$ edges of rare earths and actinides, charge transfer, crystal field, and spin flip-type electronic excitations can be observed through direct RIXS. At the same edges, indirect RIXS processes allow one to observe phonon, plasmon, and bi-magnon excitations \cite{forte08}.

An elastic peak (Rayleigh scattering) appears when the final state is the same as the initial state.
Furthermore, analogous to energy conservation, it is possible to apply momentum conservation between initial and final states, which allows to determine the momentum of the excitations created in the scattering process.

\begin{figure}
\includegraphics[width=1.0\columnwidth]{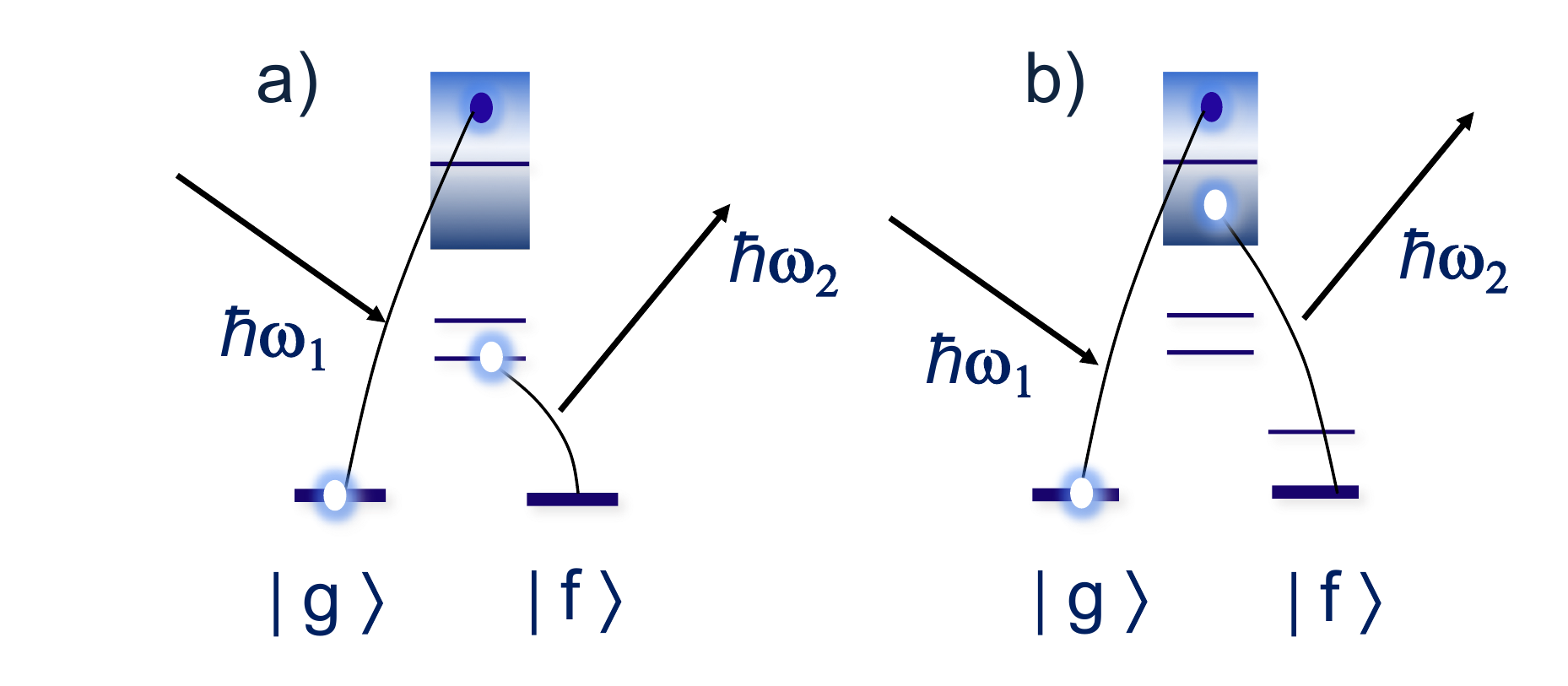}
\caption{Schematic representation of RIXS, a two-step process involving first the absorption of a photon and the promotion of a core electron into an empty valence state, followed by the emission of a photon with smaller energy and the filling of the core hole by an electron sitting either in a core level (\textit{direct core-to-core} RIXS, panel a), or in the occupied valence band (\textit{direct valence-to-core} RIXS, panel b). The process probes the convolution between empty (absorption step) and occupied (emission step) density of states and leaves an electron-hole excitation in the final state. Adapted from \cite{caciuffo21}.
\label{RIXS}}
\end{figure}

The RIXS process contains two radiative transitions, absorption and emission, which in the soft x-ray region obey
the dipole-selection rules. The total RIXS transition, however, does not follow these rules. The $ff$ excitations provide a typical
example: The dipole transition $5f \to 5f$ would be forbidden,
since $\Delta \ell = 0$, but the two steps with core $d \to 5f$ ($\Delta \ell =+1$) and $5 f \to 5d$ ($\Delta \ell = -1$)
 are allowed, making the second-order transition possible.
RIXS can therefore access transitions forbidden by dipole selection rules. As each step follows the selection rules $\Delta J = 0, \pm 1$, the RIXS process allows also $\Delta J = \pm 2$.
Under certain conditions, the orbital excitations can propagate through the crystal and give rise to an orbital wave or ``orbiton''.
Such propagating collective excitations can show dispersion that has been measured by RIXS
\cite{saitoh01a}.

RIXS also offers the possibility to explore magnetic excitations \cite{ament11}. Spin changes in the scattering process are allowed by the large spin-orbit coupling of the intermediate core-hole state. The angular momentum $\Delta L_z = 0, 1, 2$ can be exchanged through this coupling allowing magnetic excitations with $\Delta S_z = 0, 1, 2$.
The magnetic excitations are not localized but propagate in the crystal as spin-wave modes (magnons). The magnon dispersion in RIXS has shown to give comparable results as measured by inelastic neutron scattering \cite{ament09}.

\section{X-ray diffraction (XRD) experiments}\label{secXRD}

Laboratory x-ray sources are usually adequate for conventional XRD crystallography on actinide materials. However, synchrotron radiation is often necessary to perform high-resolution, low-temperature, or high-pressure measurements. Below, we review some selected examples of such XRD experiments at SR facilities.

\subsection{High pressure experiments}\label{secXRDhp}
Elemental light actinides crystallize in low-symmetry, open-packed structures that are unusual for a metal. 
However, these structures are prone to instability, and an external pressure of a few GPa is sufficient to destabilize them and induce a sequence of phase transformations. In some cases, the structural changes are accompanied by volume collapses indicating a stepwise delocalization of 5$f$ electrons and their increasing contribution to the crystal cohesion energy. High-pressure XRD experiments at synchrotron beamlines allowed one to identify the various allotropic forms assumed by the elements in the 5$f$ series, their sequence, and their range of stability. As discussed below, these experiments prompted a large amount of theoretical work, leading to the development of first-principles electronic structure calculation methods reliable also for strongly correlated systems. 

As is well known, SR has allowed optimal use of diamond anvil cells so that this field has been revolutionized by the advent of such facilities. Dedicated beamlines and special equipments have been developed at all of the major SR sources, as described, e.g., in  \cite{hirao20,liermann15,meng15,mezouar05,shen05}.
A recent review is given by \cite{mcmahon21}. A number of works have been carried out on uranium compounds, determining crystal structural variation and equations of state. Some examples are given in \cite{olsen85,yoo98,lebihan03,haire03,zvoriste13,jeffries13,dewaele13,shukla17,rittman17,
murphy18,guo19}.

\begin{figure}
   \centering
   \includegraphics[width=1.0\columnwidth]{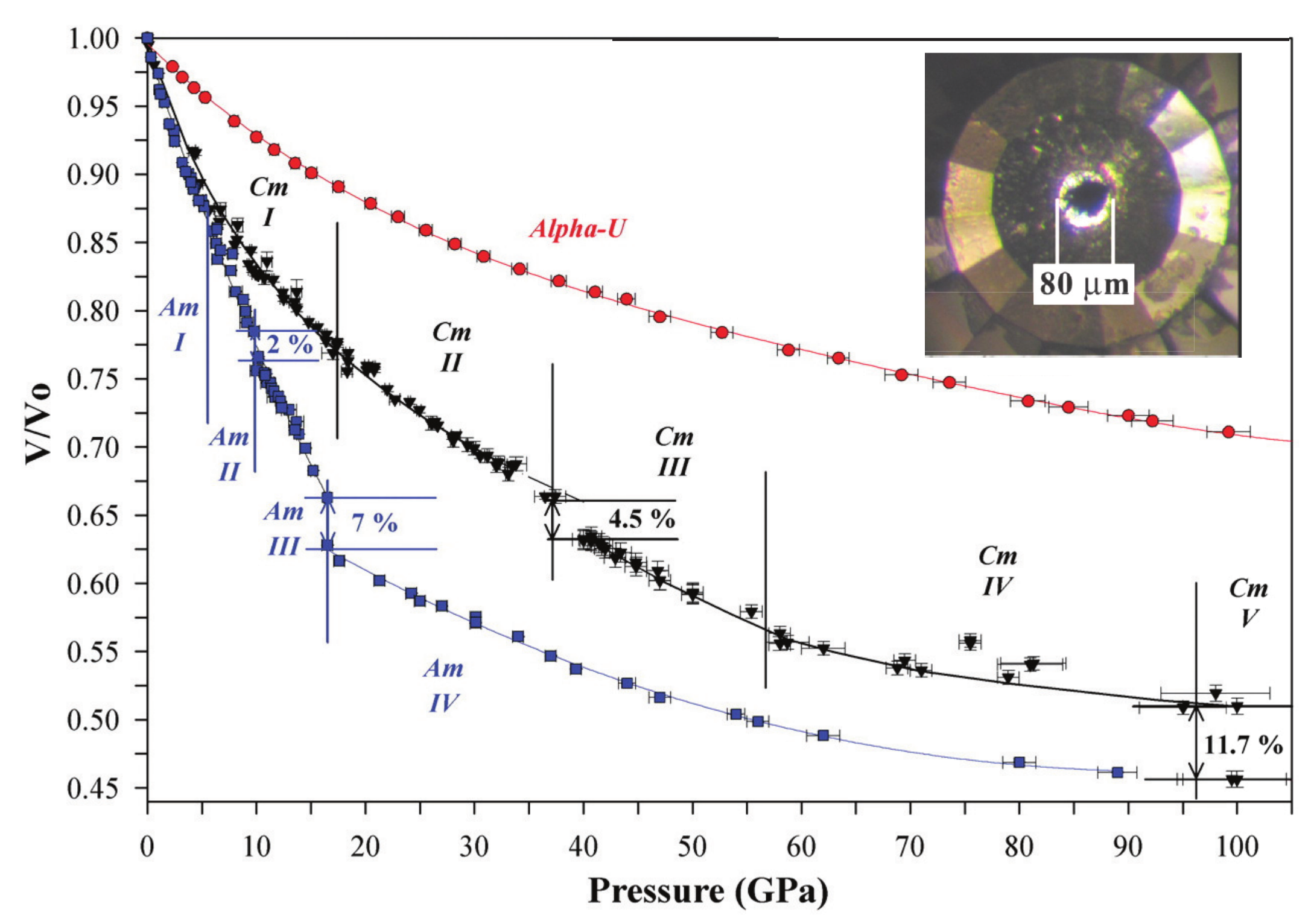}
  \caption{ Pressure dependence of the normalized unit cell volume, $V(p)/V_{0}$, for $\alpha$-uranium \cite{lebihan03}, americium \cite{lindbaum01}, and curium \cite{heathman2005}. The inset shows a typical diamond anvil cell used for angle-dispersive, high-pressure x-ray diffraction experiments.
  Adapted from \cite{heathman2005}. } 
\label{pdepAmCmU}
\end{figure}

Fig.~\ref{pdepAmCmU} shows the equation of states for $\alpha$-U, Am, and Cm as determined by high-pressure XRD. Under compression, uranium preserves its room-temperature–stable orthorhombic $Cmcm$ form up to at least 100 GPa.  A different behavior is observed for americium where the normal pressure double-hexagonal, close-packed ($P6_{3}/mmc$) structure transforms at 6.1 GPa to a face centered cubic ($Fm\bar{3}m$) lattice. At higher pressures, two lower symmetry structures appear, a face-centered orthorhombic Am III ($Fddd$) and a primitive orthorhombic structure, Am IV ($Pnma$). In the same pressure range, up to 100 GPa, curium exhibits five phases, from the double hexagonal close packed (dhcp) form of Cm I
 ($P6_{3}/mmc$) to the orthorhombic ($Pnma$) structure of Cm V. Of particular interest is the formation of the monoclinic structure with the space group $C2/c$ between about 37 and 56 GPa. Calculations based on the full potential linear muffin-tin orbital (FPLMTO) method suggest that its stabilization is driven by the magnetic correlation energy \cite{heathman2005}. The collapse of Am and Cm from simple to complex structures under pressure shows, in line with Fig.~\ref{atomicvol}, that the 5$f$ electrons have transformed from localized to itinerant under pressure.
 
Beyond Cm, the available quantities become very small, but recent work has shown that the concept  of delocalization under pressure may be a too simple picture. Examples are pressure work on Cf \cite{heathman13} and spectroscopy work on Es \cite{carter21}, both using SR techniques.

Reproducing from first-principles electronic structure calculations the observed sequence of lattice geometry and the associated evolution of physical properties in actinide elements and compounds is challenging. Simple approximations of the density functional theory, in fact, are not adequate for actinide materials, because of the strong intra-atomic correlations, the importance of scalar (first-order kinetic energy correction due to the mass variation) and non-scalar (spin-orbit coupling) relativistic effects, and the extent of the hybridization between 5$f$ and conduction electrons. In many cases a model that produces the correct crystal structure at a certain atomic volume fails to describe the electronic structure near the Fermi level, and does not reproduce the correct magnetic state. This is the case, e.~g., of conventional density functional theory (DFT) calculations in the local-spin density or generalized-gradient approximation (LDA/GGA) applied to $\delta$-Pu \cite{lashley05,clark19b}, or of static mean field correlated band theory calculations making use of different flavors of the LDA/GGA plus Coulomb U (LDA + U) method, falling short in describing the itinerant-to-localized crossover of the 5$f$ manifold in PuCoGa$_{5}$ \cite{daghero12,shick13}. During the last two decades, experiments as those reported in Fig.~\ref{pdepAmCmU} have, therefore, stimulated the development of increasingly sophisticated theoretical models that have now reached predicting capabilities close to a material-by-design level also for compounds in the 5$f$ block \cite{kotliar06,shim07,suzuki10,pezzoli11,franchini19,pourovskii21,shick21}.  A recent review on methods for computing the electronic structure of correlated materials from first principle is given by \cite{nilsson18}.

\subsection{High intensity and resolution}

The enormous increase in flux of modern synchrotron beams, as compared to conventional x-ray sources, has, of course, meant that much smaller samples can be studied with increasing resolution.

As an example of a high resolution XRD study, we show in Fig.~\ref{Pu115difpat} the results obtained for the PuCoGa$_{5}$ unconventional superconductor \cite{eloirdi17}. 
The experiment was performed at the ESRF ID22 beamline, affording a resolution $\Delta d/d$ = 10$^{-6}$ at $\lambda$ = 0.354155 \AA. Data were collected on a 4.6 mg sample, obtained by crushing a single crystal grown from metallic plutonium (99.932 wt\% $^{242}$Pu), put inside a hermetic holder providing four levels of containment.
 The absence of visible splitting or broadening of the diffraction peaks, as seen in the inset of Fig.~\ref{Pu115difpat} for the (220) Bragg reflection, indicates that the tetragonal symmetry is preserved in the superconducting phase. The temperature dependence of the refined lattice parameters shows that the thermal expansion is isotropic above $\sim$150 K. At lower temperatures, the $c/a$ ratio increases with decreasing $T$. In the same temperature range,  \cite{ramshaw15} observe a softening of the bulk modulus that they attribute to the development of in-plane hybridization between conduction electron and Pu 5$f$.

Below $T_{c}$ = 18.5 K, the critical temperature to the superconducting state in PuCoGa$_{5}$, the expansion of the unit cell volume deviates from the predictions of a simple one-phonon Gr\"{u}neisen-Einstein model (Fig.~\ref{Pu115alphaV}). The shrinking of the cell volume is similar to the one observed for the CeRu$_{2}$Si$_{2}$ Kondo system \cite{hiranaka13} and could suggest the occurrence of critical valence fluctuations at $T_{c}$ \cite{miyake14}, where the volume thermal expansion coefficient $\alpha_{V}$ has a jump larger by a factor $\sim$20 than the value predicted by the Ehrenfest relation.

\begin{figure}
\centerline{ \includegraphics[width=1.0\columnwidth]{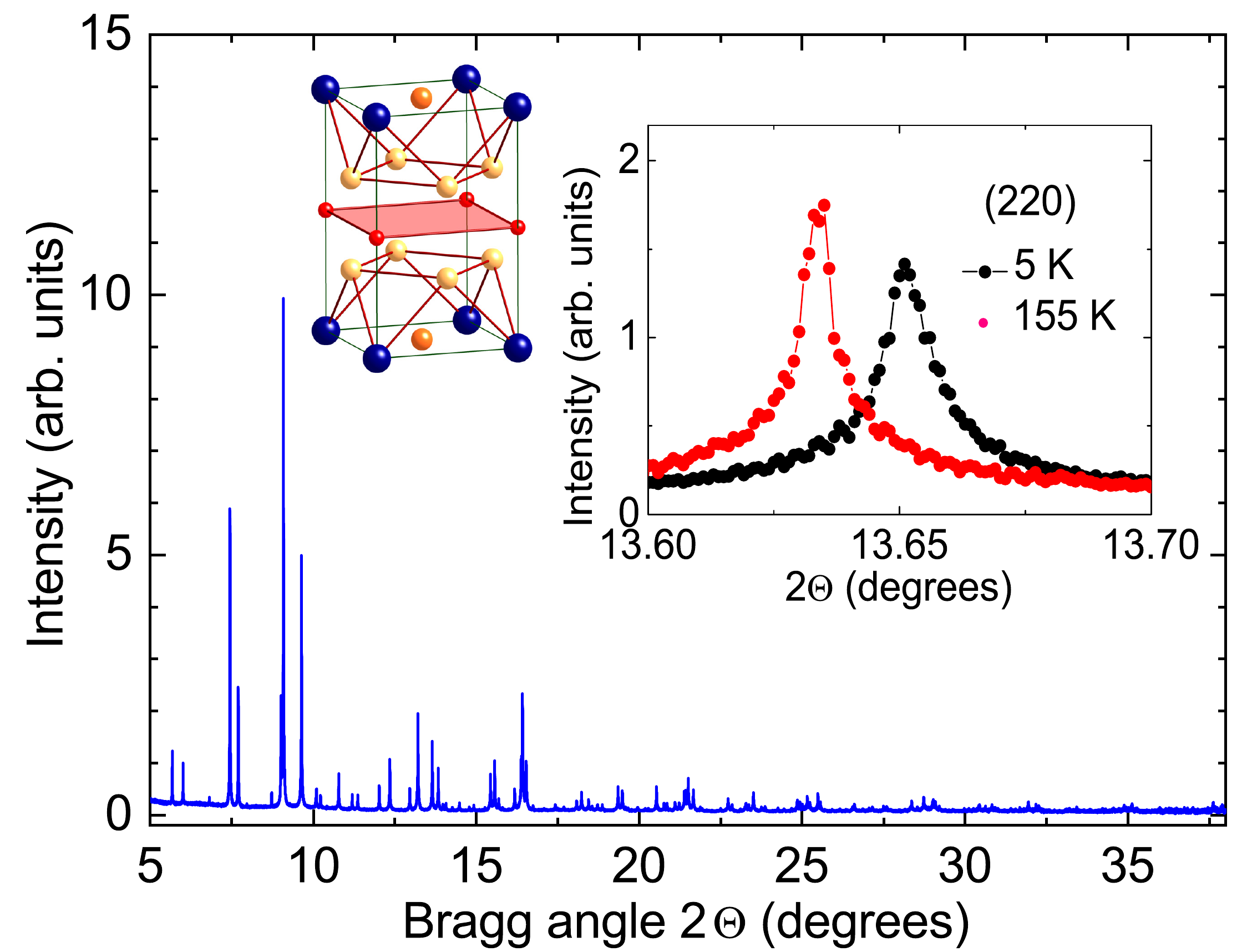}}
\caption{ X-ray diffraction pattern recorded for PuCoGa$_{5}$ at 5 K. The inset on the left shows the tetragonal crystallographic unit cell, with Pu and Co atoms represented by blue and red spheres, respectively, whilst the two inequivalent Ga atoms are shown by orange and yellow spheres. The inset on the right shows the (220) Bragg peak measured at 5 K (black circles) and 21 K (red circles).
Adapted from \cite{eloirdi17}. 
\label{Pu115difpat}}
\end{figure}
 
\begin{figure}
\centerline{\includegraphics[width=1.0\columnwidth]{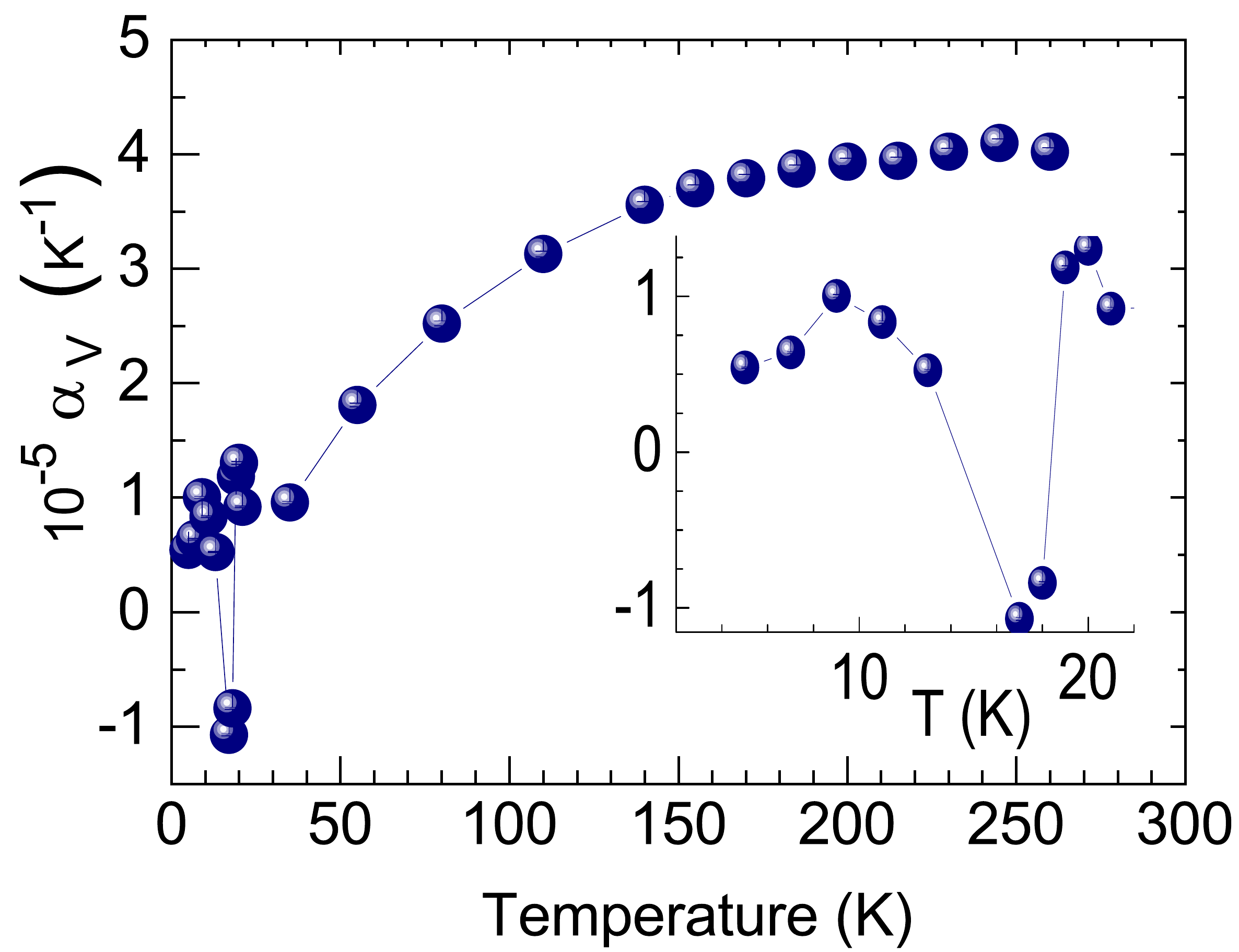}}
\caption{ Thermal expansion coefficient for the unit cell volume of PuCoGa$_{5}$ (on an expanded scale around $T_c$ in the inset). Error bars have been estimated as 5$\times$ the statistical error provided by the Rietveld refinement and are smaller than the data symbol size.
Adapted from \cite{eloirdi17}. 
\label{Pu115alphaV}}
\end{figure}

Diffraction experiments on other transuranium samples, such as NpFeAsO, have also benefitted from the additional intensity and resolution of  synchrotron beamlines \cite{klimczuk12}. Having understood the structure of the Np-compound by performing both synchrotron and neutron experiments on polycrystalline samples, for the Pu analog there was no need to use synchrotron x-rays \cite{klimczuk12b}.

In addition to looking at polycrystalline samples of complex materials, it is also possible to perform new experiments with synchrotron beams that are almost impossible to perform with conventional x-ray sources.  A good example of such a study is that examining the change in the surface morphology of a UO$_{2}$ single crystal by measuring the so-called truncation rods as a function of time while the sample is exposed to oxygen. The work by \cite{stubbs15} shows the power of this technique. Figure  \ref{Stubbs15}, taken from  \cite{stubbs15}, shows  a modulation of the layers that incorporate the extra oxygen so that they are organized to appear in every third layer down in the case of the (111) surface. The arrangement is modeled from the intensity of the truncation rods measured in the Advanced Photon Source (APS) at the Argonne National Laboratory (ANL). A very similar situation was found in the case of the UO$_{2}$ (001) surface \cite{stubbs17}.

\begin{figure}
\centerline{\includegraphics[width=1.0\columnwidth]{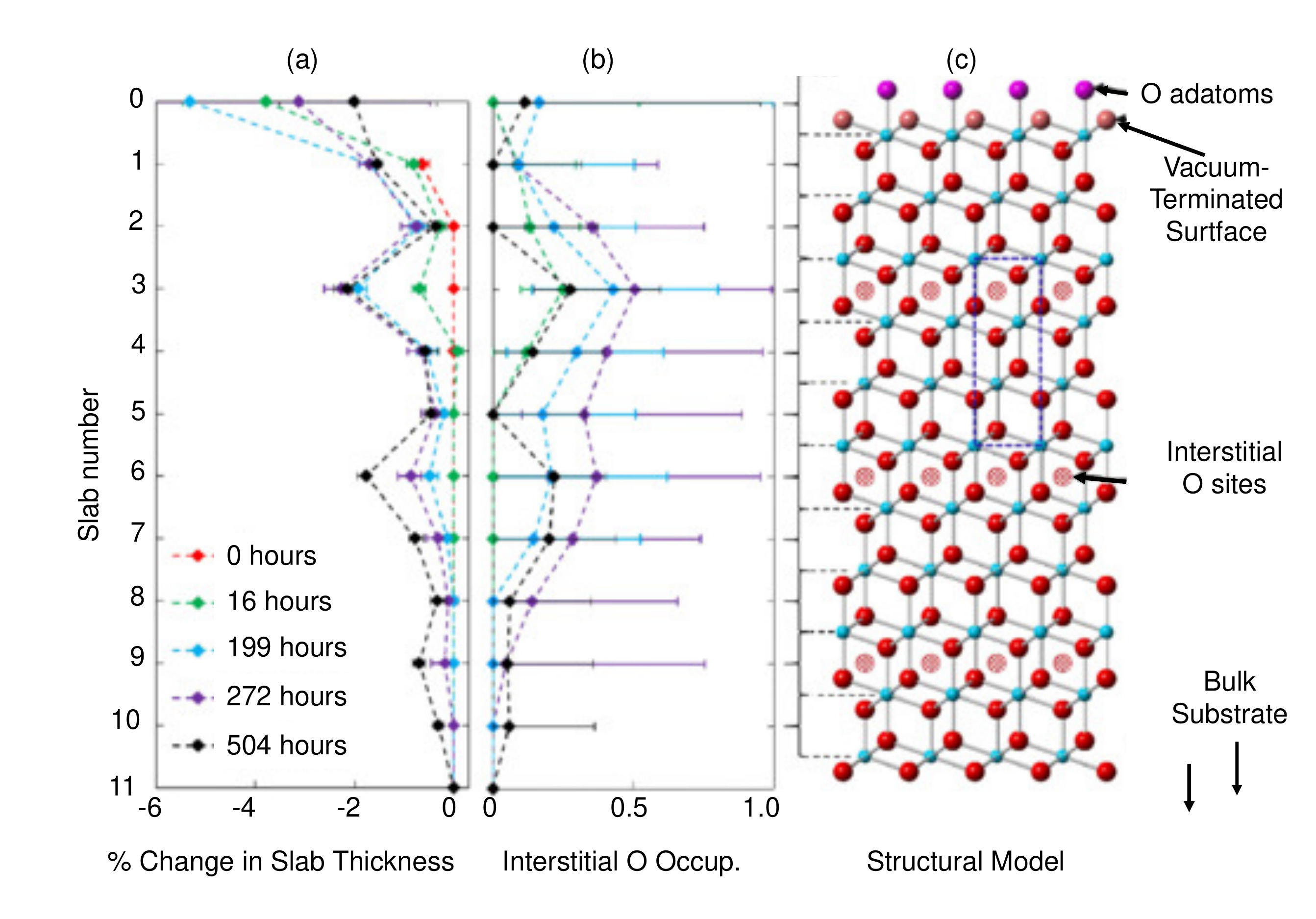}}
\caption{ Refined slab contractions (a) and interstitial occupancies (b) are plotted vs depth and show three-layer periodicity. (c) Proposed model for oxidized  UO$_{2}$ (111) surface. U is cyan, structural O is red, surface O adatoms are magenta, and interstitial O is hatched red. The surface is oxygen terminated.  $YZ$ projection of surface unit cell is indicated by blue dashed lines.
Reproduced  from \cite{stubbs15}. 
\label{Stubbs15}
}
\end{figure}

Another experiment demanding intense SR was that performed at APS on molten UO$_{2}$, which has been levitated and then heated with a laser beam, to determine the structural changes near the melting temperature (3138 K)   \cite{skinner14}. The hot solid shows a substantial increase in oxygen disorder around the
$\lambda$ transition (2670 K) but negligible U-O coordination change. On melting, the
average U-O coordination drops from 8 to 6.7 $\pm$ 0.5. Molecular dynamics models refined to
this structure predict higher U-U mobility than in 8-coordinated melts.

Finally, in this section, we cover briefly a study attempting to use non-resonant scattering to determine the individual spin and orbital moments in the uranium monopnictide (with the simple NaCl crystal structure) UAs. That x-ray scattering has a “magnetic” component was known many years ago, and pioneering experiments had already shown such weak magnetic scattering from a conventional x-ray source with the antiferromagnet Fe$_2$O$_3$  \cite{debergevin81,brunel81}. However, the signal was very weak, as it is many orders of magnitude smaller than any charge contribution. With the much higher intensity available from a synchrotron beam, together with a new formulation of the theory \cite{blume88}, an attempt was made at the ESRF on the actinide system UAs  \cite{langridge97}.  Five different antiferromagnetic satellites were observed and the final value of the ratio of the orbital to spin moment was found to be $\mu_L/\mu_S$ = $-$2.0 $\pm$ 0.5. The expected values for simple 5$f^{2}$ or 5$f^{3}$ configurations are $-$3.30 and $-$2.6, so the result suggested a 5$f^{3}$ configuration, as generally accepted. Despite using SR, this experiment turned out to be very difficult, requiring accurate intensity measurements of weak satellite reflections using photons of 8.1 keV. To our knowledge these type of experiments have not been repeated on any actinide system, perhaps because of the development of another technique, x-ray magnetic circular dichroism (XMCD), which is discussed in Sec.~\ref{secXMCD}. However, XMCD cannot be used for antiferromagnetic materials.

\subsection{Experiments at high magnetic fields}
A number of experiments using pulsed magnetic fields have been performed with SR, e.g., \cite{detlefs08a}, but the first one on an actinide system, was reported from the APS on UO$_{2}$  \cite{antonio21}. The results are shown in Fig.~\ref{antonio21}.

\begin{figure}
 \centerline{\includegraphics[width=1.0\columnwidth]{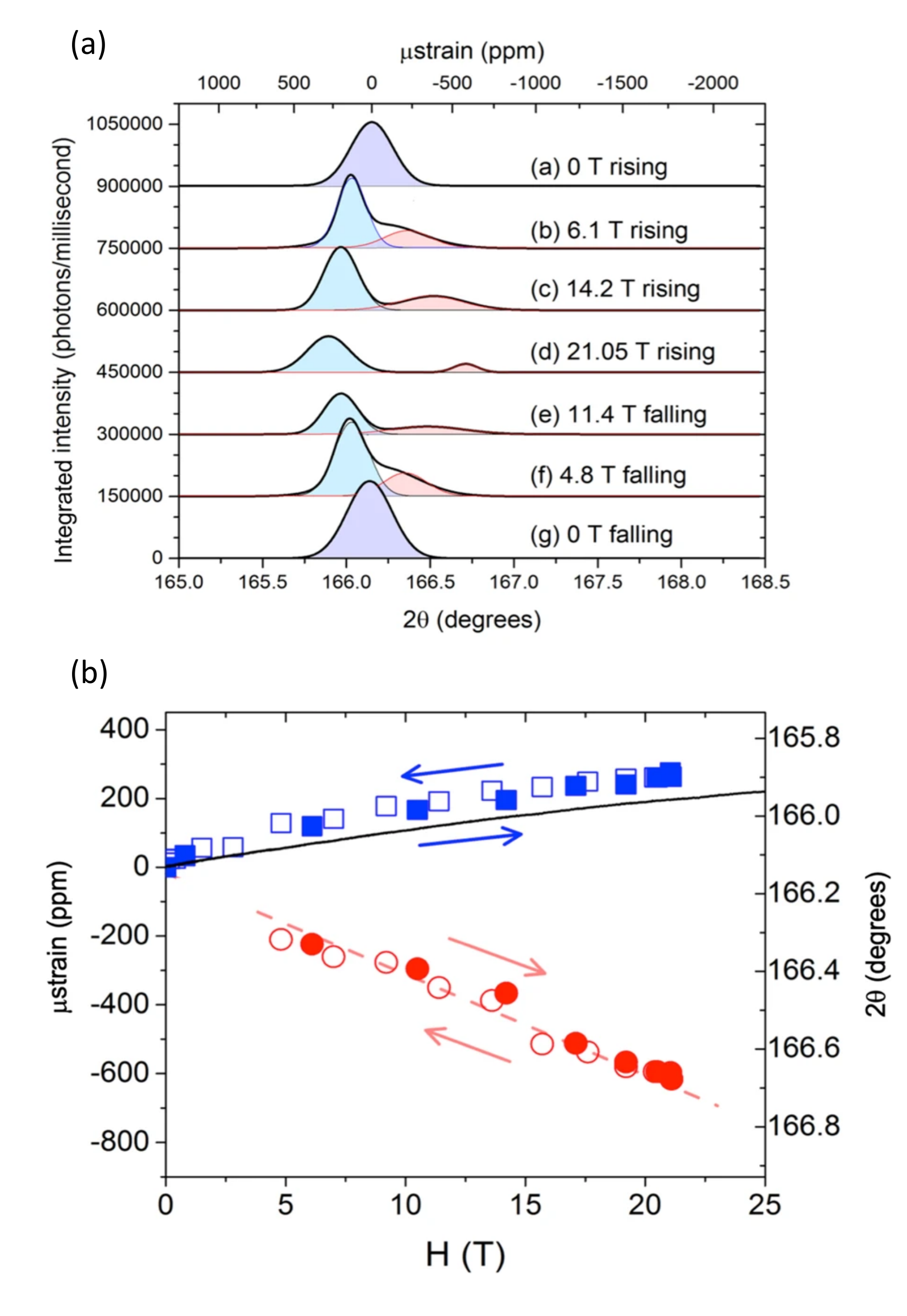}}
\caption{ (a) Waterfall plot of fits to the integrated intensity along the full diffraction angle 2$\theta$ of selected fields at 15 K.  The peak at lower 2$\theta$ (blue) corresponding to expansion is much larger than the peak at higher 2$\theta$ (red). (b) Peak positions of the two peaks at 15 K, with filled symbols corresponding to rising field and open symbols to falling field. Red circles represent the smaller peak and blue squares the larger one. The blue peak and its positive magnetostriction can be seen to match magnetostriction measurements using the fiber Bragg grating technique (black line), with $\sim$150 ppm expansion at the maximum 21.1 T applied field. The red dashed line is a guide to the eye.
Reproduced from \cite{antonio21}. 
\label{antonio21}
}
\end{figure}

As discussed further on, UO$_{2}$ is an antiferromagnet (AF) with $T_N$ = 30.8 K and orders with a so-called 3-$\mathbf{k}$ AF structure, as determined many years ago. Recently, work at high magnetic field at Los Alamos National Laboratory has shown that interesting magnetoelastic effects occur when high magnetic fields are applied along the
$\left<111\right>$ direction with the samples in the AF state \cite{jaime17}. Further information of these effects was obtained by performing an experiment at the APS synchrotron with applied magnetic fields up to 25 T \cite{antonio21}. Because of the constraints of the experiment only one reflection, the (888), could be measured, which, for $T > T_N$, was a single reflection. However, as shown in Fig.~\ref{antonio21}, in the AF state, two reflections were observed, and the process was reversible. The authors have discussed this in terms of piezo-magnetic domains \cite{antonio21}, although they have not explained how such domains having different $d$-spaces to give the two reflections. A simpler quantitative explanation exists if one assumes the crystal undergoes a rhombohedral distortion along a domain with its $\left[111\right]$ $||$ $\mathbf{H}$. Other $\left<111\right>$ type directions in the crystal (which are not parallel to $\mathbf{H}$) will have a $d$-space inequivalent to those with $\mathbf{H}$ $||$
$\left[111\right]$. This hypothesis, described in \cite{lander21b}, explains the presence of two reflections (Fig.~\ref{antonio21}), and their relative slopes and intensities as a function of applied H. A complete understanding of this effect and how it relates to earlier work still remains to be determined.

\subsection{Materials science and microstructural studies}\label{materialsscience}

The intensity and micron-dimensions of the x-ray beams  has also allowed a multitude of experiments to be performed in what might be loosely called “materials science”. We give four examples in this section, and there is little doubt that other experiments will be done in the future.  Normally, these experiments use high-energy x-rays to ensure deep penetration into the material being examined. There is also use of tomography techniques (that we mention here, although not related to diffraction) for determining the pore (or impurity) structure inside the sample,  i.e.,  the method is non-destructive.

An interesting experiment performed on thin films and following the diffraction pattern as a function of time (and hydrogen dose) was that by \cite{darnbrough18} at the ESRF. The sample started with an epitaxial film of $\alpha$-U and then a layer of polycrystalline UO$_{2}$ was either deposited on top of the epitaxial metal film, or an oxide film was allowed to grow naturally by exposing the film to air. After this, the diffraction pattern was taken over a period of time when hydrogen was bled into the chamber with the sample being also heated. The U-H reaction is known to be strongly exothermic, and these experiments exhibit also an anisotropy; certain faces in the U-metal react more rapidly with H$_{2}$ than others. The entry of hydrogen is seen to reduce the UO$_{2}$ lattice and expand that of the metal. It appears that the reaction is strongly localized at grain boundaries, so that amorphous or nanocrystalline UH$_{3}$ is formed at such boundaries and no diffraction peaks from this material are observed. More experiments need to be performed, but this represents an interesting first test.

Tomography has turned out to be a key feature that can be performed with high-energy synchrotron beams and can achieve micron-size resolution. With  the new optic features of recent synchrotron upgrades, this resolution will reach sub-micron. We feature three papers below that have been exemplary in opening up new areas in materials science. The first \cite{bonnin14} explores the impurities (normally around the 1\%) level in the atomization process that is used to prepare U-Mo droplets, from which pellets are prepared for the new generation of low-enriched uranium fuels for research reactors. The impurities are UO$_2$ and grains of U(C,O). The experiments, done on ID11 at the ESRF, used a beam of $\sim$30 keV and a focal spot size of 0.11 $\times$ 0.14 $\mu$m$^{2}$. The results showed that the UO$_2$ deposits on the surface of the droplets (not surprisingly) but the U(C,O) impurity phases, which have the \textit{fcc} NaCl structure, grow directly from the \textit{bcc} U-Mo main component. For example, it was found that the diameter of the \textit{bcc} U-Mo particles is much larger in the regions where U(C,O) impurities were found. Evidence shows that the onset of U(C,O) grain crystallization can be described by a precipitation mechanism, since one single U-Mo grain has direct orientation relationship with more than one surrounding U(C,O) grains. In addition, the porosity could be directly observed, and the relationship with the outer UO$_2$ coating determined.

\begin{figure}
\centerline{\includegraphics[width=1.0\columnwidth]{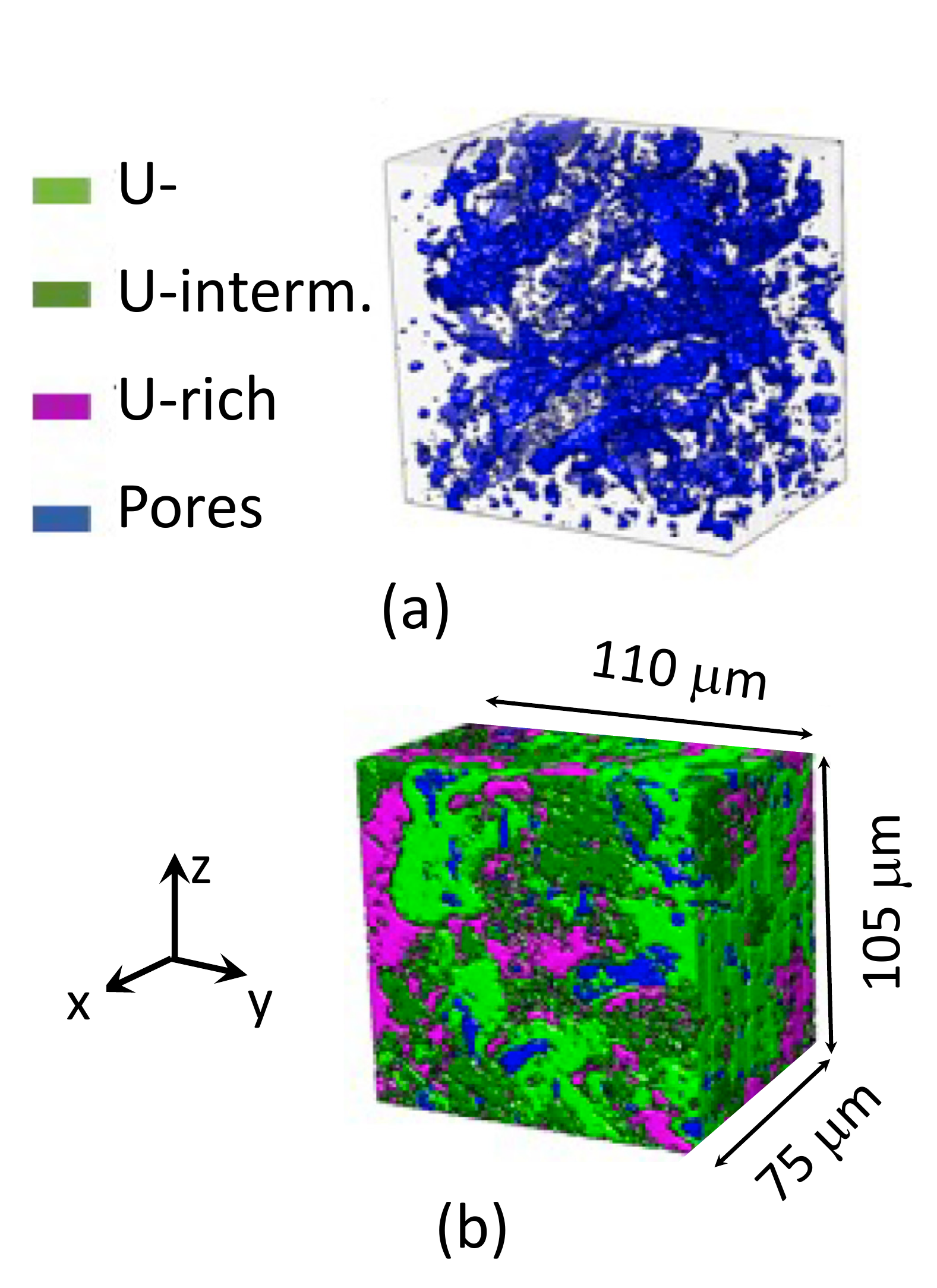}}
\caption{Three-dimensional segmentation of the propagation-based phase contrast enhanced tomography images of U-10Zr FIB cuboid indicating the (a) pores, and (b) composite image of pores and all three phase regions.
Reproduced from \cite{thomas20}. \label{thomas2020}
}
\end{figure}

The second example examined an irradiated U-Zr fuel \cite{thomas20}.  The sample had a volume of
$\sim$125$\times$90$\times$100 $\mu$m$^{3}$ which was prepared by using a focused-ion beam (FIB) method from a sample of U-Zr 10\% wt fuel pin that was irradiated to 5.7 at.\% burnup in a reactor. The experiments were performed on beamline 1-ID-E  at the APS with a beam of incident energy of 71.68 keV. The total radiation inventory is not given, but the safety conditions included the specification that the sample had to give less than 5 mR/h at 30 cm, which was achieved. On contact the sample measured 40 mR/h. The principal result is a detailed mapping of the U concentration and the pores in this irradiated material, as shown in Fig.~\ref{thomas2020}.

A clear indication of the pore structure in such an irradiated fuel has not been obtained before, so there is much to be analyzed here and compared with estimates of such pore sizes by numerical methods. Interestingly, the localized microscopic swelling of the fuel was estimated from these results as just under 8\%, whereas the macroscopic swelling has been measured as a function of burnup, and has been found to be about twice this value, i.e., $\sim$16\%. This large discrepancy is not totally understood, but the authors give a number of reasons that may play a role.

The third example is a x-ray tomography experiment conducted at room temperature and 1000$^{\circ}$C on unirradiated Tristructural Isotropic (TRISO) nuclear fuel particles under uniaxial compression \cite{liu20}. The TRISO particles were part of a neutron irradiation experiment \cite{knol12} made at the High Flux Reactor (HFR) in Petten (the Netherlands) to study the impact of high temperature fast neutron irradiation on the thermo-mechanical properties of various coating materials for TRISO-coated particle fuel. X-ray tomography measurements were performed at the beamline 8.3.2 of the Advanced Light Source (ALS) at the Lawrence Berkeley National Laboratory, USA, using a monochromatic beam of 25 keV.  The results allowed the authors to correlate microstructure and dimensions of the individual layers of each particle to the load-displacement behavior and provided real time information of the evolution of porosity, crack formation, and layer thickness as a function of stress and temperature \cite{liu20}.

These last three experiments use tomography, a technique requiring vast arrays of data, so that capabilities go hand in hand with advances in image reconstruction. We can expect more of these type of experiments which feed directly into questions of importance to the nuclear industry in terms of fuel fabrication, utilization, and (later no doubt) fuel treatment and waste disposal.

In this field it is clear that the advances in synchrotron lattice upgrades, optics, and resolution will further enhance capabilities, and using free-electron lasers  the beams will be even smaller.

Another technique that we do not describe in detail in this review but one that has considerable potential in the study of colloidal and nanostructured materials is small angle x-ray scattering (SAXS), i.e., the diffuse scattering produced around the incident beam by sample inhomogeneities (scattering length fluctuations) with sizes several times larger than the nearest-neighbors interatomic distances. A SAXS experiment is sensitive to microstructural details up to a few thousand \AA\ and provide information on the surface, volume, and shape of the inhomogeneities, on the distribution of their sizes, and on the inter-particle correlation. In actinide science, SAXS has been used, for instance, to study the formation of atomic clusters \cite{burns05}, polynuclear colloidal particles \cite{zhai22}, the interactions between natural colloidal organic matter and actinides in solutions \cite{tian20}, and the solution speciation of metal-oxo clusters \cite{nyman17}.
\\
\\

\section{REXS experiments}\label{secREXS}

\subsection{Electric quadrupole order}\label{secREXSEQ}
Resonant elastic x-ray scattering (REXS) occurs when a photon is absorbed promoting a core electron to an empty state, and is subsequently re-emitted when the electron and the core hole recombine. The process can introduce an anisotropic contribution to the x-ray susceptibility tensor, whose amplitude strongly increases when the photon energy is tuned to an atomic absorption edge. The scattering amplitude also depends on the initial and final polarization of the photons. Measurements are usually performed with incident photons linearly polarized along the direction perpendicular to the scattering plane
($\sigma$ polarization), whilst optionally a polarization analyzer is used to detect photons linearly polarized either along the same direction ($\sigma$$\sigma$ channel) or parallel to the scattering plane ($\sigma$$\pi$ channel). In REXS experiments performed without energy analysis of the scattered photons, a careful check of the inelastic background should be performed, in particular when dealing with weak and broad peaks in the reciprocal space.

The first REXS measurements were performed with photons tuned to the $L$ edges of holmium metal \cite{gibbs88}, and it was soon realized that to observe the maximum magnetic effects one had to tune to energies that were associated with empty states of the spin-polarized electrons, i.e., $L_{2,3}$ edges for 3$d$ systems, and $M_{4,5}$ edges for 4$f$ and 5$f$ systems. At the $M_4$ edge of uranium in UAs, an antiferromagnetic compound with the rocksalt \textit{fcc} crystal structure, an increase in intensity of six orders of magnitude was observed for the magnetic Bragg peak in the ordered phase \cite{isaacs89}. 

For an electric-dipole transition (E1), involving the excitation 3$d_{3/2,5/2} \rightarrow$ 5$f$ at the $M_{4,5}$  absorption edges of an actinide atom, the resonant x-ray scattering amplitude contains a scalar term, probing the electric charge, a rank-1 tensor odd in time reversal symmetry, probing the magnetic dipole moment, and a rank-2 tensor even under time reversal, probing the electric quadrupole moment 
\begin{equation}
\begin{aligned}
f_{\mathrm{E1}} = { } & \frac{2}{3}( \bm{\varepsilon}'^\ast \cdot \bm{\varepsilon}) [F_{11} + F_{1-1} + F_{10}]   \\
& - i (\bm{\varepsilon}'^\ast \times \bm{\varepsilon}) \cdot \bm{z}  [F_{11} - F_{1-1}]   \\
& + \frac{1}{3}( \bm{\varepsilon}'^\ast \cdot \mathbf{T} \cdot \bm{\varepsilon}) [2F_{10} - F_{11} - F_{1-1}],
\label{ResAmpl}
\end{aligned}
\end{equation}
\noindent
where $\bm{\varepsilon}$ ($\bm{\varepsilon}'$) are unit polarization vectors of the incident (scattered) photons, $\bm{z}$ is the direction of the magnetic dipole moment, 
$\mathbf{T}$ is a rank-2 tensor proportional to the electric quadrupole operators or arising from an intrinsic asymmetry of the crystal lattice, and $F_{1q}$ ($q = 0, \pm 1$) are resonant energy factors \cite{hill96}. 
In resonant scattering, where the excitation is from core states that are close to the nucleus, the scatterers can be taken as points in space. This implies that the scattering will have almost no $Q$ dependence, unlike non-resonant x-ray scattering with an atomic form factor. In this sense there is an analogy with neutron scattering from the nucleus, which can also be regarded as a point source.

If magnetic-dipole moments or electric-quadrupole exhibit long-range order, and the photon energy is large enough for diffraction to occur, the interference of the anomalous scattering amplitudes lead to the appearance of Bragg peaks at positions $\mathbf{Q}$ forbidden by the crystallographic space group.  Their intensity depends on the incident photon energy across the $M_{4,5}$ actinide absorption edge, on the polarization of the incident and diffracted photons, and on the sample rotational orientation around the scattering vector $\mathbf{Q}$ (azimuthal angle $\rho$). 

In case of electric quadrupole order, the structure factor is obtained from the electric quadrupole operators in cartesian components, $\mathbf{T}_n = (J_{i}J_{j}+J_{j}J_{i})/2$, ($ijk=xyz$) and $\mathbf{J}$ being the angular momentum operator, as 
\begin{equation}
\mathbf{F'}(\mathbf{Q}) =  \sum_{n} \mathbf{T}_n \exp (i \mathbf{Q} \cdot \mathbf{r}_n)  ,
\label{strucfact}
\end{equation}
\noindent
where the sum runs over all the atoms in the unit cell, at positions $\mathbf{r}_n$, and the scattering amplitude at resonance conditions is $F(\mathbf{Q}) = \bm{\varepsilon}' \cdot \mathbf{F'}(\mathbf{Q}) \cdot \bm{\varepsilon}$.  

\begin{figure}
\centerline{ \includegraphics[width=1.0\columnwidth]{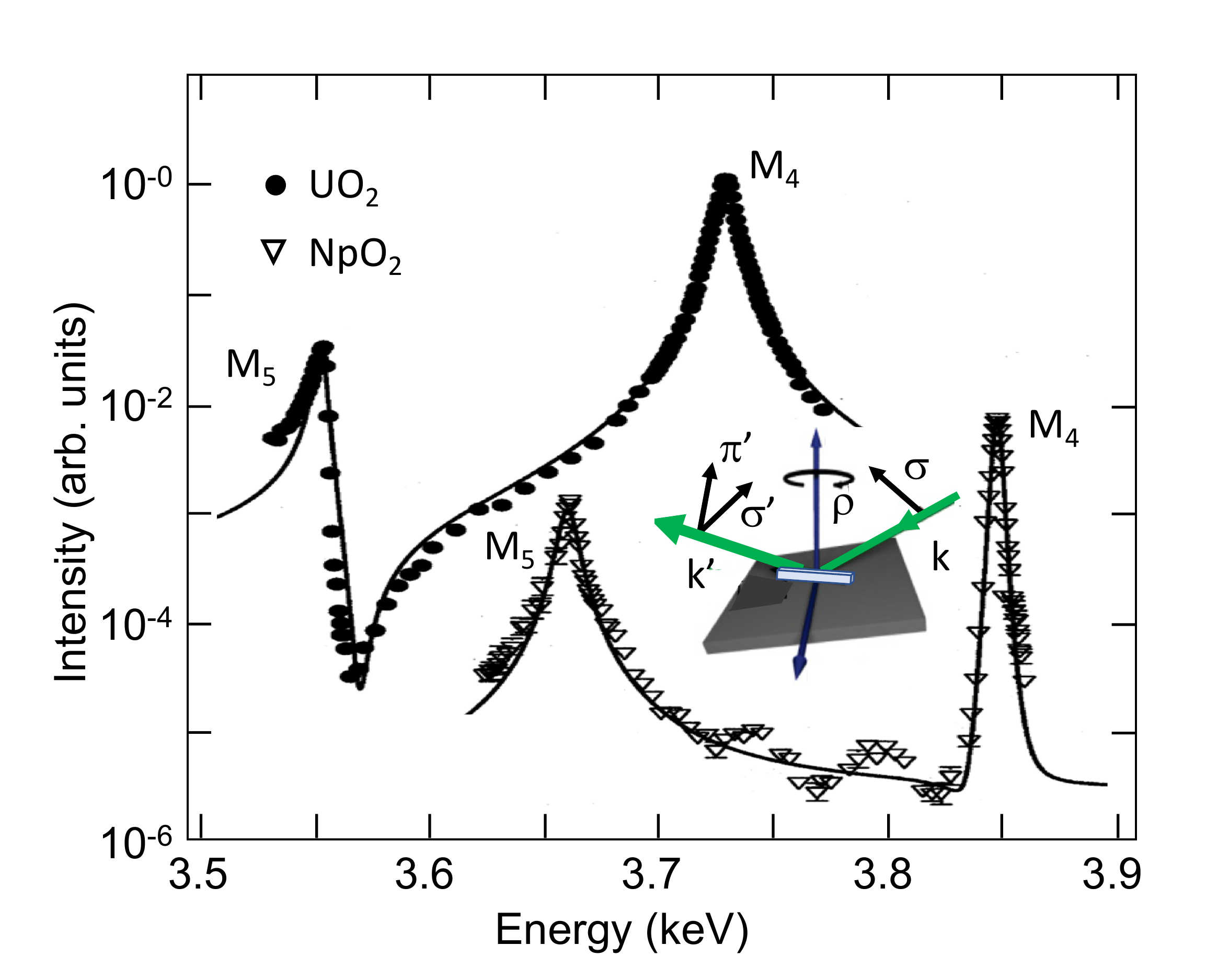}}
\caption{Integrated intensity of the (003) superlattice Bragg peak as a function of the photon energy around the $M_{5}$ and $M_{4}$ absorption edge of U and Np in UO$_{2}$ and  NpO$_{2}$, respectively. Data were collected in the ordered phase, with the incident beam polarized perpendicularly to the diffraction plane ($\sigma$) and the scattered photon beam polarized in the diffraction plane ($\pi$). The intensity data are corrected for self-absorption. The maxima of the intensity enhancement occur at the electric-dipole transition threshold energy and are associated with 3$d_{3/2,5/2}$ $\rightarrow$ 5$f$ transitions. Note that the shape of the $M_{4}$ resonance in NpO$_{2}$ is a Lorentzian squared, as predicted in \cite{nagao05}. The inset is a sketch of the scattering geometry used to measure the integrated intensity as a
function of azimuthal angle $\rho$, describing the rotation of the sample about the scattering vector. The arrows $\varepsilon(\sigma,\pi)$ correspond to
the polarization direction of photons polarized perpendicularly ($\sigma$) or parallel ($\pi$) to the scattering plane. Adapted from \cite{caciuffo21}.
\label{RXD1}}
\end{figure}

The capability of the REXS technique to detect higher-order electric multipoles has, of course, been of great interest and some examples are given in \cite{santini09}. Important early work was done on UPd$_3$ \cite{mcmorrow01,walker06,walker08c,walker11a} and extended experiments were carried out to investigate the ordered ground state of actinide dioxides. Notably,
REXS experiments provide direct evidence for the ordering of electric quadrupole moments in UO$_2$ \cite{wilkins06}, NpO$_2$ \cite{paixao02,caciuffo03a}, and in mixed U$_{1-x}$Np$_{x}$O$_{2}$ solid solutions \cite{wilkins04}. These oxides crystallize in the {\it{fcc}} fluorite structure, but in the ordered phase resonant superlattice Bragg peaks appear at positions that are forbidden in the $Fm\bar{3}m$ space group, such as $\mathbf{Q}$ = ($0 0 \ell$), $\ell = 2n+1$. The nature of the order parameter can be established by analyzing the azimuthal angle dependence of their intensity in different polarization channels. As shown by Eq.~(\ref{ResAmpl}), the term probing magnetic dipoles rotates the photon polarization by
$\pi/2$. If measurements are performed with $\sigma$-polarized incident photons, magnetic scattering appears only in the $\sigma$$\pi$ channel, whereas quadrupole scattering will contribute to both $\sigma$$\pi$ and $\sigma$$\sigma$ channels. The intensity modulation of the \textit{forbidden} peaks provides information on the relative orientation of the moments carried by the atoms in the base of the unit cell.

UO$_2$ orders at $T_N$ = 30.8 K. The primary order parameter is the magnetic dipole, whilst electric quadrupoles act as secondary order parameters. A 3-$\mathbf{k}$, type-I, transverse antiferromagnetic structure with propagation vector $\mathbf{k}$ = (001) becomes stable below $T_N$ \cite{burlet86,blackburn05}. The symmetry of the lattice is reduced to $Pa\bar{3}$ and the uranium sublattice becomes simple cubic. Each of the four atoms in the base ($C_{2h}$ point group) carries an electric quadrupole moment given by a linear combination of the three $\Gamma_5$ quadrupoles transforming as $xy$, $xz$, and $yz$. The resulting order is also transverse 3-$\mathbf{k}$. The two possible symmetry-equivalent S domains are shown schematically in panels (b) and (c) of Fig.~\ref{3korder}. A visual inspection of  the figure makes it immediately evident that 
for a transverse configuration (panels (b) and (c)), and not for the longitudinal configuration (panel (a)), 
the order of the electric quadrupoles must be accompanied by an internal distortion of the oxygen sublattice as a consequence of the perturbed electrostatic interaction between the oxygen anions and the asymmetric 5$f$ electronic cloud around the uranium ions. This distortion of the oxygen atoms in UO$_2$ was first reported  in 1975 using neutron scattering \cite{faber75}, but the reason for the distortion was not understood at the time. Neutrons cannot observe the electric quadrupoles.

\begin{figure}
\centerline{\includegraphics[width=0.5\columnwidth]{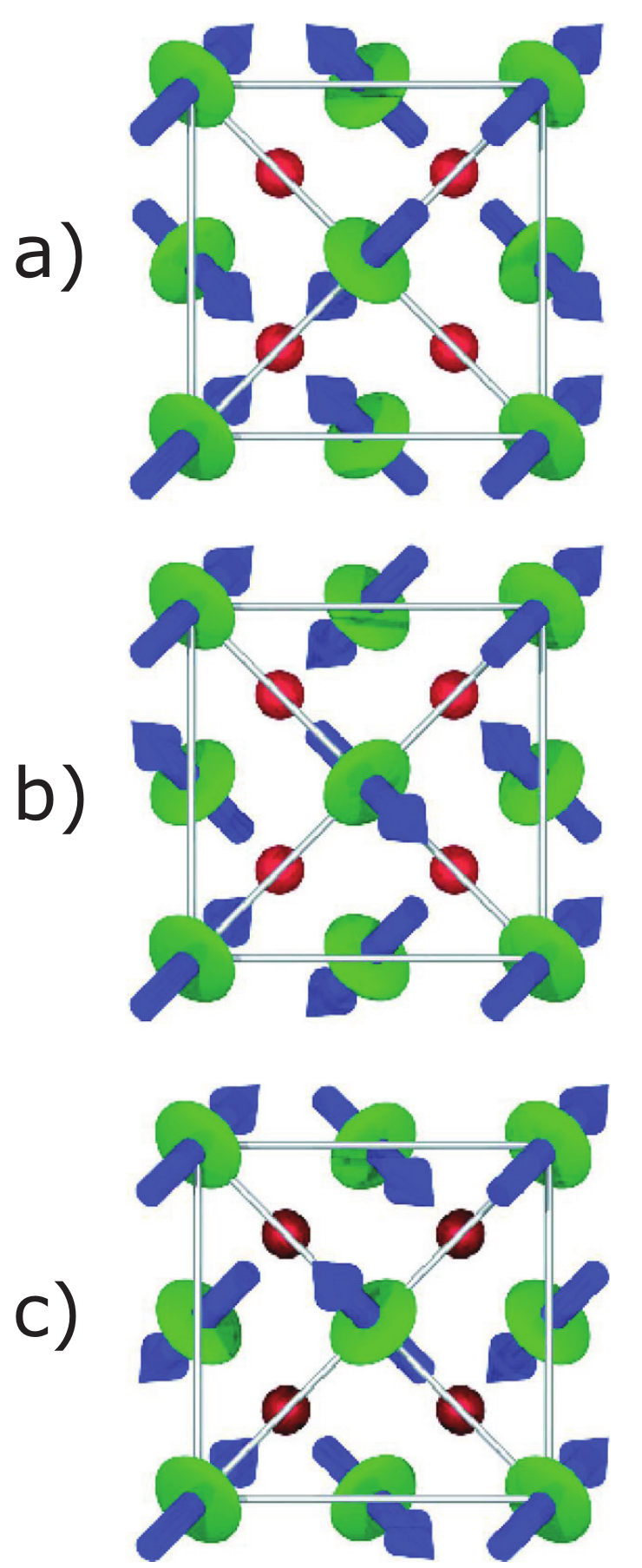}}
\caption{ Schematic representation of the projection
onto the $a$-$b$ plane of the 3-$\mathbf{k}$ magnetic and electric-quadrupole
ordering for the longitudinal (a) configuration and the two
S-domains of the transverse configuration (b), (c). The magnetic
dipole moments are represented by blue arrows whereas the
electric-quadrupole moments are shown as green ellipsoids. The
red spheres represent oxygen atoms. UO$_{2}$ adopts the transverse structure, whereas the electric quadrupole longitudinal order is realized in NpO$_{2}$ with zero magnetic dipole moment.  
Adapted from \cite{wilkins06}.
\label{3korder}}
\end{figure}

In NpO$_{2}$, a second-order phase transition is observed at $T_0$ = 25 K. In this case, the crystallographic structure is preserved and neither external nor internal distortions are observed \cite{boeuf83,caciuffo87}. The crystal field potential was established by inelastic neutron scattering experiments, indicating a $\Gamma_8$ quartet ground state \cite{fournier91,amoretti92}. The $\Psi$ dependence of the ($001$) and ($003$) Bragg peaks intensity has been measured with the sample kept at $T$ = 10 K at the maximum of the $M_4$ absorption edge ($E$ = 3.846 keV) \cite{mannix99,paixao02}. Data collected in the $\sigma$$\pi$ and
$\sigma$$\sigma$ channels have been used to obtain the Stokes parameters
\begin{equation}
\begin{aligned}
P_{1} = { } &\frac{|F_{\sigma\sigma}|^{2}-|F_{\sigma\pi}|^{2}}{|F_{\sigma\sigma}|^{2}+
|F_{\sigma \pi}|^{2}} , \\
P_{2} = &\frac{|F_{\sigma \sigma}+F_{\sigma \pi}|^{2}-
|F_{\sigma \sigma}-F_{\sigma \pi}|^{2}}
{|F_{\sigma \sigma}+F_{\sigma \pi}|^{2}+
|F_{\sigma \sigma}-F_{\sigma \pi}|^{2}} .
\label{Stokes}
\end{aligned}
\end{equation}
\noindent

Figure~\ref{NpO2azimuth} shows the experimental results. The lines in the figure correspond to calculations assuming the longitudinal 3-$\mathbf{k}$ order of $\Gamma_5$ electric quadrupoles shown in panel (a) of Fig.~\ref{3korder}, assuming a zero-ordered magnetic-dipole moment \cite{caciuffo03a}. 

For a  $\mathbf{Q}$ = (0 0 $\ell$) reflection, choosing $\rho = 0$ where the
$[100]$ vector is in the scattering plane with a component parallel to
the incident photon beam, the azimuthal and polarization dependence of the
REXS amplitude for the considered structure are   
\begin{equation}
\begin{aligned}
F_{\sigma \sigma} & =4\Phi \sin2\rho  ,  \\
F_{\sigma \pi} & = 4\Phi \sin \theta \cos2\rho.
            \label{ampli003}
            \end{aligned}
\end{equation}
\noindent
where $\Phi$ is the quadrupole-order parameter. Therefore,
\begin{equation}
\begin{aligned}
P_{1} ={ }&\frac{\sin^{2}2\rho -\sin^{2}\theta \cos^{2}2\rho}
{1-\cos^{2}\theta \cos^{2}2\rho}  , \\
P_{2} =  &\frac{\sin\theta \sin 4\rho}{1-\cos^{2}\theta \cos^{2}2\rho} .
            \label{P1P2}
            \end{aligned}
\end{equation}
\noindent

\begin{figure}
\centerline{ \includegraphics[width=1.0\columnwidth]{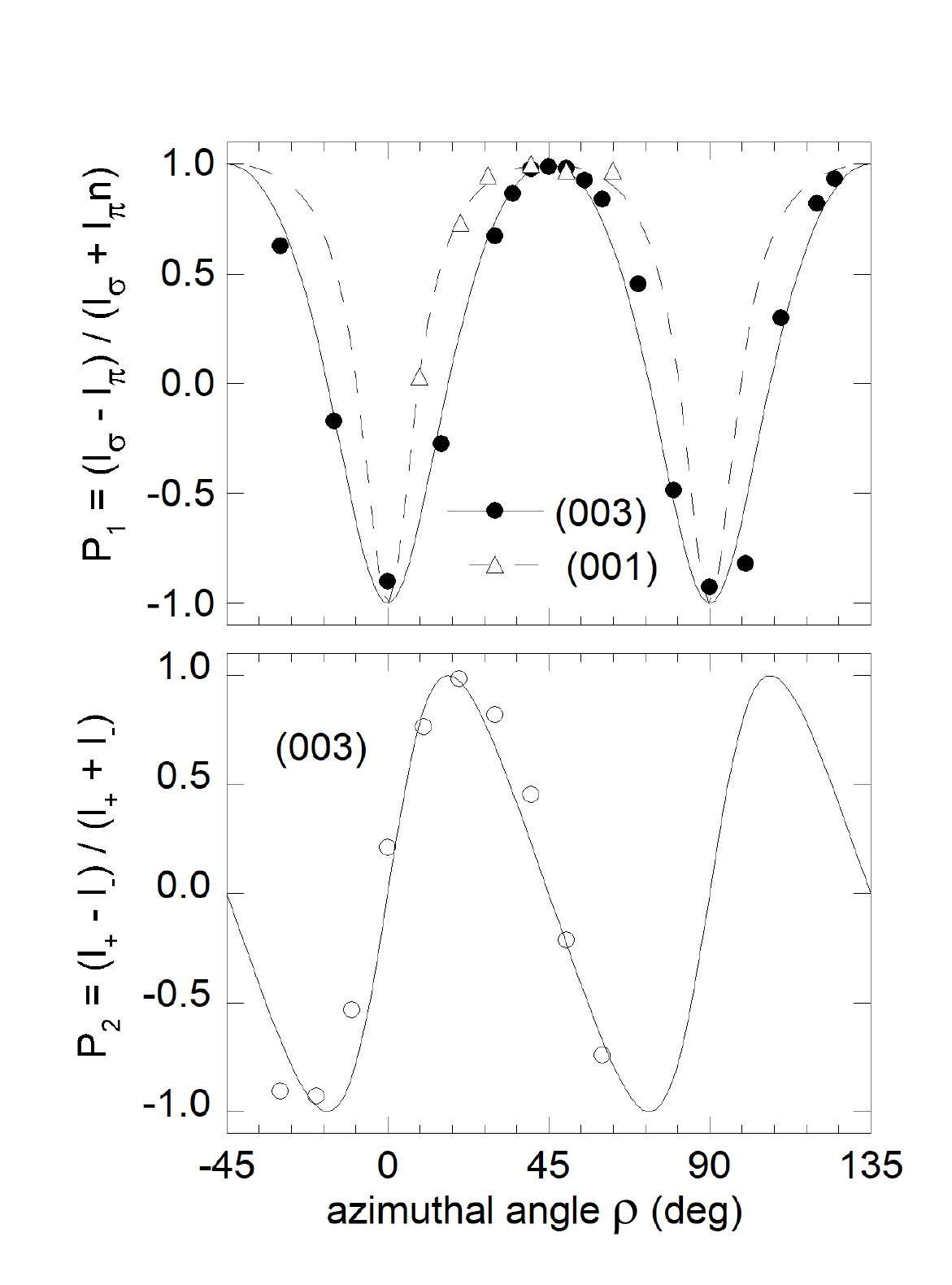}}
\caption{ Azimuthal angle dependence of the Stokes parameters
$P_{1}$ and $P_{2}$ for the ($001$) and ($003$) superlattice reflections measured in NpO$_2$ at 10 K with $E$ = 3.846 keV. The origin of the azimuthal angle $\rho$ corresponds to the $a$-axis lying in the scattering plane. Lines are calculations based on a longitudinal,
3-$\mathbf{k}$ order of $\Gamma_5$ electric quadrupoles with zero ordered magnetic dipole moment. No fitting parameters are involved.
Adapted from  \cite{caciuffo03a}.  }
\label{NpO2azimuth}
\end{figure}

The agreement between experimental and calculated data in Fig.~\ref{NpO2azimuth} is excellent, considering that no fitting parameters are used. As a consequence of the quadrupole order, the point symmetry at the Np site is reduced to $D_{3d}$ and the space group is lowered to $Pn\bar{3}m$. However, with Np ions on $4b$ and O ions on $2a$ and $6d$ Wyckoff positions, the crystallographic extinction rules remain the same as those of the $Fm\bar{3}m$ space group. Also for NpO$_2$, quadrupoles are not the primary order parameter. Indeed, by probing the dynamics of the ordered state by inelastic neutron scattering it appears that the driving order parameter is provided by the rank-5 magnetic triakontadipoles \cite{santini06,magnani08}.

Searches for quadrupolar order by REXS experiments has been performed also on URu$_{2}$Si$_{2}$. This intermetallic compound has been widely investigated in the attempt to explain the nature of its phase transition at $T_{0}$ = 17.5 K \cite{mydosh11,mydosh20,broholm91,mason90,walker93,oppeneer11,santini94,santini00,amitsuka10,walker11,chandra15,wang17}.  The puzzle arises from the difficulty in reconciling the tiny value of the ordered magnetic moment ($\mu_0$ = $\sim$0.03 $\mu_{\mathrm{B}}$ along the $c$-axis of the tetragonal unit cell) with the large macroscopic anomalies observed at $T_{0}$ \cite{broholm91,mason90,walker93}. For instance, if the order parameter were the magnetic dipole moment, the anomaly in the specific heat  should be $\sim$30 times smaller. This indicates that macroscopic anomalies are not associated with $\mu_0$, but rather with a \textit{hidden} order parameter not directly coupled to scattering probes. Among a number of theoretical models \cite{oppeneer11,mydosh20}, also a staggered ordering of electric quadrupoles  have been suggested to occur in URu$_{2}$Si$_{2}$ \cite{santini94,santini00}.

REXS experiments at the U $M_4$ absorption edge have not confirmed this hypothesis \cite{amitsuka10,walker11}. Data have been collected on high-quality single crystals of URu$_{2}$Si$_{2}$ with low residual stress, cut with a [$101$] direction specular, spanning an extended region of the reciprocal space plane [$H0L$], including the nesting vector ($1.4, 0, 0$) suggested by neutron scattering experiments. The results of both studies \cite{amitsuka10,walker11} exclude electric quadrupoles of any symmetry as a hidden-order parameter with a propagation vector in the explored region. Indeed, as shown in the top panel of Fig.~\ref{URu2Si2RXS}, forbidden Bragg peaks emerging in the ordered state have non-zero intensity only in the $\sigma$$\pi$ polarization channel, with an azimuthal angle dependence corresponding to an ordered magnetic dipole moment along the crystallographic $c$-axis. The bottom panel of Fig.~\ref{URu2Si2RXS} shows that the $\rho$ dependence of the superlattice peak intensity is very sensitive to the inclusion of a component of the ordered magnetic moment in the $a$-$b$ plane, whose estimated upper limit is $\sim$0.003 $\mu_{\mathrm{B}}$. This experimental result essentially eliminates the idea of hastatic order proposed in \cite{chandra15}.

An attempt to detect multipolar order parameters in URu$_{2}$Si$_{2}$ with REXS measurements at the U $L_3$ edge with the E2 electric quadrupole transition was performed on the 6-ID-B beamline at the APS  \cite{wang17}, but no evidence of multipolar moments was found within the experimental uncertainty.

\begin{figure}
\includegraphics[width=0.8\columnwidth]{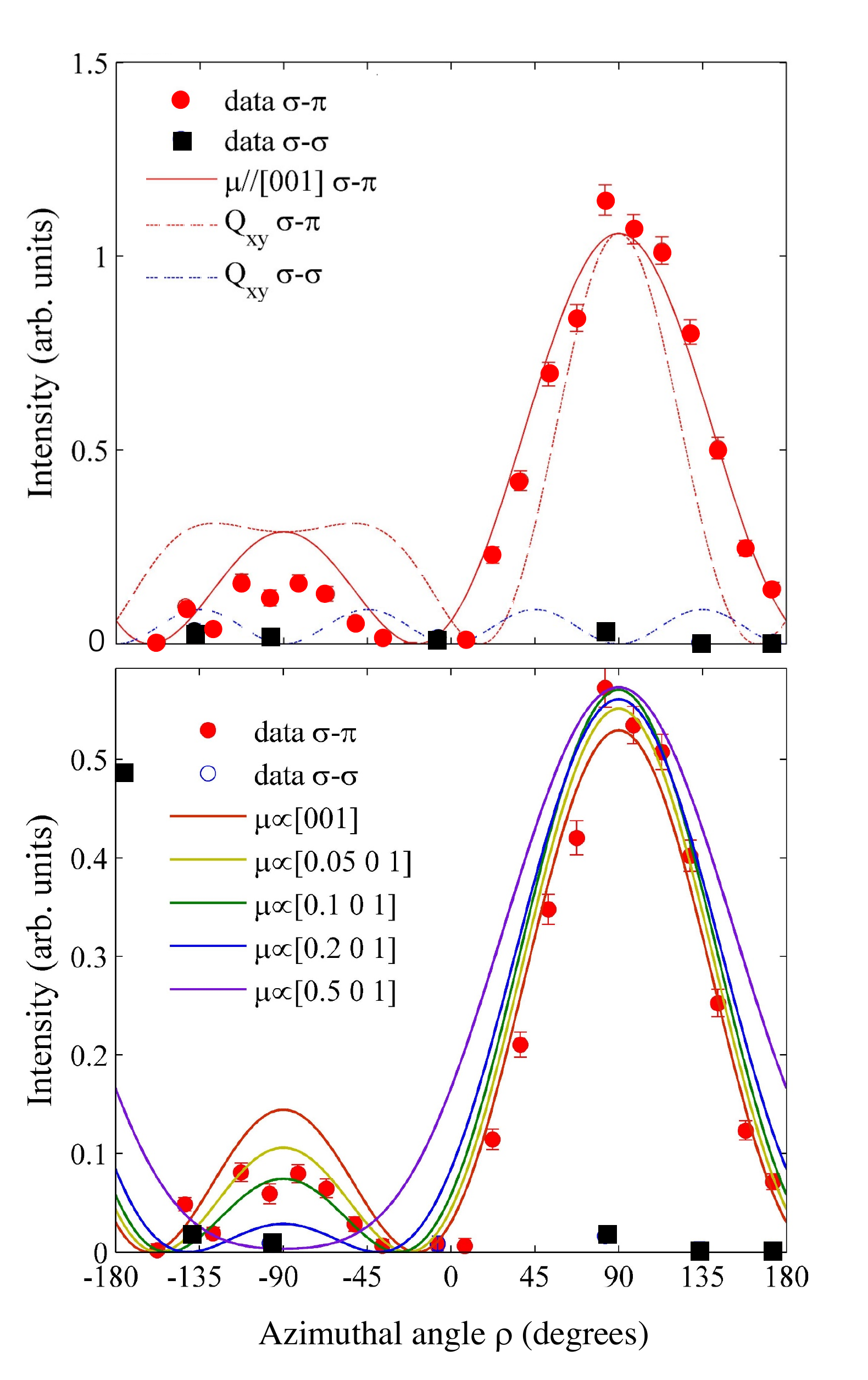}
\caption{Top panel: azimuthal dependence of the ($201$) Bragg reflection in URu$_2$Si$_2$. Solid (red) circles: $\sigma$$\pi$ channel; black squares: $\sigma$$\sigma$ channel. Full (red) line is the theoretical intensity variation for magnetic dipoles ordered along [$001$]. The dashed red (dashed blue) line is the $\sigma$$\pi$ ($\sigma$$\sigma$) $\rho$ dependence of the intensity expected for $xy$ quadrupole order. Bottom panel: experimental data compared with theoretical predictions for a dipole magnetic moment with an increasing component in the $a$-$b$ crystallographic plane.
Adapted from \cite{walker11}.
  \label{URu2Si2RXS}
}
\end{figure}

\subsection{General trends in REXS studies}

The enormous enhancement of the magnetic signal at the actinide $M$ edges has also been of interest in many actinide materials to learn more about the magnetic structure itself. Early work, for example, determined in greater detail the nature of the multi-$\mathbf{k}$ structures found in materials with transuranium ions \cite{langridge94,langridge94b,normile02,normile02b,lidstr00}, as well as those with uranium \cite{longfield02,bernhoeft04}, and thin films \cite{bernhoeft98,bao13}.

Some important progress was made with REXS on the question of multi-$\mathbf{k}$ magnetic configurations, which are frequently found in actinide systems. Experiments showed that for more than a single $\mathbf{k}$-vector a satellite related to the third term in Eq. (\ref{ResAmpl}) occurred in the magnetic diffraction pattern at the positions described by the vectors $\langle k_1, k_2, 0 \rangle$ added to any allowed magnetic lattice point, thus uniquely identifying either a 2-$\mathbf{k}$ or 3-$\mathbf{k}$ magnetic configuration with a term that is quadrupole in nature representing the coupling of components
\cite{longfield02}. Surprisingly, for the 3-$\mathbf{k}$ systems additional satellites were found at the positions $\langle k_1, k_2, k_3 \rangle$ added to any allowed lattice point, but the scattering process giving rise to these peaks, which appears to be dipole in nature, is not totally clear \cite{bernhoeft04,bdetlefs07}. These weak magnetic peaks, which have been called $\langle kkk \rangle$ peaks, were also observed in a tour-de-force experiment using neutrons. Such experiments prove the existence of the 3-$\mathbf{k}$ magnetic configuration and confirm the dipole nature of the scattering, although the momentum dependence is not presently understood \cite{blackburn06,blackburn06b}. Such weak peaks cannot be observed in systems such as UO$_2$ or USb, which both have
3-$\mathbf{k}$ configurations with $|k|$ = 1, because they superimpose on strong lattice reflections.

The large intensity also allowed experiments to show that the magnetic ordering in URu$_2$Si$_2$ was not fully long-range \cite{isaacs90}, and to probe surface magnetism in UO$_2$ showing new examples of surface ordering not previously measured \cite{watson96,watson00,langridge14}. 

An experiment was also performed early on at the Synchrotron Radiation Source (SRS) at  Daresbury Laboratory, UK, showing that there is enough intensity at the resonant U $M_4$ energy to observe the antiferromagnetic peaks in a polycrystalline sample of UO$_2$ \cite{collins95}. Surprisingly, this technique has not been used for other actinide antiferromagnets, perhaps because the moment can not be reliably determined by this method or perhaps because the momentum resolution is necessarily poor to obtain the intensity needed to observe the peaks. An attempt to use this method was made with Cm \cite{lander19} but failed as the sample had lost its crystallinity due to self-radiation. One scattering experiment has also been performed at the uranium $L_{2,3}$ edges, where the enhancement is relatively weak \cite{wermeille98}.

Similarly, the large intensity led to the discovery of an interesting effect that appears to occur when a magnetic material disorders, namely a small apparent shift in the position of the peak arising from the disordered diffuse scattering. This is not caused by the diffuse scattering becoming incommensurate, which would result in the appearance of two peaks about the commensurate position. No satisfactory explanation of this effect has yet been proposed \cite{bernhoeft04a}. After a considerable effort this small effect was also seen with neutrons in the antiferromagnet MnF$_2$, showing that it is not limited to actinides, and is not related to a surface effect \cite{prokes09}.

Resonant scattering effects also occur in other atoms, whose electrons interact with the magnetic dipoles (or electric quadrupoles) of the actinide ions.  This was first observed in the relatively simple antiferromagnets UGa$_3$ and UAs \cite{mannix01}. Below the ordering temperature $T_N$ when the x-ray energy is tuned to the $K$-edge of Ga or As, a large signal is present and disappears at $T_N$. Similar observations were made at the Ga $K$ edges in UTGa$_5$ (T = Ni, Pd, and Pt) \cite{kuzushita06}, as well as in NpRhGa$_5$ and NpCoGa$_5$ \cite{detlefs08}. 

\begin{figure}
\includegraphics[width=0.8\columnwidth]{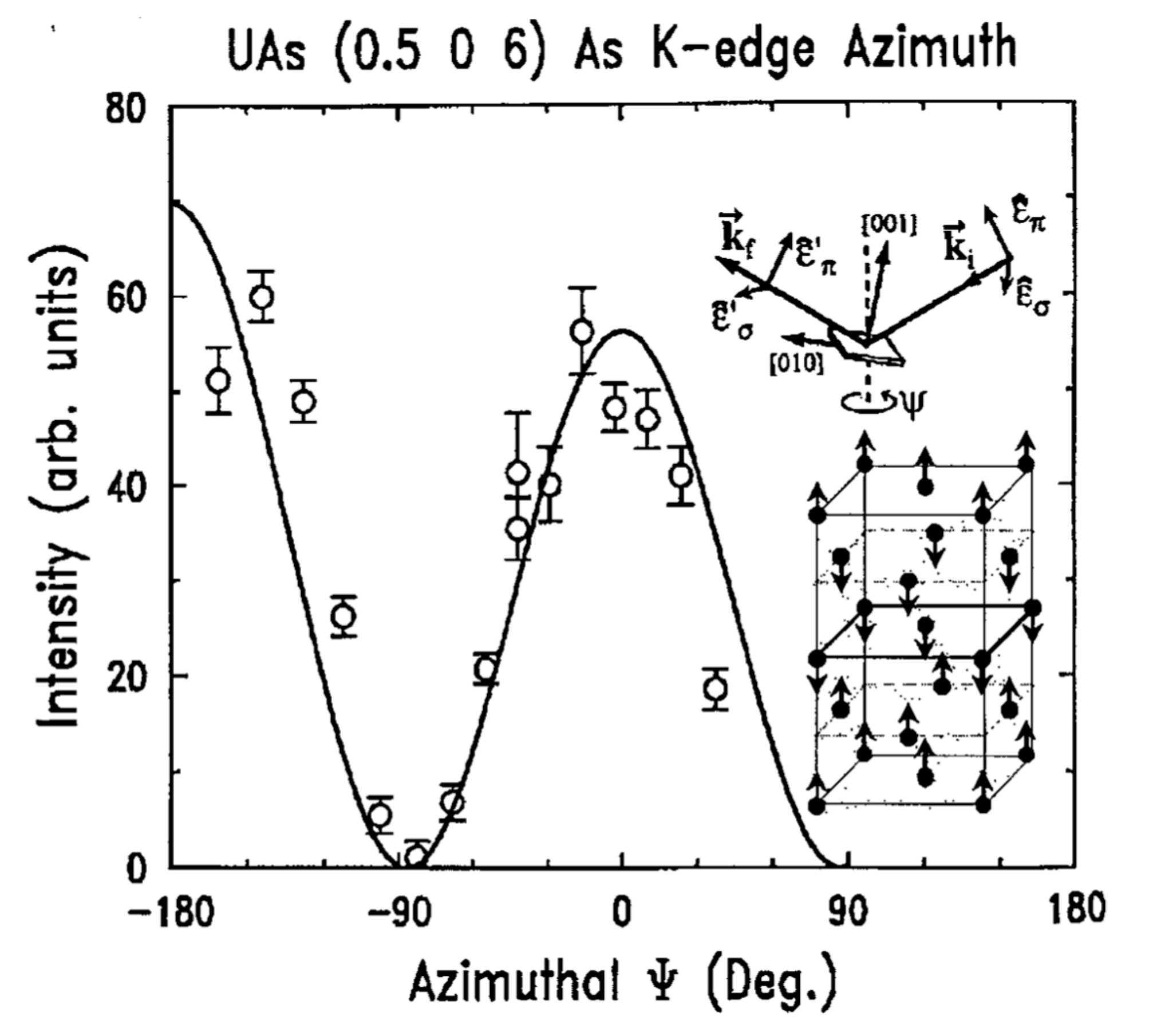}
\caption{Measurements at the As $K$ edge of the integrated intensity in the $\sigma  \pi$ channel of the $(0.5, 0, 6)$ reflection in UAs at $T$ = 20 K as a function of the azimuthal angle around the normal to the $\left[001\right]$. See insets for geometry and schematic of the Fourier component of the magnetic structure of UAs that gives rise to this reflection, where the propagation vector of the structure is $\left[100\right]$. The azimuthal angles $\Psi$ = 0$^{\circ}$, 180$^{\circ}$ correspond to the $\left[100\right]$ axis lying in the scattering plane, defined by incident and scattered wave vectors $\textbf{k}$ and $\textbf{k}'$, respectively.
Reproduced from  \cite{mannix01}.  
 \label{mannix01}
}
\end{figure}

Figure \ref{mannix01} shows the azimuthal dependence of the signal at the As $K$ edge in UAs. It  completely agrees with the known magnetic structure  for UAs, so suggests a small moment appears on the As ions. The signal arises from the hybridization of the actinide 5$f$/$6d$ and anion 2$p$ states, and results in a small polarization of the 2$p$ anion electrons, thus appearing as a “magnetic moment”. This would not occur if the 5$f$ states were fully localized.  It is not possible (as yet) to relate the size of the signal to any magnetic moment, either for the active ions themselves, or for induced effects such as measured here. \cite{sanchez00}  used M\"ossbauer spectroscopy to suggest that the induced polarizations of the 2$p$ electrons give a net moment of the order of 0.02 $\mu_{\mathrm{B}}$ at the Ga site in UGa$_3$. There have also been theory papers explaining these large resonances on non-magnetic ions \cite{veenendaal03,usuda04}.

At resonant energies additional scattering effects can be observed that are not connected to magnetism. An early example was reported by \cite{templeton94}, followed by observations in other systems, as summarized in \cite{kokubun94}. Resonant scattering depends on the energy
as well as the x-ray polarization where the scattering factor
must be treated as a tensor quantity.
The effects can be observed in crystals with screw axes or glide planes, but are not observable in higher symmetry systems, e.g., UO$_2$. So far, almost all the effects have been observed at transition metals $K$-edges, but these are often complex to understand as the absorption process involves both an $s \rightarrow p$ transition in the dipole channel (E1) and $s \rightarrow d$ transition in the quadrupole channel (E2), complicating the interpretation. However, in the cubic compound U$_2$N$_3$
\cite{lawrence19} found the first example of Templeton scattering involving the $M_4$ U resonance. This transition is 3$d$ $\rightarrow$ 5$f$ and is certainly an E1 transition, as the quadrupole transition is to possible $g$ states with a very small matrix element. These forbidden reflections are independent of temperature and give information about the bonding in the system.

The large enhancement of the magnetic scattering at the U $M_4$ edge allows one to measure magnetic Bragg peaks with good statistics even when a pinhole aperture is used to increase the transverse coherence length of the beam up to tens of $\mu$m. This is necessary to perform speckle spectroscopy, i.e., x-ray photon correlation spectroscopy (XPCS). This technique uses the intensity-intensity time correlations between coherently diffracted x-rays to access dynamics on the 10$^{-6}$ to 10$^{3}$ second timescales \cite{grubel04}. Pioneering studies were performed at the ESRF, France, on UAs \cite{yakhou01} and at the Diamond Light Source, UK, on the isostructural compound USb \cite{lim14}. The latter is a longitudinal 3-$\mathbf{k}$ antiferromagnet, with $T_N$ = 218 K and spins pointing along local $\left<111\right>$-type directions, with three equivalent wavevectors of the form $\left[001\right]$ \cite{jensen81,caciuffo07}. Inelastic neutron scattering experiments showed that the spin wave energy in USb soften approaching a temperature $T^{\ast}$ = 160 K, and no collective excitations are observed for
$T^{\ast} <  T  < T_N$  \cite{hagen88,lim13}, although the magnetic Bragg peaks appear unaffected by the change in the dynamics. The XPCS results reported by \cite{lim14} reveal a change in the static and dynamical speckle patterns that show an increase in
fluctuations and decrease in domain size around $T^{\ast}$, suggesting that  changes in the magnetic domain structure can be at the origin of the observed anomalous behavior.
\\

\section{XAS experiments}\label{secXAS}

XAS of actinides at SR sources started in the late 1980s, exploring both the $L_{2,3}$ (2$p$ $\rightarrow$ 6$d$) \cite{kalkowski87a,bertram89} and the $M_{4,5}$, $N_{4,5}$, and $O_{4,5}$
($nd$ $\rightarrow$ 5$f$; $n$ = 3, 4, 5) \cite{kalkowski87} absorption edges. The potential of XAS for elucidating the electronic structure of actinides was immediately recognized and a number of studies on actinide speciation in compounds and minerals followed \cite{silva95,conradson98,denecke06}. 

XAS experiments have provided important information about the oxidation state of actinide elements in minerals and molecular compounds, their coordination with neighboring atoms, and the structure of their local environment. The technique can be applied to samples in solid or liquid forms, including nanoparticles \cite{gerber20} and colloidal forms \cite{micheau20}, with element sensitivity down to tens of ppm. In combination with small angle x-ray scattering (SAXS) measurements, XAS provides detailed information on the local structure, size, shape, and interfacial properties of atomic clusters and intrinsic colloidal particle  
\cite{nyman17,tian20,micheau20,zhai22}.

Besides the scientific importance, actinide speciation studies are key to understanding extraction mechanisms and separation processes in fuel reprocessing \cite{moeyaert21,dressler22,pruessmann22},  the corrosion and the reduction-oxidation behavior of nuclear waste forms in storage facilities \cite{husar22,glasauer22},  the  mechanisms by which actinides can contaminate the biosphere
\cite{cot-auriol21,estevenon21,pidchenko20,vitova20}, their mobility in contaminated sites and how best they can be removed from the environment \cite{lepape20,jegou22,dumas22}. The clean up of the Rocky Flats Nuclear Weapons Plant, near Denver, is exemplary of the practical importance of these studies \cite{clark06}. Redox/solid-liquid interface reactions can be studied at metal ion concentrations of specific relevance for contaminated sites, as shown in \cite{schacherl22a,schacherl22} where $M_5$ XAS spectra are reported for 1 ppm Np adsorbed on clay.

Today, XAS experiments at state-of-the-art spectrometers have been extended to the heaviest actinides available in macroscopic quantities, as recently demonstrated by a study of a coordination complex of $^{254}$Es that used less than 200 nanograms of this highly radioactive isotope (half-life of $\sim$275.7 days) \cite{carter21}. XAS techniques, in combination with other chemical physics methods, are assuming growing importance also in the nuclear toxicology field and are enabling the development of effective actinide decorporation agents with high complexation affinity, high tissue specificity, and low biological toxicity \cite{ye21,zurita21,zurita22}. In a recent study, XANES and X-ray fluorescence chemical imaging have been used to characterize trace impurities in fuel pellets and assess the potential of these techniques as a forensic tool for the investigation of interdicted special nuclear materials \cite{vanveelen22}.

From the point of view of physics, XAS experiments at the $M_{4,5}$ and $N_{4,5}$ absorption edges of actinides have been important to probe the relativistic nature of the 5$f$ electrons, thanks to the application of simple spin-orbit sum rules relating the  branching ratio of the core-valence transitions to the expectation value of the angular part of the 5$f$ spin-orbit interaction per hole \cite{thole88,vanderlaan04,shim09,caciuffo10a}. The information obtained is analogous to that provided by electron-energy loss spectroscopy  (EELS) \cite{moore09}, a technique that proved to be very useful in elucidating the change between $\alpha$ and $\delta$ plutonium \cite{moore06}, the magnetic stabilization in curium metal \cite{moore07}, or to study rare metals only available in small quantities \cite{dieste19}.

\begin{table}[t]
\caption{Binding energies (BE), x-ray attenuation lengths (1/$e$) at the continuum maximum, and core level linewidth of the various absorption edges in solid uranium (density 18.92 g\,cm$^{-3}$). BE and attenuation length values, obtained from \cite{henke93}, do not explicitly include the white line intensity, but give a good impression of the x-ray penetration depth in the solid just above the white line. Attenuation lengths for U compounds roughly scale with the inverse of the partial U density. Core level linewidth (Full Width at Half Maximum) are taken from \cite{raboud99} and from \cite{caciuffo10} for the $O_{4,5}$ edges.}

\begin{tabular}{l|l|l|l|l}
\hline \hline
Edge & Core level &BE (eV) & Attenuation& Core level \\ 
  &   &  & length (nm) & width (eV)\\ \hline
$K$ & 1$s$		&115606 &  108000  & \\
$L_2$  & 2$p_{1/2}$		& 20948 &  5786 & 10.0 $\pm$ 0.1\\
$L_3$  & 2$p_{3/2}$		&   17166   & 4798 & 8.4 $\pm$ 0.2	\\
$M_4$ & 3$d_{3/2}$		&   3728   &  394 & 3.2 $\pm$ 0.1\\
$M_5$ & 3$d_{5/2}$		&    3552  & 444 & 3.3 $\pm$ 0.1\\
$N_4$ & 4$d_{3/2}$		&   778   & 72 & 4.7 $\pm$ 0.1\\
$N_5$ & 4$d_{5/2}$		&    736  & 78 & 4.2 $\pm$ 0.6\\
$O_{4,5}$ & 5$d_{3/2,5/2}$	&  100    &  12 & $\sim$8\\
\hline \hline
\end{tabular}
\label{attlengths}
\end{table}

The importance of self-absorption corrections in XAS studies of actinides must be duly emphasized. Such corrections should take into account the chemical composition, the density, the practically infinite thickness of the bulky sample if the fluorescence detection mode is used, the various background contributions (fluorescence of subshells and matrix as well as coherent and incoherent scattering), the
angle of incidence of the x-ray beam, and finally the solid angle of the detector \cite{goulon82,troger92,pfalzer99}. The actinide edge-jump intensity ratio $M_5$/$M_4$ (defined as the ratio between the occupation numbers for the two spin-orbit-split core levels $j$ = 3/2 and 5/2) must be normalized according to the statistical edge-jump ratio. This ratio is equal to 1.5, very close to the value tabulated in the XCOM tables by \cite{berger10}. A deviation of
$\pm$10\% in the $M_5$/$M_4$ XAS edge-jump normalization would affect the branching ratio $B$
and the $5f$ occupation numbers of the $j$ = 5/2 and $j$ = 7/2
subshells by $\sim \pm$2.5\%. A discussion of the importance of making accurate self-absorption corrections at the actinide $M$ edges is given in  \cite{janoschek15}.

Calculations of attenuation lengths based on isolated-atom models \cite{parratt54,cromer70} are available for most of the elements in the periodic table, including actinides. Table \ref{attlengths} gives the attenuation length at different absorption edges of uranium. It is important to notice, however, that data for bare atoms reproduce only the gross spectral features but neglect the local environment around the resonant atom and therefore fail to reproduce important effects such as the appearance of a large \textit{white line} at the resonance energy \cite{bohr90}. In actinides, these effects are dramatically large, as at the $M_4$ edge of uranium where the white line is about 7.5 times larger than the step height \cite{cross98}. 
For the $M_{4,5}$ edges, where the transitions involve the partially occupied 5$f$ states, the work of  \cite{cross98} suggests that the attenuation length including the white line is $\sim$100 nm for the $M_4$ and only $\sim$50 nm for the $M_5$ edge. Similar reductions would be expected for the $N_{4,5}$ and $O_{4,5}$ edges listed in Table  \ref{attlengths}, with an attenuation length at the U $N_{4,5}$ edges of the order of 10 nm as provided by calculations based on the Cowan code \cite{cowan81}. Computer codes for general-purpose \textit{ab initio} calculations of the attenuation lengths for embedded atoms are available. An example is the FEFF code \cite{rehr91,zabinsky95} that takes into account both atomic and photoelectron scattering contributions from the neighbor atoms, providing a dramatic improvement in the simulation of the resonant spectral features for actinides \cite{cross98}.

\subsection{Examples of XAS studies on actinide systems}
An example of work at the Pu $L_{3}$ edge, which is far easier to work at than at the lower energy $M$ edges, given that the samples must be at least doubly encapsulated, is shown in Fig.~\ref{PuHandbook}, \cite{conradson04a,clark19b}.

\begin{figure}
 \includegraphics[width=0.8\columnwidth]{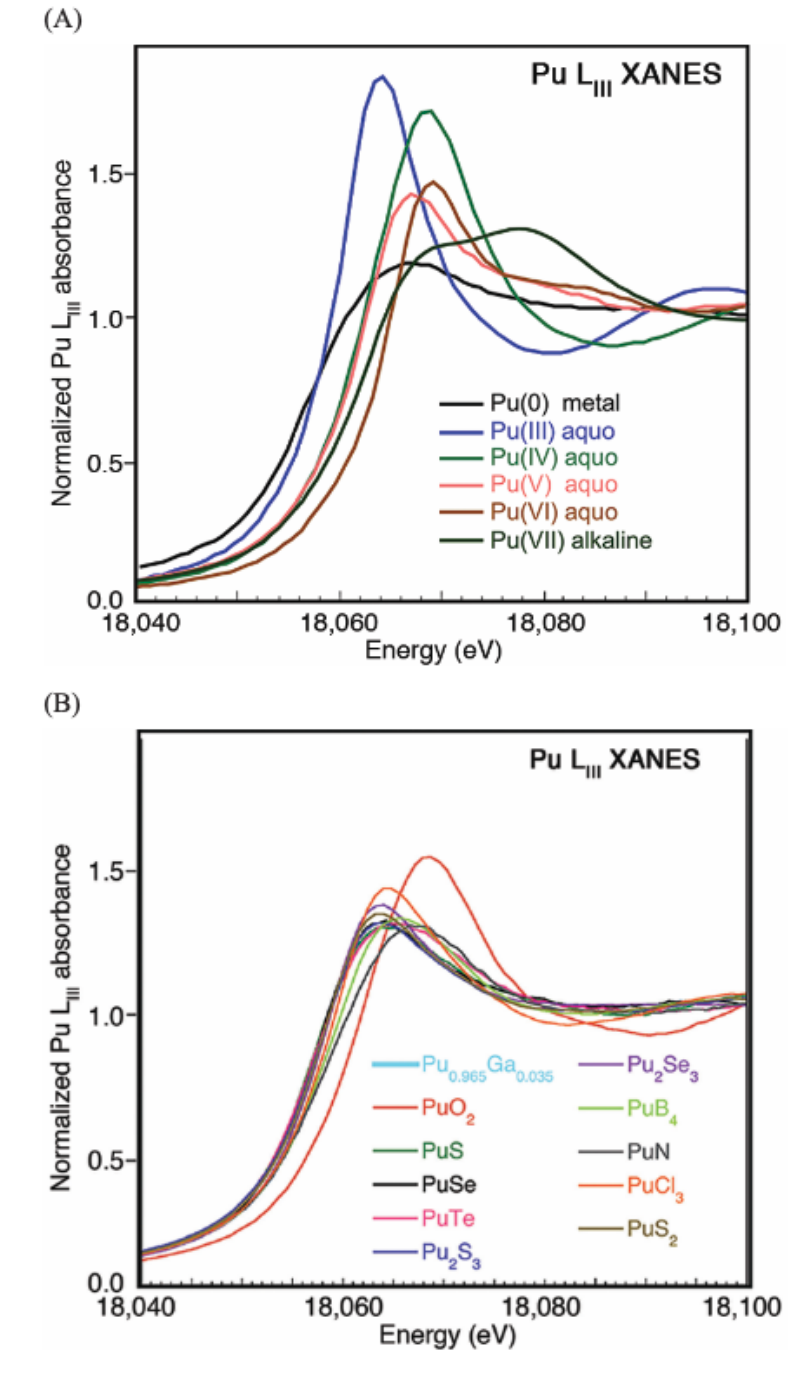}
\caption{Pu $L_{3}$ edge XANES (normalized absorbance) for a series of plutonium molecules and compounds in different formal oxidation states. (A) Gallium stabilized $\delta$-Pu metal compared to the plutonium aqua-ions in oxidation states III–VI and the Pu(VII) oxohydroxide. Spectra of Pu(0)–Pu(VI) are replotted from original data reported by \cite{conradson04a} and data for Pu(VII) in 2 M NaOH solution is courtesy of M.~Antonio, ANL. (B) The peak amplitude of the $\delta$-Pu alloy is suppressed by self absorption in this opaque sample. XANES spectra of gallium stabilized $\delta$-Pu metal compared to a number of binary plutonium compounds, including PuO$_2$, which represents Pu(IV). Adapted with permission from \cite{conradson04a,clark19b}.
\label{PuHandbook}
}
\end{figure}

The $L_{3}$ absorption edge involves transition from the 2$p$ core states to the unoccupied 6$d$ states around the Pu atom. The energy sensitivity of the observed spectra comes because the screening of the 6$d$ hole depends on the number of 5$f$ electrons in the partially occupied 5$f$ shell. In Fig.~\ref{PuHandbook}(A) there are no 6$d$ electrons present in any of the ionic materials, so this gives a simple interpretation of the energy shift of the Pu $L_3$ maximum. On the other hand, in Fig.~\ref{PuHandbook}(B) we show a number of spectra from the metal and other semi-metallic compounds, and within experimental error, they all have approximately the same peak value. This shows that the simple interpretation of ionic complexes (A) cannot be used for the intermetallics (B), as the presence of a conduction band containing $spd$ partially occupied states complicates the simple interpretation useful in (A). One can also observe in Fig.~\ref{PuHandbook} that the spectral linewidths are considerable. This arises from the short core-hole lifetime of the 2$p_{3/2}$ state, which results in spectral full width at half maximum (FWHM) of at least 7.5 eV.

There are two ways to improve this resolution. The first is to move to the actinide $M$ edges, which involves directly the 5$f$ electrons, and where the intrinsic lifetimes give a resolution of FWHM $\sim$3.3 eV. A difficulty of working at these energies is that the x-ray beam is strongly absorbed in air. The second way to improve the resolution is to construct a spectrometer that analyzes specific emission lines, for instance the $M_{\beta}$ (4$f_{5/2} \rightarrow$ 3$d_{3/2}$). The total spectral width may then be reduced to below 1 eV as shown in \cite{kvashnina13,kvashnina22}. The resulting spectra are then referred to as high energy resolution fluorescence detection (HERFD) XANES and are shown in Fig.~\ref{kvashnina13} for some of the oxides of uranium. It must be noticed that conventional XAS essentially integrates over all outgoing channels. By selectively detecting outgoing channels with limited opening angle to gain a line width narrowing, one measures something different than XAS, and the results will depend on the specific edges used  because of selection rules and final state effects
\cite{carra95}. For instance, for most actinide dioxides, a good agreement between XAS and HERFD spectra is found at the $M_4$ edge, whereas additional transitions at the $M_5$ edge produce a different spectral profile for XAS and HERFD measurements \cite{butorin20}.

An M4 edges demonstrate overall better agreement between the HERFD and XAS spectra 

\begin{figure}
\includegraphics[width=0.8\columnwidth]{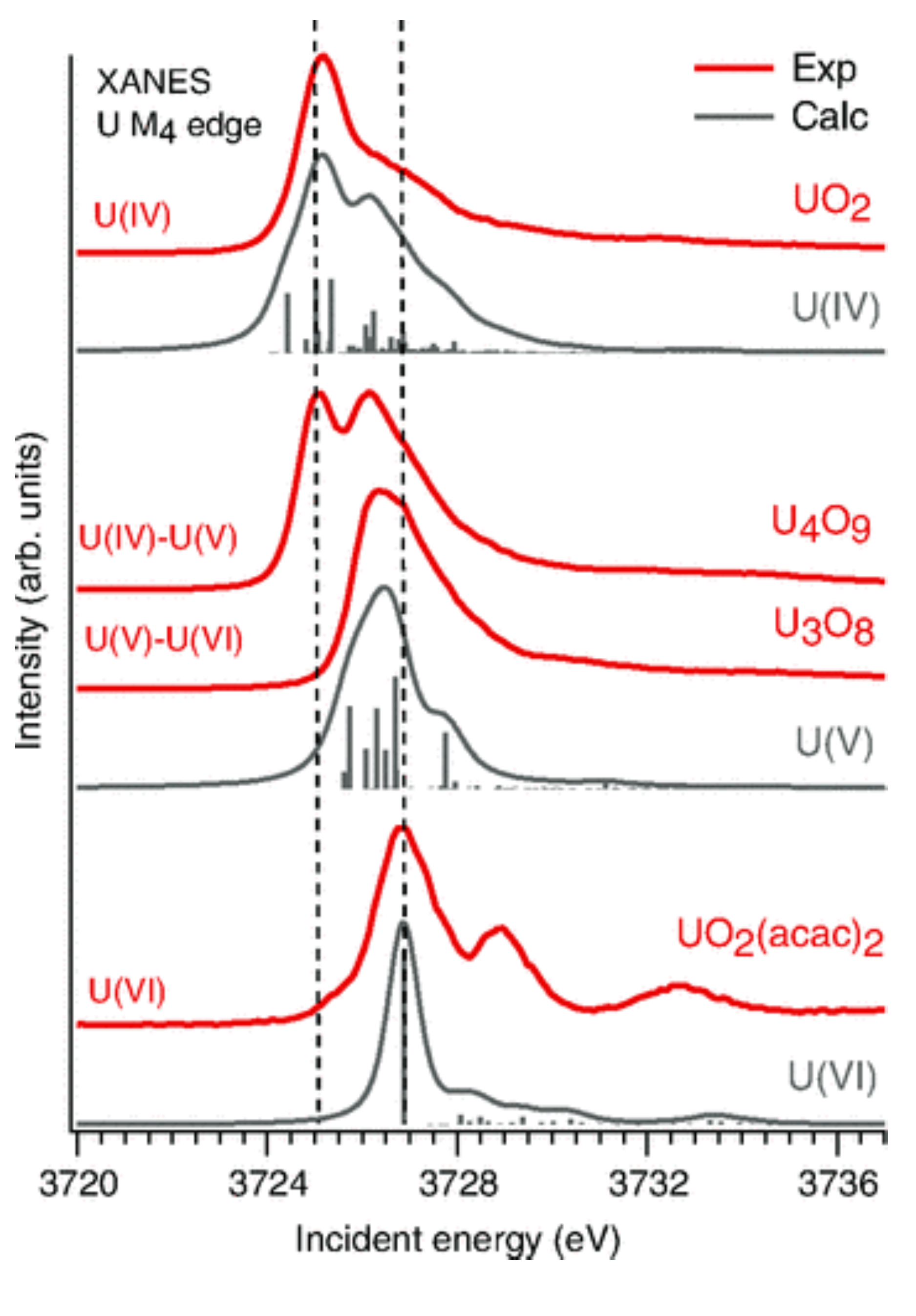}
\caption{Experimental and calculated HERFD-XANES spectra at the U  
$M_{4}$ edge of  U$_4$O$_9$ and U$_3$O$_8$ compared with those of the reference systems  UO$_2$ and UO$_2$(\textit{acac})$_2$. Dashed lines indicate the energy position of the main peaks corresponding to uranium(IV) and (VI), respectively. Reprinted from \cite{kvashnina13}.  
\label{kvashnina13}
}
\end{figure}

This work shows a clear shift of $\sim$2 eV from U(IV)  to U(VI)  with a resolution of $\sim$1 eV  \cite{kvashnina13}. Somewhat more information may be obtained by multi-edge measurements \cite{bes22} and by examining the full diagram of incident energy vs emitted energy \cite{kvashnina22}, but still for intermetallic systems the same problems arise as was evident from the early work on Pu systems (see Fig.~\ref{PuHandbook}), despite the much higher resolution now achieved with better $M$-edge spectrometers \cite{kvashnina17}.
It is noteworthy that these results for uranium oxides agree well with an alternative technique used in x-ray photoelectron spectroscopy at the core 4$f$ emission lines \cite{ilton11,gouder18}.

The high sensitivity of HERFD-XANES to the arrangement of atoms around the absorber is demonstrated by measurements at the U $M_4$ edge in hexavalent uranium in various local configurations \cite{amidani21} and at the thorium $M_4$ edge in ThO$_2$ nanoparticles (NPs) \cite{amidani21a}. These experiments show the possibility to correlate spectral details with local symmetry breakdowns around the absorbing atoms sitting close to the NPs surface.

Absorption spectroscopy may also give valuable information if performed at the anion $K$ edges, e.g., oxygen, carbon, or chlorine. At the oxygen $K$-edge (532 eV), polarized x-ray absorption and emission spectra have been reported for a single crystal of Cs$_2$UO$_2$Cl$_4$ \cite{denning02}. The absorption data led to the assignment of the relative energies of the empty molecular orbitals and of their character (mainly uranium 5$f$ and 6$d$). The emission spectra provided the energy of excitations to these orbitals from filled valence-shell orbitals, showing that valence excitations from the $\sigma_u$ occupied valence orbitals occur at substantially lower energies than those from the
$\sigma_u$, $\pi_g$, and $\pi_u$ orbitals. The presence of a charge-transfer transition in emission  established the participation of the pseudo-core 6$p$
shell in the covalent bonding, with a strong hybridization between 5$f_{\sigma}$ and 6$p_{\sigma}$ states.
Other examples at the oxygen $K$ edge were reported on PuO$_2$ \cite{modin11} and on uranium systems \cite{kvashnina22}. At the chlorine $K$ edge (2822 eV), \cite{minasian12} studied the relative $f$ and $d$ orbital contributions to the U-Cl covalency in UCl$_{6}^{2-}$ and UOCl$_{5}^{-}$. An example at the carbon $K$ edge (284 eV) is given by the study of 5$f$ covalency in $\left[U(C_{7}H_{7})_{2}\right]^{-}$, reported by \cite{qiao21}. These spectra are complex and cannot be interpreted without substantial modeling, but they do give important information on covalency \cite{ganguly20,sergentu22}. 

\cite{tobin15} used spectroscopy at the $L_{3}$ edge and have also measured branching ratios at the $N_{4,5}$ edges using soft x-rays at the ALS (Berkeley, CA), and found that UO$_2$ has an $f^{2}$ configuration (which is expected) and UO$_3$ has an $f^{1}$ configuration, which is a surprise, as a straightforward valence argument would expect a U(VI) valency, hence $f^{0}$. Although UO$_{3}$ was not specifically examined in the HERFD work of  \cite{kvashnina13}, a number of such oxides were examined by x-ray core-hole photoemission, and there appears strong evidence that UO$_{3}$ consists of hexavalent U atoms \cite{ilton11}. Work has also been reported on the branching ratios of UF$_4$, UO$_2$, and UCd$_{11}$ systems at the $M_{4,5}$ with a view to determining the occupied and unoccupied density of states \cite{yu11,tobin15b,tobin19,tobin21}.

Absorption spectroscopy has also played an important role in probing the electronic structure of Pu metal. \cite{terry02} reported early experiments on the absorption at the $O_{4,5}$ edge (involving 5$d \rightarrow$ 5$f$ transitions) in both the $\alpha$-Pu and $\delta$-Pu forms taken at the ALS using soft x-rays. These experiments have been analyzed in different ways by discussing the electronic structure of Pu \cite{tobin03,tobin05,tobin08}. This is a complex subject that needs the results of experiments using photoemission, EELS, XPS core-level spectroscopy, and even neutron inelastic scattering. We refer the reader to articles covering this subject in depth \cite{moore09,clark19b}.

High-energy resolution XANES spectra have been measured at the Pu $M_{5}$ edge in PuO$_2$ ~\cite{vitova13,vitova18,bahl17,bagus21} at the CAT-ACT wiggler beamline at KARA in Karlsruhe, Germany \cite{zimina17} and the Pu $M_{4,5}$ edges at the ESRF \cite{kvashnina19}.  In \cite{bagus21}, fully relativistic quantum chemical computations of the electronic structure of the isolated Pu$^{4+}$ cation and for Pu$^{4+}$ in PuO$_2$ are used for the interpretation of the experimental spectra, giving insight into the level of covalency of the Pu 5$f$ valence orbitals and their role in chemical bonding. The authors emphasize the need of taking many-body effects into account to properly describe the wavefunctions, for both the initial and excited states of the XANES transitions. They show that a single-determinant representation of the Pu states is not a sufficient approximation for calculating the Pu $M_{4,5}$ XANES spectra, which implies that dipole selection rules must be applied between the total wavefunctions for the initial and excited states. The spectral broadening due to the angular momentum coupling of the open-shell electrons dominates over spin-orbit and ligand field splittings of the 5$f$ shell orbitals. As a consequence, energy splittings and energy shifts of the excited multiplets in PuO$_2$ are similar to those calculated for the isolated Pu$^{4+}$ cation, and the predicted $M_{4,5}$ edges for the two cases are very similar. Furthermore, it appears that both the $M_4$ and $M_5$ edge XANES spectra probe mainly $J  = 7/2$ states and that the Pu-O bond covalency does not change between the ground and excited configurations \cite{bagus21}. The possibility to exploit HERFD-XANES for probing crystal fields and covalency effects in actinide compounds is discussed in \cite{butorin16,butorin16a,butorin20}. Many of the techniques proposed are interesting, but probably need better resolution before they can match standard RIXS techniques at the $M$- and $N$-edges \cite{butorin16b,kvashnina22}.

Recently, HERFD-XANES measurements at the Pu $M_4$ and $L_3$ absorption edges, combined with theoretical calculations, have been used to investigate the formation of PuO$_2$ NPs from oxidized
Pu(VI) under alkaline conditions, revealing the formation of a  stable intermediate Pu(V) solid phase, similar to NH$_4$PuO$_2$CO$_3$ \cite{kvashnina19}. Further studies reported in \cite{gerber22} show that the Pu(IV) oxidation state dominates in all NPs formed at pH 1–4, with Pu(III) and Pu(VI) present in addition to Pu(IV) due to the redox dissolution of PuO$_2$ NPs under acidic conditions. The results of these studies are important to better understand the colloid-facilitated transport governing the migration of plutonium in a subsurface environment.
\\

\subsection{Examples of EXAFS studies on actinides}

EXAFS has been greatly useful to the actinide community as a tool for probing in situ gas, solution (liquid), or solid phases. An extensive review on earlier studies is given in \cite{shi14}. As mentioned above, beamlines at synchrotrons have been constructed to exploit this technique for actinide materials \cite{dardenne09,zimina17,scheinost21,solari09}.   An excellent  example of the power of the technique is \cite{clark06}. The main reason for its success is that the actinides, with so many electrons, give a very strong signal so that very small quantities can be successfully examined \cite{carter21}. For the most part, it is probably safe to say that EXAFS tackles materials that are of interest to chemistry and materials science problems of concern for nuclear fuel and nuclear waste disposal \cite{richmann01,hubert06,walter06,prieur08,degueldre11,degueldre13,gaona13,nastren13,ding21}, molten salts \cite{volkovich05,smith19}, and extraction chemistry
\cite{antonio01,bolvin01,ikeda-ohno08,skanthakumar08, boubals17,ferrier18,bhattacharyya19}. 
Nevertheless, studies of interest to condensed matter physics are not rare. For instance, \cite{bridges20} reports the results of an EXAFS investigation of the local structure of Fe-doped URu$_2$Si$_2$, suggesting the occurrence of a local orthorhombic distortion with B$_{1g}$-like symmetry below 80-100 K, and \cite{booth07} uses EXAFS to characterize the structural damage from self-irradiation in the plutonium unconventional superconductor PuCoGa$_5$. An excellent history of the early years \cite{lytle99} shows that all the original work, dating back to the 1920s, relies on physics but, as so often is the case, once the technique and analysis have been developed, then the main applications are in other condensed matter subfields. 

Here we shall consider one important question that has concerned the actinide community for at least 50 years. What is the distribution of oxygen and uranium atoms in the series of materials characterized with the formula UO$_{2+x}$ (where $0 < x  <  1$)? Although there are earlier papers on this subject, the best summary of earlier work is that of \cite{willis87}. Since no new diffraction peaks are observed for samples up to $x$ = 0.25, these samples have disordered oxygens and the subsequent short-range order of these O atoms (and any accompanying relaxation of the nearest U atoms) are then ideal problems for the EXAFS technique. Willis had earlier proposed a series of clusters (the so-called 2:2:2 cluster proposed for $x$ = 0.12 being the most famous) and this has been discussed in a more recent paper using theoretical modeling \cite{wang14}. At $x$ = 0.25 the compound U$_4$O$_9$ is formed and that structure was reported in \cite{bevan86} and a second modification in  \cite{cooper04}. None of these structures have a U-O distance less than $\sim$2.2 \AA, whereas in the cubic fluorite structure (UO$_2$, $x$ = 0) the shortest U-O distance is 2.37 \AA.

In a series of papers using EXAFS (at the actinide $L_3$ edges) and pair distribution function (PDF) methods (using data taken with neutron diffraction) as the primary analysis technique,  \cite{conradson04,conradson04b,conradson05,conradson13} have claimed that there are unusual aspects of the UO$_{2+x}$ and PuO$_{2+x}$ systems that had not been observed previously. There is much in these papers that is beyond the scope of the present review, but (for x $>$ 0) essential features on which we wish to concentrate here, are the proposed U-O bond of $<$ 2 \AA, and the amorphous-like nature of a part of the sample. Since the U-O distance in UO$_2$ is $\sqrt{3} a_{0}/4$ = 2.37 \AA\, (where $a_{0}$ is the lattice parameter) a U-O distance of $<$ 2 \AA \, represents a major structural change, and has not been reported for any previous work on the UO$_{2+x}$ system - see, for example, recent modeling work \cite{wang14}. The reason for this short bond was ascribed to some U atoms having a $f^{0}$ (U$^{6+}$) configuration as the uranyl-type U-O bond is 1.97 \AA. Also for PuO$_{2+x}$ \cite{conradson04b} the nearest-neighbor O atoms were found in a multisite distribution, with the excess O  at Pu-O distances $<$ 1.9 \AA. This was attributed  to multiply bound oxo-type ligands, as  found in molecular complexes of Pu(V) and Pu(VI) \cite{conradson04b}.

\begin{figure}
\centerline{\includegraphics[width=1.0\columnwidth]{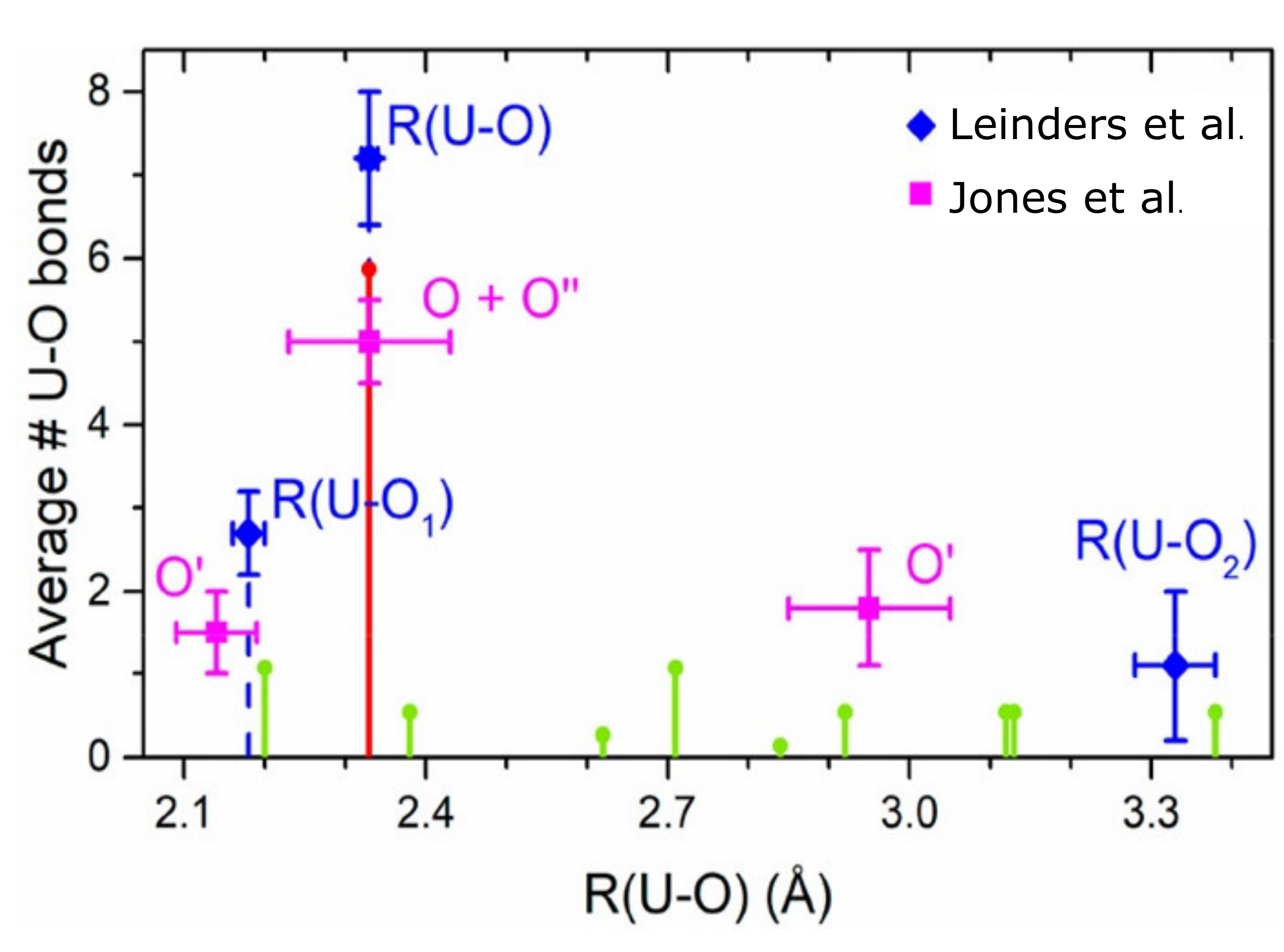}}
\caption{ Distribution of U-O bonds in the average unit cell of U$_3$O$_7$ at 18 K. The structural model assumes a regular geometry of cubo-octahedral oxygen clusters. The histograms in red and green correspond to the fluorite-type and additional U-O bonds, respectively. Blue diamonds and magenta squares are EXAFS results reported in Refs.~\cite{leinders20} and \cite{jones86}, respectively. 
Reproduced from \cite{leinders20}, https://pubs.acs.org/doi/10.1021/acs.inorgchem.9b03702. Further permission related to the material excerpted should be directed to the ACS.
\label{leinders20}}
\end{figure}

More recent work has questioned both of the premises emerging from the above EXAFS work. First, one should recall that in the range $0 < x < 0.20$ the samples are only single phase at temperatures above $\sim$673 K, and that below this temperature the sample will consist of mixtures of UO$_2$ and U$_4$O$_9$ ($x$ = 0.25). Second, using neutron diffraction \cite{garrido06} showed that by carefully analyzing a sample of U$_4$O$_9$ neither of these two premises were correct. Further evidence comes from a paper by \cite{palomares19}, again using neutron diffraction of UO$_{2.07}$, with data taken at between 873 and 1273 K, showing that, although the O-O distance does approach 2 \AA \, for the solid solutions, the closest U-O distance remains above 2 \AA\, and close to that in UO$_2$. These short O-O distances are key features of the Willis clusters \cite{willis87}. In the case of PuO$_{2+x}$, the short Pu-O distance $<$ 1.9 \AA\, has been attributed to multi-electron excitation effects in the EXAFS spectra \cite{rothe04}.

\cite{leinders20} have published an exhaustive study of U$_3$O$_7$ ($x$ = 0.33) using HERFD-XANES at both the $L$ and $M$ edges, as well as EXAFS at the $L$ edges. They are able to determine the different uranium valencies in these materials with x $>$ 0, and that the U-O bonds are spread over values from 2.25 to 2.80 \AA, see Fig.~\ref{leinders20}.  The valencies for the U atoms are almost 50\% each of U$^{4+}$ and U$^{5+}$ for $x$ = 0.33, and there is no sign of U$^{6+}$. These results are in agreement with \cite{kvashnina13}, see Fig.~\ref{kvashnina13}, where it is clear that the U$^{6+}$ state first appears in U$_3$O$_8$ ($x$ = 0.67).

Finally, several EXAFS studies have been dedicated to the characterization of the local structure in actinide nanoparticles (NPs) and colloids \cite{gerber20,micheau20,hudry14,dalodiere17,plakhova19,bonato20,gerber21}. EXAFS spectra on PuO$_{2}$ and CeO$_{2}$ NPs, measured  at the Structural Materials Science beamline \cite{chernyshov09} of the Kurchatov Synchrotron Radiation Source (Moscow, Russia),  are reported in \cite{romanchuk22}, where  a new core-shell model for NPs is discussed and size effects on oxygen disorder and metal-metal coordination number are studied.

\subsection{Examples of XMCD studies on actinides}\label{secXMCD}

XMCD is associated with time-reversal symmetry breaking by a magnetic field and involves electric-dipole or electric-quadrupole transitions promoting an electron in a spin-orbit split core state to an empty valence state of the absorbing atom. The technique provides an element- and shell-specific probe for studying the electronic structure of a wide range of materials \cite{vanderlaan13,vanderlaanCCR14,rogalev15}.  In actinides materials, XMCD spectra are conveniently measured at
the $M_{4,5}$ (3$d_{3/2,5/2} \rightarrow$ 5$f$) absorption edges that directly interrogate the 5$f$ shell. The XMCD signal is obtained as the difference
$\Delta I_{M_{4,5}} = \mu^-(E) - \mu^+(E)$ of the absorption spectra of circularly polarized photons with helicity parallel ($\mu^+(E)$) and antiparallel  ($\mu^-(E)$) with respect to a magnetic field applied along the incident beam direction. The spectra must be corrected for self-absorption effects and for the incomplete polarization of the incident beam emerging from the crystal monochromator. Standard protocols have been developed for performing such corrections \cite{goulon82,troger92,pfalzer99}.

 The power of the technique comes from the simplicity of the sum rules relating orbital $\langle L_z \rangle$ and spin $\langle S_z \rangle$ moments of the absorbing atoms to linear combinations of the dichroic signals integrated over the spin-orbit-split absorption edges,
normalized to the isotropic x-ray absorption spectrum 
\cite{thole92,carra93,vanderlaan96,vanderlaan04}.
The  orbital and spin sum rules are given in Eqs.~(\ref{eq:orbital-rule}) and (\ref{eq:spin-rule}), respectively.
%
%
%
%
The orbital and spin components of the total magnetic moment of the 5$f$-shell $\mu = -(\langle L_{z} \rangle + 2\langle S_{z} \rangle)$ can then be obtained from  XMCD spectra, together with an estimate of $\langle T_{z} \rangle$, if the value of the total moment $\mu$ and the occupation number of the $5f$ shell are known.

Furthermore, the expectation value of the angular part of the valence states spin-orbit operator, $\langle\psi| {\bm{\ell}} \cdot \bm{s}|\psi\rangle$, can be obtained from the XAS branching ratio, $B = \rho_{M_{5}}/(\rho_{M_{5}}+\rho_{M_{4}})$, using Eq.~(\ref{eq:BR}), as \cite{thole88}
\begin{equation}\label{so}
\frac{2\langle {\bm{\ell}} \cdot \bm{s}\rangle}{3n_{h}}-\Delta = -\frac{5}{2}(B-\frac{3}{5}),
\end{equation}
where $\Delta$ is a correction factor dependent on the electronic configuration \cite{vanderlaan04} (for Np$^{3+}$, e.g., $\Delta$ = $-$0.005).
Interestingly, Eq.~(\ref{so}) has also been used to determine the $f$-counts of various actinides \cite{moore09}}.

The presence of orbital moments in actinide systems was recognized many years ago, but the coupling scheme between the spin and orbital components was controversial for many years. Whereas this is known to follow Russell-Saunders coupling for the lanthanide 4$f$ systems, there was some question as to whether this would be the case for the 5$f$’s in the actinides, where the spin-orbit coupling is larger than in the lanthanides, or whether it would follow the intermediate regime or even the $jj$ scheme. XMCD promised to be a tool that would be able to resolve this question, so that relatively early on a number of experiments were undertaken. Up to that time, the only experimental  method that could give information on the orbital and spin contributions was neutron scattering, and it relied on a model (for the magnetic form factor) to resolve the components \cite{lander93}. The XMCD method is certainly more direct, and does not require single crystals as is the case with neutron scattering, but it does have the weakness that although the orbital moment can be determined directly, the deduction of the spin moment requires the value of
$\left< T_z \right>$, see Eq.~(\ref{eq:spin-rule}), 
which can in principle be determined by measuring the angular dependence of the XMCD  \cite{stohr95,vanderlaan98a}.

The first XMCD experiments on actinides were done on ferromagnetic US at the Daresbury Synchrotron Radiation Source (SRS) and published in \cite{collins95a}. They showed results consistent with those obtained with neutron scattering, but it was not easy from either measurement to determine whether the configuration was 5$f^{2}$ or 5$f^{3}$. Earlier neutron results had shown a peculiar situation for the U atom in the compound UFe$_2$ (cubic Laves phase), where there appeared to be an almost complete cancellation between the orbital and spin moments of the 5$f$ states, so that the resulting moment was very small \cite{wulff89}. This naturally spurred a number of XMCD efforts, but work at both the $M_{4,5}$ \cite{finazzi97} and $N_{4,5}$ edges \cite{okane04,okane06} confirmed the neutron results. The effect is thought to occur via a strong mixing of the Fe 3$d$ and U 5$f$ states, and a consequent reduction in both the iron and uranium moments, with the U orbital moment being severely suppressed \cite{antonov03}.

\begin{table*}[t]
\caption{Values of orbital ($\mu_L$) and spin ($\mu_L$) moments per atom, their ratios, and the value of the magnetic dipole term (see Eq. (\ref{eq:spin-rule})) 3$\left<T_z\right>$. All values are in intermediate coupling, see Table VIII of   \cite{vanderlaan96}. Note that by convention the total moment $\mu$ =
$\mu_L$ + $\mu_S$. However, as shown in the theory section $\mu_L = – \left<L_z\right>$, and
$\mu_S  = - 2 \left<S_z\right>$. Notice that all values correspond to the ground $J$ state, except for the $f^{6}$ configuration. Normally, all numbers would be zero for the $f^{6}$ ground state. However, in a strong magnetic field there is mixing with the $J$ = 1 excited state, and these are the values calculated (and measured) for Am$^{3+}$ in AmFe$_2$.}
\begin{center}
\begin{ruledtabular}
\begin{tabular}{l|l|l|l|l|l|l}

$f^{n}$ &Configuration &$\mu_L$ & $\mu_S$ & $\mu_L$/$\mu_S$ & 3$\left<T_z\right>$ &
3$\left<T_z\right>/\left<S_z\right>$ \\ \hline
1	&U$^{5+}$	&+2.856	&$-$0.714	&$-$4.00	&$-$1.714	&+4.80\\
2	&U$^{4+}$, Np$^{5+}$&	+4.698&	$-$1.397&	$-$3.36&	$-$2.428&	+3.48\\
3	&U$^{3+}$, Np$^{4+}$, Pu$^{5+}$& $-$5.571&	$-$2.140&	$-$2.60&	$-$1.978&	+1.85\\
4	&Np$^{3+}$, Pu$^{4+}$&	+5.406&	$-$2.810&	$-$1.92&	$-$0.781&	+0.56\\
5	&Pu$^{3+}$, Am$^{4+}$&	+3.885&	$-$2.767&	$-$1.40&	+0.302&	$-$0.22\\
6	&Am$^{3+}$&	+0.47	&$-$0.94&	$-$0.50&	$-$0.51&	$-$1.1\\
7	&Cm$^{3+}$	& +0.330&	+6.343&	+0.052	& $-$0.225	& $-$0.071\\
\end{tabular}
\end{ruledtabular}
\end{center}
\label{moments}
\end{table*}

\begin{figure}
\centerline{\includegraphics[width=1.0\columnwidth]{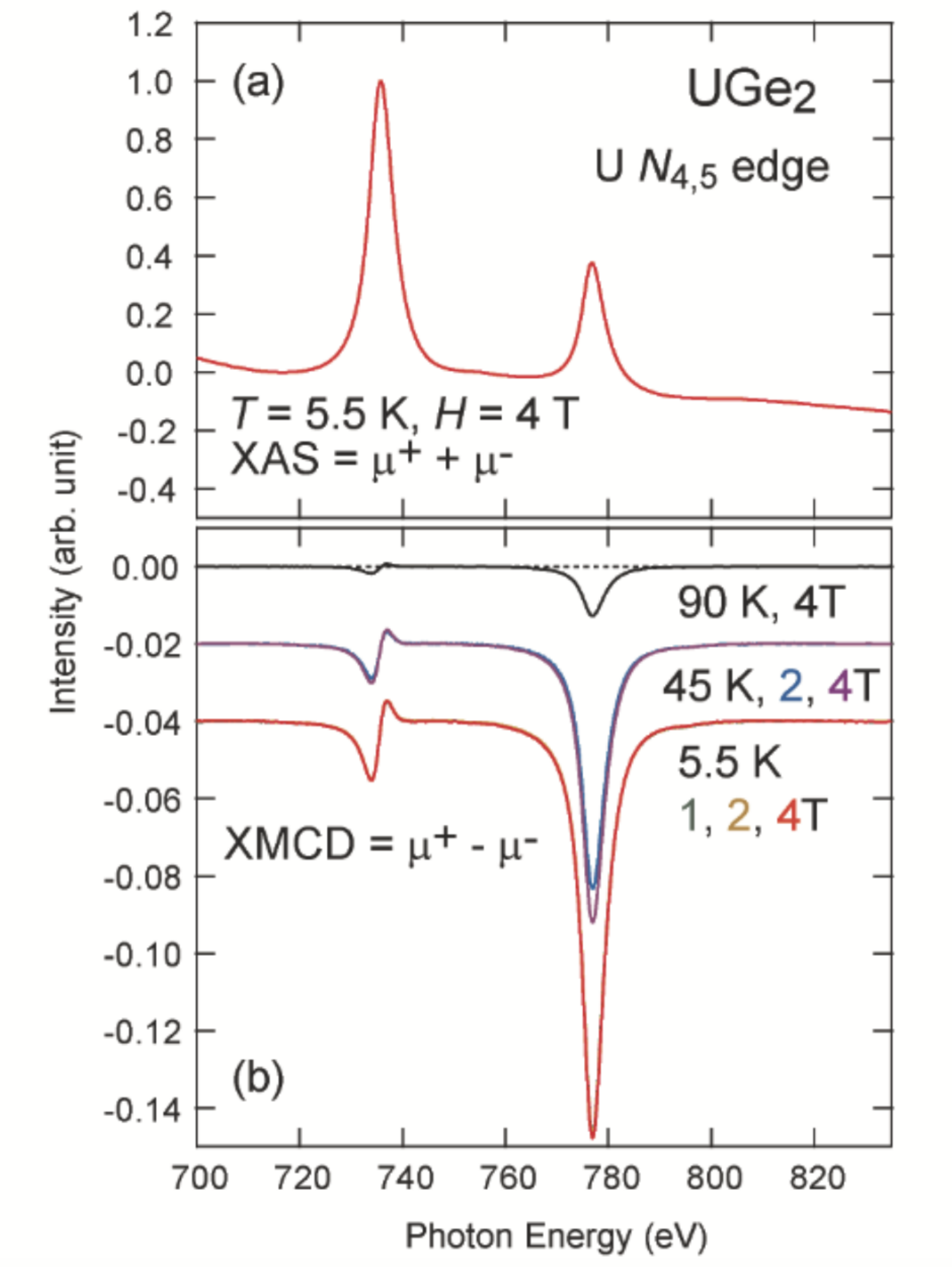}}
\caption{ (a) XAS spectra of UGe$_2$ measured at the U $N_{4,5}$ absorption edges at 5 K in a magnetic filed of 4 T. (b) XMCD signal measured at different temperatures and applied magnetic fields.
Reproduced with permission from \cite{takeda18}. $\copyright$ (2018) Atomic Energy Society of Japan.}  
\label{UGe2}
\end{figure}

Many other U ferromagnets were examined \cite{taupin15,butchers15,dalmas97,dalmas98,kernavanois01a}, as well as heavy-fermion compounds \cite{yaouanc98,dalmas99} and compounds of the formula 1:1:1, e.g., UTAl, with T = transition metal \cite{kucera02,antonov03b,taupin15}, and there was an effort to compare the observed orbital and spin moments, and their ratio, to results from band-structure calculations. Results were also reported using the $N_{4,5}$ edges from the SPring-8 synchrotron \cite{okane04,okane06,okane06b,okane08}. A major difficulty with using such low-energy photons ($<$ 1 keV) is that the self-absorption corrections are larger than at the $M_{4,5}$ edges, and thus more difficult to make.

We show in Fig.~\ref{UGe2} the results from the study of UGe$_2$ (a ferromagnetic superconductor at modest pressure) studied at SPring-8 \cite{okane06}. The results showed that the U ion was in an 5$f^{3}$ state, in agreement with an extensive neutron investigation reported earlier \cite{kernavanois01}. For more on UGe$_2$ see theory of XMCD \cite{shick04}. Note that this use of the sum rules at the $N_{4,5}$ edges implicitly assumes that the spin-orbit splitting of the core 4$d$ level, which is $\sim$50 eV, is far larger than other interactions, including the core-valence Coulomb and exchange interactions. This is certainly the case with the $M_{4,5}$ edges, where the 3$d$ core levels are split by $\sim$200 eV, but the authors have shown that any correction at the $N$ edges is less than 5\%.

XMCD has been used as a general tool to understand the magnetism in a number of actinide systems. Early work showed the sensitivity to small amounts of material in multilayers of U/T, where T = Fe, Co, Ni, and Gd \cite{wilhelm07,springell08}. An induced moment ($\sim$0.1 $\mu_{\mathrm{B}}$) was found on the U atom in U/Fe multilayers, but an even smaller moment was found on the U atoms in the other ferromagnetic systems. This can be understood by considering the overlap of the bands of the T atoms with the uranium 5$f$ density-of-states. In the case of U/Gd the induced moment is small, but shows an oscillatory behavior, whereas in the transition-metal systems, the induced U moment appears to decay rapidly as the distance from the U-Fe interface increases. A good review has been published on the XMCD on U compounds at the ESRF ID12 beamline and contains more information than given here \cite{wilhelm18}.

One of the aspects of this XMCD work was the limitation of not knowing the value of
$\left< T_z \right>$. Thus, the orbital moment can be measured uniquely, but the
$\left< T_z \right>$ operator means that only an effective $\left<S_{\mathrm{eff}}\right>$ = $\left<S_z\right>$ + 3$\left< T_z \right>$ can be determined from the XMCD experiment. As shown in Table \ref{moments}, which gives some of these values for intermediate coupling, the values of $\left< T_z \right>$ for $5f^n$ configurations up to 3 electrons are large, so it became of interest to perform experiments on transuranium materials to see whether the predicted decrease in the effect of the $\left< T_z \right>$ term was observed consistent with intermediate coupling, as shown in  Table \ref{moments}  \cite{vanderlaan96}. As the section below shows these results on transuranium materials were successful in showing that the theory for $\left< T_z \right>$ in intermediate coupling can be assumed to apply whether the sample is ionic or band-like.

The XAS and XMCD experiments reviewed below were carried out at the ID12 beamline of the ESRF using samples encapsulated in an Al holder with Kapton windows of 60-$\mu$m thickness in total. The cryomagnet available at ID12 affords magnetic fields up to 17 T and a base temperature of $\sim$2 K.
The minimum mass of the sample depends on the magnetic susceptibility of the system. For elemental curium, data have been collected on a 0.55 mg sample of $^{248}$Cm  \cite{lander19}, whereas a sample mass of 16 $\mu$g was sufficient for Am in a sample of AmFe$_2$ \cite{magnani15}.

\begin{figure}
\centerline{\includegraphics[width=1.0\columnwidth]{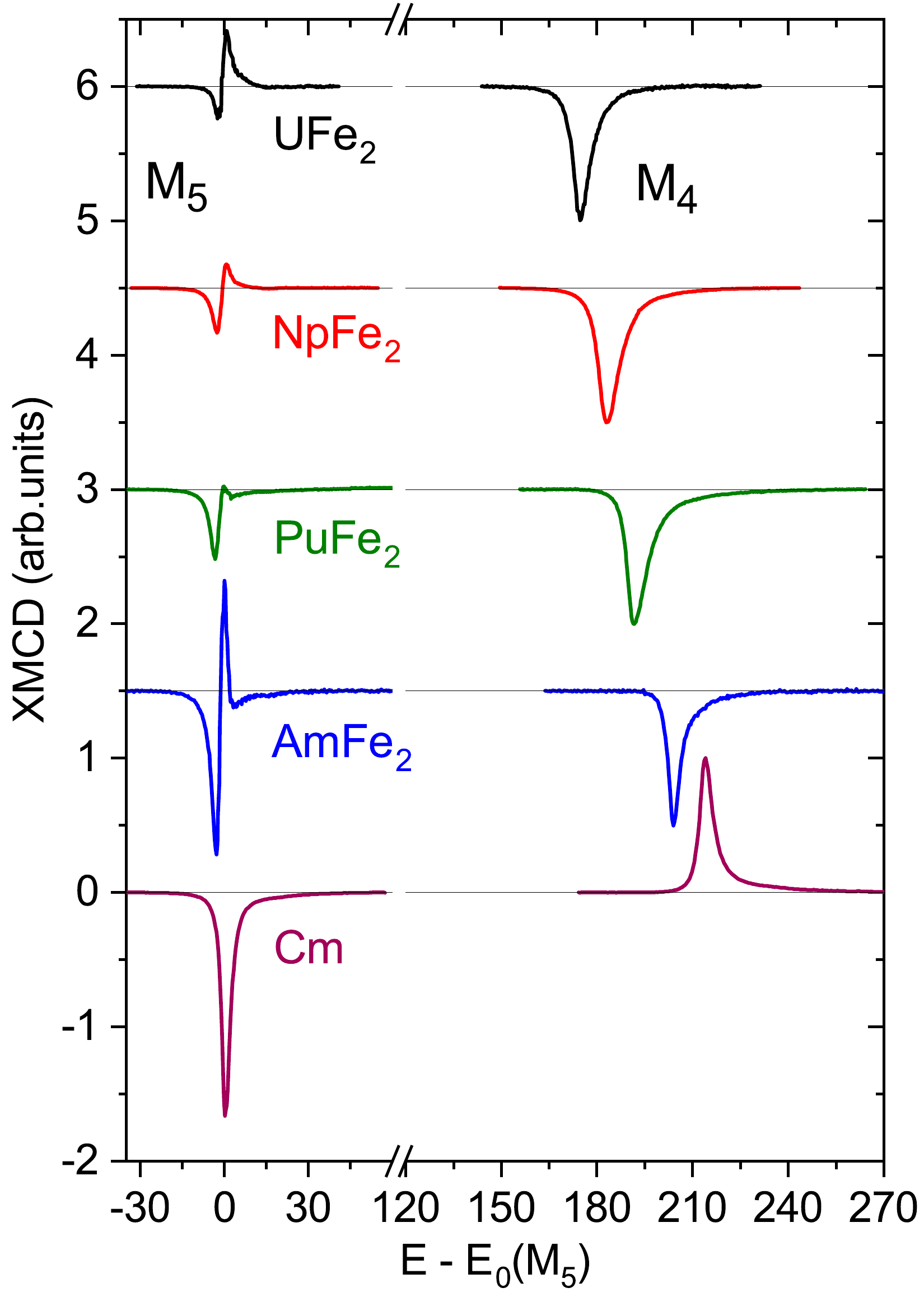}}
\caption{ XMCD spectra as a function of photon energy through the An $M_{4,5}$ edges in AnFe$_2$ (An = U, Np, Pu, Am) and elemental Cm. The spectra have been corrected for self-absorption effects and incomplete circular polarization of the incident beam.
Adapted from  \cite{lander19}.
  \label{XMCDAnFe2}}
\end{figure}

Figure \ref{XMCDAnFe2} shows an overview of the XMCD signals measured at the $M_5$ and $M_4$ edges in ferromagnetic AnFe$_2$ (An = U, Np, Pu, Am) compounds  \cite{wilhelm13,magnani15}. These spectra are compared with the XMCD signal measured for elemental curium \cite{lander19}. 
For convenience, the photon energy is set to zero at the $M_5$ edge and the amplitude at the $M_4$ edge is normalized to unity. It must be noticed that the spectral intensity at the $M_4$ actually becomes smaller with increasing atomic number from uranium to americium, as the number of $5f$ holes in the $j$ = 5/2 subshell decreases. Moreover, in absolute units the XMCD signal for AmFe$_2$ is smaller than for NpFe$_2$, reflecting the difference between the magnetic moments in the two compounds, whereas the narrower line-width of the $M_4$ line in AmFe$_2$ indicates the presence of localized 5$f$ states, as observed in PuSb \cite{janoschek15}.

In the case of curium, a visual inspection of the XMCD spectra is sufficient to realize that the integrated intensity of the features at the $M_5$ and $M_4$, opposite in sign, are not equal. From the orbital sum rule, this means that the orbital moment is not zero. The value provided by the experiment at 70 K is $\mu_L  = - \langle L_z \rangle$ = 0.10(1) $\mu_{\mathrm{B}}$,  with a ratio $\mu_L/\mu_S$ = +0.06 close to the calculated value of +0.052 for the $f^{7}$ configuration in intermediate coupling \cite{vanderlaan96}. The fact that $\mu_L$ and $\mu_S$ are parallel in Cm metal is the reason why the sign of the $M_4$ line suddenly changes in comparison to the earlier elements (Fig. \ref{XMCDAnFe2}). 

\begin{figure}
\centerline{\includegraphics[width=1.0\columnwidth]{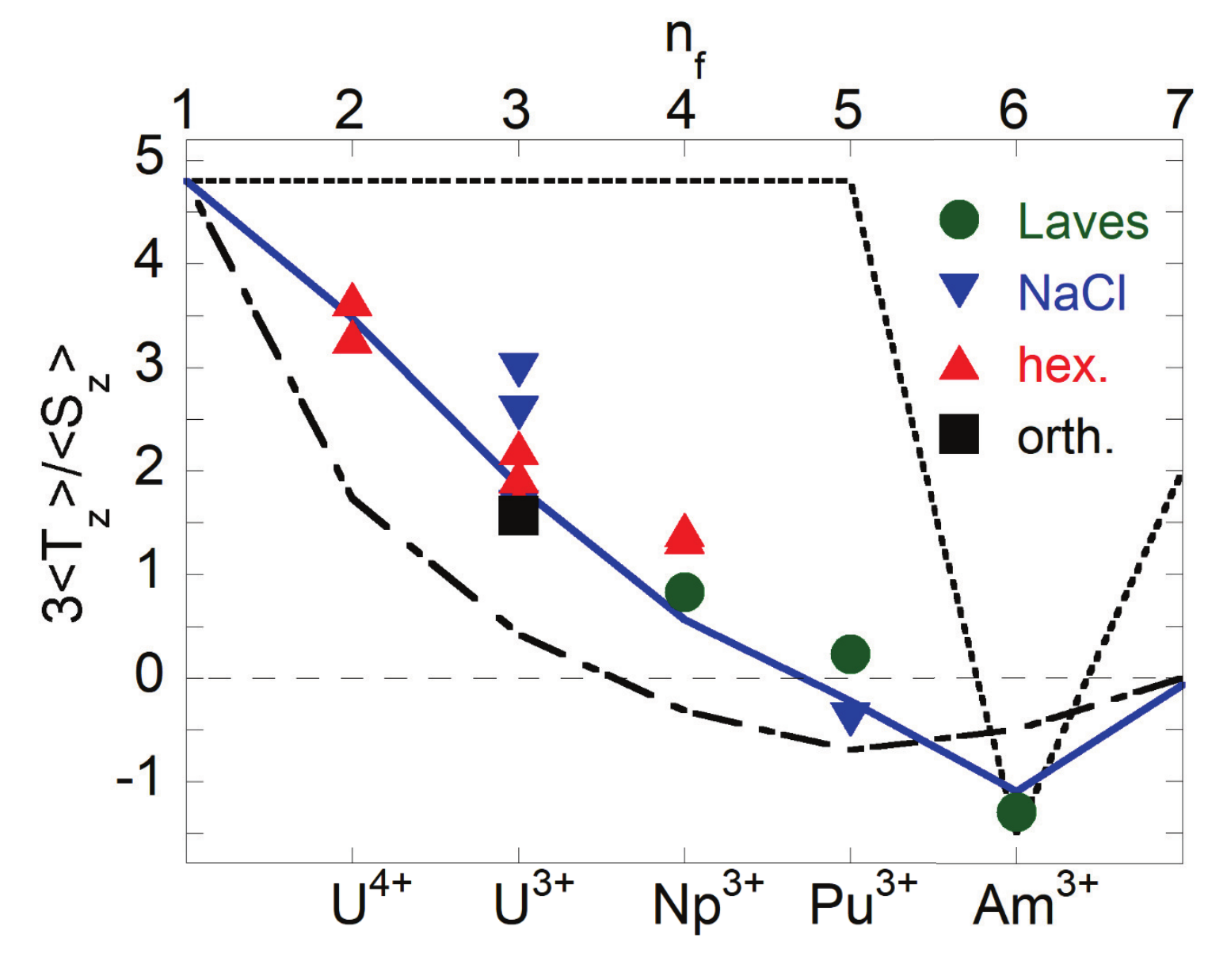}}
\caption{The ratio 3$\left< T_z \right>$/$\left<S_z\right>$ between the expectation values of the magnetic dipole term and the spin moment along the quantization direction as a function of the 5$f$ occupation number n$_f$. Symbols are experimental data obtained from an analysis of XMCD spectra at the $M_{4,5}$ actinide edges for  compounds with different physical properties and crystallographic structure (circles for C-15 cubic Laves phases, down triangle for the NaCl-type structure, up triangles for hexagonal lattice, and square for orthorhombic cells).  Theoretical estimates are shown for intermediate coupling (IC) (solid lines), Russell-Saunders (LS) (dashed line), and $jj$ (dotted line) coupling approximations.  All the calculated values refer to the free-ion ground state, except for the $f^{6}$ configuration, where $J$ mixing with the first excited state is taken into account; otherwise,  the ratio can not be determined since all quantities would be zero for the ground state. 
Adapted from \cite{magnani15}. 
\label{XMCDToverS}}
\end{figure}

Equation (\ref{eq:spin-rule}) shows that the spin component of the magnetic moment can only be obtained if the value of $\langle T_{z}\rangle$ is known. This quantity cannot easily be measured directly. However, it can be immediately obtained from an analysis of the XMCD spectra if an independent measurement of the total magnetic moment $\mu$ = $\mu_S$ + $\mu_L$ is available, for instance from neutron diffraction experiments or, in the case of Np compounds, from $^{237}$Np M\"ossbauer spectroscopy. Figure \ref{XMCDToverS} shows the 3$\langle T_{z}\rangle$/$\langle S_{z}\rangle$ ratio obtained for several compounds for which such information was available. By changing the 5$f$ occupation number $n_f$, the ratio 3$\langle T_{z} \rangle$/$\langle S_{z}\rangle$ varies as predicted by atomic calculations in the intermediate coupling (IC) approximation. The correlation is very convincing and suggests that  $\langle T_{z}\rangle$ can be reliably estimated by IC calculations when the total magnetic moment $\mu$ is not known.

\begin{figure}
\centerline{\includegraphics[width=1.0\columnwidth]{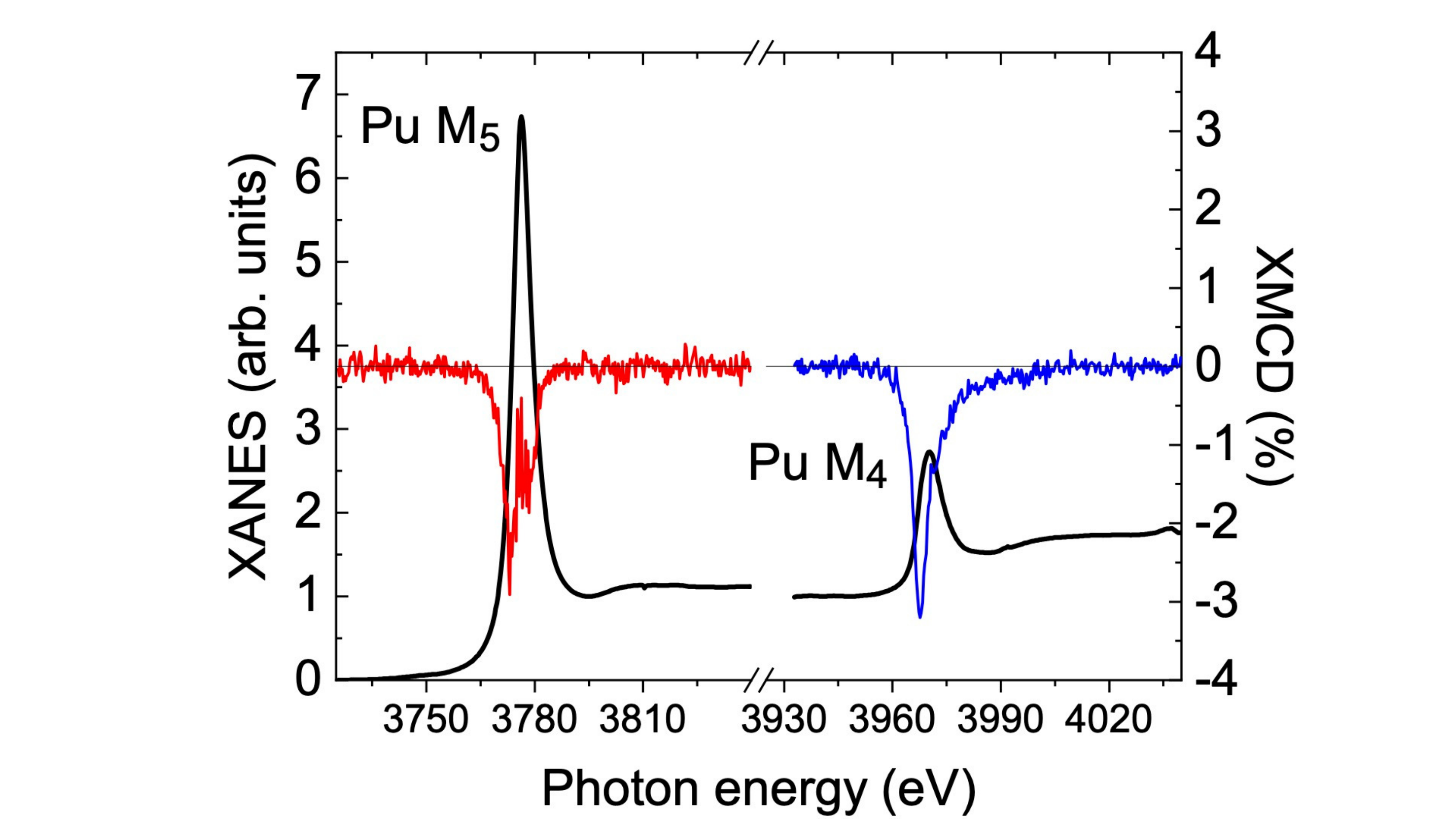}}
\caption{ XAS (solid black lines) and XMCD spectra as a function of photon energy through the Pu $M_5$ (red line) and $M_4$ (blue line) edges in PuCoGa$_5$. 
Adapted from \cite{magnani17}.
\label{XMCDPu115}}
\end{figure}

This is, e.g., the case when the absorbing atom is located inside a vortex in a type-II superconducting compound. As an example, Fig.~\ref{XMCDPu115} shows XAS and XMCD spectra measured in PuCoGa$5$ \cite{magnani17}, a tetragonal heavy fermion superconductor with a critical temperature $T_c$ = 18.5 K \cite{sarrao02}. The origin of superconductivity in this compound remains puzzling, despite years of intensive investigations \cite{jutier08,graf15,ramshaw15}. The Pu ground state is non-magnetic \cite{hiess08}, and the superconducting order parameter has $d$-wave symmetry \cite{daghero12}.

XMCD spectra have been measured above and below $T_c$ on a 2 mg single-crystal sample of $^{242}$PuCoGa$_5$ (99.99 wt\% $^{242}$Pu) \cite{magnani17}. The $c$-axis of the crystal was oriented along the incident photon beam, and therefore parallel to the applied magnetic field. The critical field $B_{c2}$ $(T=0)$ along the $c$-axis is 63 T, so that a vortex phase is present in an applied field smaller than 15 T. The 5$f$ shell of the Pu atoms inside the vortex cores is polarized by the external field and the XMCD response is different from zero. Applying the sum rules, the moments and their ratio can be extracted from the experimental data. The values obtained are in excellent agreement with dynamical mean field theories (DMFT) predictions of a non-magnetic ground state for Pu in PuCoGa$_5$ (see Table I in  \cite{magnani17}). This surprising result is a consequence of two effects: intermediate valence driven by 5$f$-6$d$ Coulomb interaction mixes a magnetic 5$f^{5}$ sextet with a non-magnetic $5f^{6}$ singlet, reducing the magnetic moment; a complete quenching is then produced by a Kondo-like screening  promoted by the hybridization between 5$f$ and conduction electron states \cite{pezzoli11,shick13}.

A good example of the utility of the XMCD technique in considering a complex material, about which little was known previously, is the study of Np$_2$T$_{17}$ (T = Co, Ni), where the unit cell contains two different Np sites and site-selective magnetic order can occur \cite{halevy12,hen15}. By combining XMCD with M\"{o}ssbauer $^{237}$Np spectroscopy, magnetization measurements, and first principles electronic structure calculations a deep insight on the ground state of these Np intermetallics was obtained. For the Ni analogue, a nontrivial situation was observed below the ordering temperature $T_N$ where a large ordered moment (2.25 $\mu_{\mathrm{B}}$) appears in one Np site while the other  carry only an induced moment of about 0.2 $\mu_{\mathrm{B}}$. Experiments were also performed on the Haucke intermetallic compound NpNi$_5$ \cite{hen14}, showing a positive and large ratio between the expectation value of the magnetic dipole operator and the spin magnetic moment, as predicted for localized 5$f$ electrons, and an expectation value of the angular part of the spin-orbit interaction operator in good agreement with the value calculated in intermediate coupling approximation for Np ions.

More recently, a rather complete experiment including applying pressure up to 7 GPa has been reported on the ferromagnet UGa$_2$ \cite{kolomiets21}. This investigation has also used absorption spectroscopy (HERFD-XANES as discussed in Sec.~\ref{secRIXS}) as well as XMCD, and includes considerable theory. This is an intermetallic system with a large ferromagnetic moment on the uranium of 3.0(2) $\mu_{\mathrm{B}}$, but the authors claim through the spectroscopy that the 5$f^{3}$ ground state is not localized, and it should probably be considered as a band system. Interestingly, one of the challenges for theory, already for many years, on intermetallic actinide systems has been to get the correct values for the orbital moments. In  \cite{kolomiets21} the authors show that by using the LDA+DMFT they get a value for 
$\mu_L / \mu_S \approx -$2.5, which is very close to the $-$2.6 in Table \ref{moments}. There is also a cautionary tale from this study, as experimentally they find only a moment of 1.87 $\mu_{\mathrm{B}}$, whereas both neutron and magnetization give 3.0 $\mu_{\mathrm{B}}$. Even though they have a single crystal, they ascribe this problem to surface contamination by UO$_2$. This again shows the surface sensitivity of this technique, even at the $M_{4,5}$ edges. Ironically, the experiments on trans-uranium samples \cite{magnani15,wilhelm13,janoschek15,magnani17} have not suffered from this problem, as safety requirements demand airtight encapsulation.

\begin{figure}
\centerline{\includegraphics[width=1.0\columnwidth]{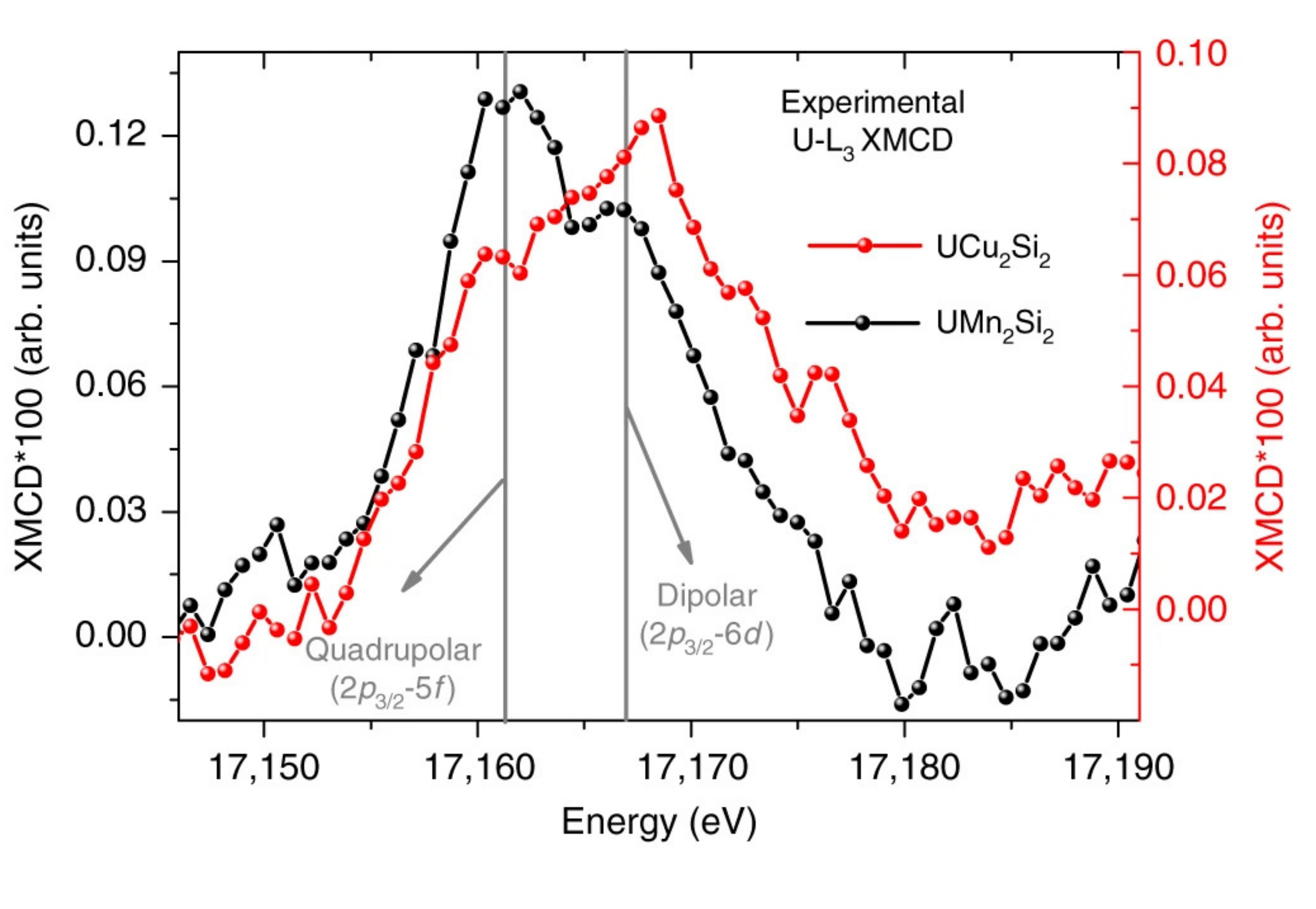}}
\caption{XMCD measurements for UCu$_2$Si$_2$ and UMn$_2$Si$_2$ performed at temperatures of 10 K and 300 K, respectively. The positions of the electric dipole and quadrupolar transitions are marked.
Taken from \cite{dosreis17}. 
 \label{reis}}
\end{figure}

Whereas the vast majority of XMCD experiments on actinide systems have used the $M_{4,5}$ edges, as we have shown above, a few studies (all at SPRING-8 in Japan) have been done at the $N_{4,5}$ edges \cite{okane06,okane08}, and a small number also at the $L_{2,3}$ edges \cite{dosreis17}. The latter edges do not couple directly to the 5$f$ electrons, as their electric-dipole contribution corresponds to 2$p \rightarrow$ 6$d$ transitions, whereas only the weaker electric-quadrupole term connects
2$p \rightarrow$ 5$f$ states. However, these transitions are known to be separated in energy by about 5 eV, so it is possible to measure both transitions. This has been done at the Brazilian synchrotron (with additional measurements at the APS) by \cite{dosreis17} on the uranium intermetallic compounds UCu$_2$Si$_2$ and UMn$_2$Si$_2$. 
The authors show a much larger quadrupolar peak in the case of UMn$_2$Si$_2$ and are able to model important differences between the 5$f$-6$d$ hybridization in the two different compounds. The experiments on the Mn compounds above are taken at 300 K, but it is known from neutron work that the U moments order only at 100 K, whereas the Mn moments order at 377 K. On lowering the temperature below 100 K a large increase was observed for the quadrupole transition, which was also observed at the uranium $M_4$ edge, proving that the assignment of the electric-quadrupole transition to the 5$f$ states is correct. Although these interpretations do depend on modeling, it is perhaps surprising that more work at the $L$ edges has not been reported.

\begin{figure}
\centerline{\includegraphics[width=1.0\columnwidth]{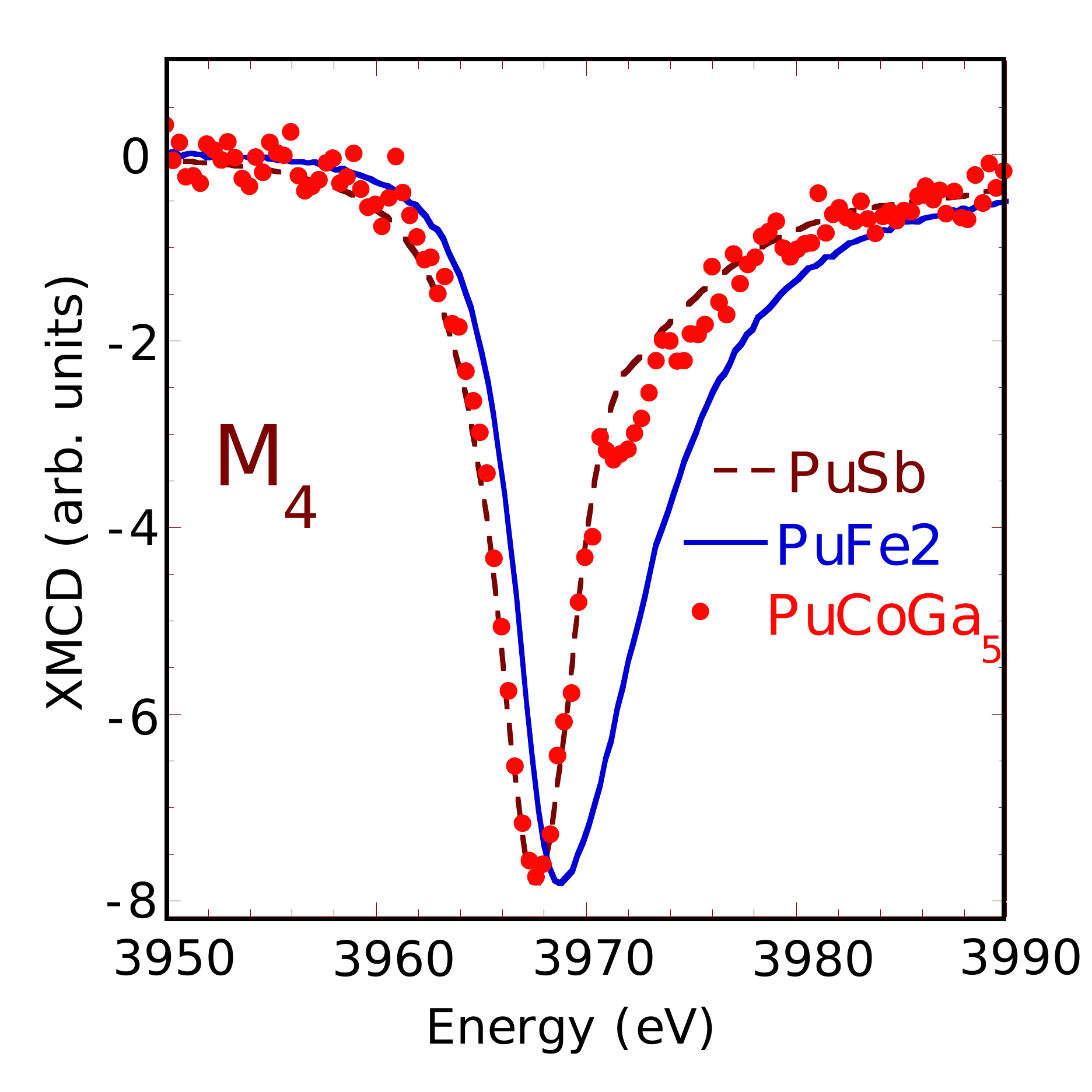}}
\caption{ XMCD signal at the $M_4$ edge from three different Pu materials: the localized system PuSb \cite{janoschek15}, the heavy-fermion superconductor PuCoGa$_5$ \cite{magnani17}, and the band-like system PuFe$_2$ \cite{wilhelm13}. \label{wilhelm13}}
\end{figure}

As far as we know, there is no quantitative theory addressing directly the linewidths of the $M_{4,5}$ XMCD signals, but it is clear from an examination of Fig.~\ref{XMCDAnFe2} that the spectra are much sharper (longer lifetime for the transitions) when the systems are localized. The spectra for localized AmFe$_2$ and Cm metal are both narrower than the other Laves-phase materials, which are band-like. This is related to the transitions giving rise to the absorption edge into the unoccupied 5$f$ states, so is a measure of the 5$f$ bandwidth above $E_{\mathrm{F}}$, the Fermi level. In an itinerant system, the 5$f$ hybridization with the conduction states will increase the bandwidth of the unoccupied states, so we expect a broader transition in energy. Lifetime effects will also broaden the transitions due to additional decay channels. The $M_4$ transition involves only the 3$d_{3/2}$ core subshell so there is only a single transition to the 5$f_{5/2}$ subshell, and the differences in some Pu materials are shown in the Fig.~\ref{wilhelm13}. The $M_4$ lifetime is given as $\sim$4 eV, so the localized system PuSb \cite{janoschek15} is very close to this value. On the other hand, it is a surprise that the heavy-fermion PuCoGa$_5$ superconductor  \cite{magnani17} is also so narrow. In contrast, the band-like ferromagnet PuFe$_2$ \cite{wilhelm13} has an energy full-width at half maximum that is at least twice that of the intrinsic lifetime. 

Such effects may also be observed at the $M_5$ edges, although there the situation is more complex as both the 3$d_{3/2} \rightarrow$ 5$f_{5/2}$ and the 3$d_{5/2} \rightarrow$ 5$f_{7/2}$ transitions may be observed, and they differ in energy by a few eV. So, for a localized system, as PuSb \cite{janoschek15}, two separate peaks may be observed, whereas the broadening of the spectra in PuFe$_2$ does not allow the observation of two such peaks. Interestingly, the double-peak feature was first observed in resonance diffraction experiments on PuSb \cite{normile02}, but not understood at that time.

The case of AmFe$_2$ is worth a further discussion. For a localized Am$^{3+}$ ion the ground state with six 5$f$ electrons has the $j$ = 5/2 subshell full, so the quantum ground state is $J_z$ = $L_z  +  S_z$ = 0. In AmFe$_2$ the Am ions are in a very strong magnetic field induced by the surrounding magnetized Fe atoms, but $\left<J_z\right>$ is still zero. The finite spin and orbital moments arise from mixing with the $J$ = 1 excited state \cite{magnani15}. Since the orbital moment $\mu_L$ = $- \left<L_z\right>$, whereas the spin moment 
$\mu_S  = -2 \left<S_z\right>$, and they are opposite to each other, one can see that, as the induced field is increased, the $\mu_S$ begins to dominate the $\mu_L$ moment, as the spin moment will be twice the orbital moment and of  opposite sign. We have the counter-intuitive situation that the spin moment is opposed to the molecular field induced by the Fe moments, which are easily aligned by an external field and increases with the applied field! This is a direct consequence of the quantum mechanics defining the $J_z$ = 0 ground state. The actual value of the Am moment in AmFe$_2$ was found to be $-$0.44(11) $\mu_{\mathrm{B}}$ \cite{magnani15}. This is in excellent agreement with the $\sim$ $-$0.4 $\mu_{\mathrm{B}}$ reported from an early experiment using polarized neutrons on polycrystalline samples, although the reason for the negative moment on the Am site was not understood at that time \cite{lander77}.

\section{RIXS and resonant x-ray emission spectroscopy (RXES) experiments}\label{secRIXS}

RIXS experiments consist in measuring the energy dependence of the emission line after creating a core hole by photons tuned around an atomic absorption edge \cite{kotani01,ghiringhelli04,braicovich10,rueff10}. One of the main applications of RIXS is measuring the energy dependence of electronic and magnetic excitations and their tensorial character
\cite{veenendaal96,borgatti04,ament11}, although this has not yet been applied to actinide materials.

In a RIXS experiment, the scattering cross-section is measured by scanning the incident and emitted photon energy plane in a region around the selected absorption edge and emission transition. In the case of actinides, the combinations used in the majority of the experiments performed so-far are the $L_3$-$L \alpha_{1} (L\beta_{5}$) and the $M_{4,5}$-$M(\alpha,\beta)$ ones. The former, spanning the incident energy range $\sim$17-19 keV and an energy transfer range $\sim$3-4 keV, implies an absorption transition 2$p^{6}$6$d^{n}$ $\rightarrow$ 2$p^{5}$6$d^{n+1}$ followed by the emission transition 3$d^{10}$(5$d^{10}$)6$d^{n+1}$ $\rightarrow$ 2$p^{6}$3$d^{9}$(5$d^{9}$)6$d^{n+1}$. The latter, in the incident energy range $\sim$3.5-4.3 keV and energy transfer range $\sim$0.3-0.4 keV, probes the 3$d^{10}$5$f^{n} \rightarrow$ 3$d^{9}$5$f^{n+1}$ absorption followed by a 4$f^{14}$5$f^{n+1}$ $\rightarrow$ 3$d^{10}$4$f^{13}$5$f^{n+1}$ emission. 

As written in previous sections, incident energy scans at the maximum of the emission line are called HERFD-XANES spectra. On the other hand, measurements at fixed incident energy and varying final energy are referred to as `resonant x-ray emission spectroscopy' (RXES) \cite{rueff10,vitova10,walshe14,kvashnina14}. Compared to conventional XANES, the HERFD-XANES offers the advantage of an increased resolution in the absorption process, since the final transitions are from intermediate states with larger intrinsic core-hole interaction lifetime. At the U $L$ edge, e.g., the energy resolution is improved from $\sim$8 to $\sim$4 eV, whilst energy resolutions of 0.4-0.3 eV can be obtained at the U $M_{4,5}$ edges \cite{kvashnina14}.

At the $L_3$ edge and the 
$L\alpha_1$ emission line, HERFD-XANES and RXES (see, e.g., Fig.~\ref{kvashnina13} and discussion there) have been used to study the 
multiconfigurational nature of 5$f$ orbitals in several thorium, uranium, and plutonium intermetallics and oxides  \cite{rueff07,booth14,boehler14,bao18,honda20,kawamura20}, demonstrating that the edge position can be related to the density of states at the Fermi level and that the contributions from different 5$f$ electron configurations can be estimated from an analysis of the spectral features \cite{booth12,tobin15,booth16}.

As both incident and scattered photon energies are high in a RXES measurement at the $L_3$ edge of actinides, this technique has been applied to study the evolution of the 5$f$ states upon compression. Being sensitive to final-state energy distributions in presence of a core-hole, RXES is a good probe of mixed-valency \cite{rueff10}. Experiments have been performed on localized systems (UPd$_3$) and heavy fermion (UPd$_2$Al$_3$) compounds, showing for the first a stable U$^{4+}$ valence state, for the second a progressive transformation upon compression from a mixed valent U$^{4-\delta}$ state to a U$^{4+\delta}$ configuration \cite{rueff07}. This kind of measurements in diamond anvil pressure cells requires a very small amount of material, usually $\mathcal{O}$(1 $\mu$g). This is particularly welcome when working on transuranium materials because it eases the management of the safety risk. For instance, an experiment at the $L_3$ edge of elemental americium (18.518 keV) was performed using a wide-aperture, membrane-type diamond-anvil cell with a $\sim$5 $\mu$m thick foil of $^{243}$Am metal (with a mass of
$\sim$ 600 ng) loaded into the hole (130 $\mu$m diameter) of an inconel gasket \cite{heathman10}. The energy dependence of the $L\alpha_1$ emission line (14.625 keV) was measured for pressure ranging from 4 GPa (double-hexagonal, closed-packed Am phase I) to 23 GPa, into the orthorhombic Am IV phase (see Fig. \ref{pdepAmCmU}).  5$f$ electrons are supposed to be localized in Am-I and itinerant in Am-IV \cite{heathman00,griveau05}. Many-body electronic structure calculations, based on local-density approximation and dynamical mean field theory (LDA+DMFT), suggested that the localization-delocalization edge is approached by a mixing the non-magnetic 5$f^{6}$ Am ground state with the magnetic 5$f^{7}$ configuration, combined with hybridization with the 6$d$ valence band \cite{savrasov06}. The experiment, however, does not support such a scenario as no evidence of a mixed valence character emerges at high pressure \cite{heathman10}. The experimental results seem rather in agreement with the predictions of a variant of the LDA+DMFT method that takes into account not only the correlations among the 5$f$ electrons, but also the feedback of these correlations on the rest of the system by means of an appropriate adjustment of the electronic charge density \cite{kolorenc12}.

RXES measurements at the $M_{4,5}$ edges have the advantage over measurements at the $L$ edge of interrogating directly 5$f$ states. The lower energies involved (few keV), however, imply a reduced photon penetration depth (few hundreds of nm, see Table \ref{attlengths}) and, therefore, attention must be paid to avoid any surface oxidation of the measured sample.  Such experiments have been performed
to study coordination complexes \cite{vitova10,bes16}, uranium intermetallic compounds \cite{kvashnina17,gumeniuk15,gumeniuk18}, and
to elucidate the gradual conversion of the U oxidation state in mixed uranium oxides \cite{kvashnina13}.

\begin{figure}
\centerline{\includegraphics[width=1.0\columnwidth]{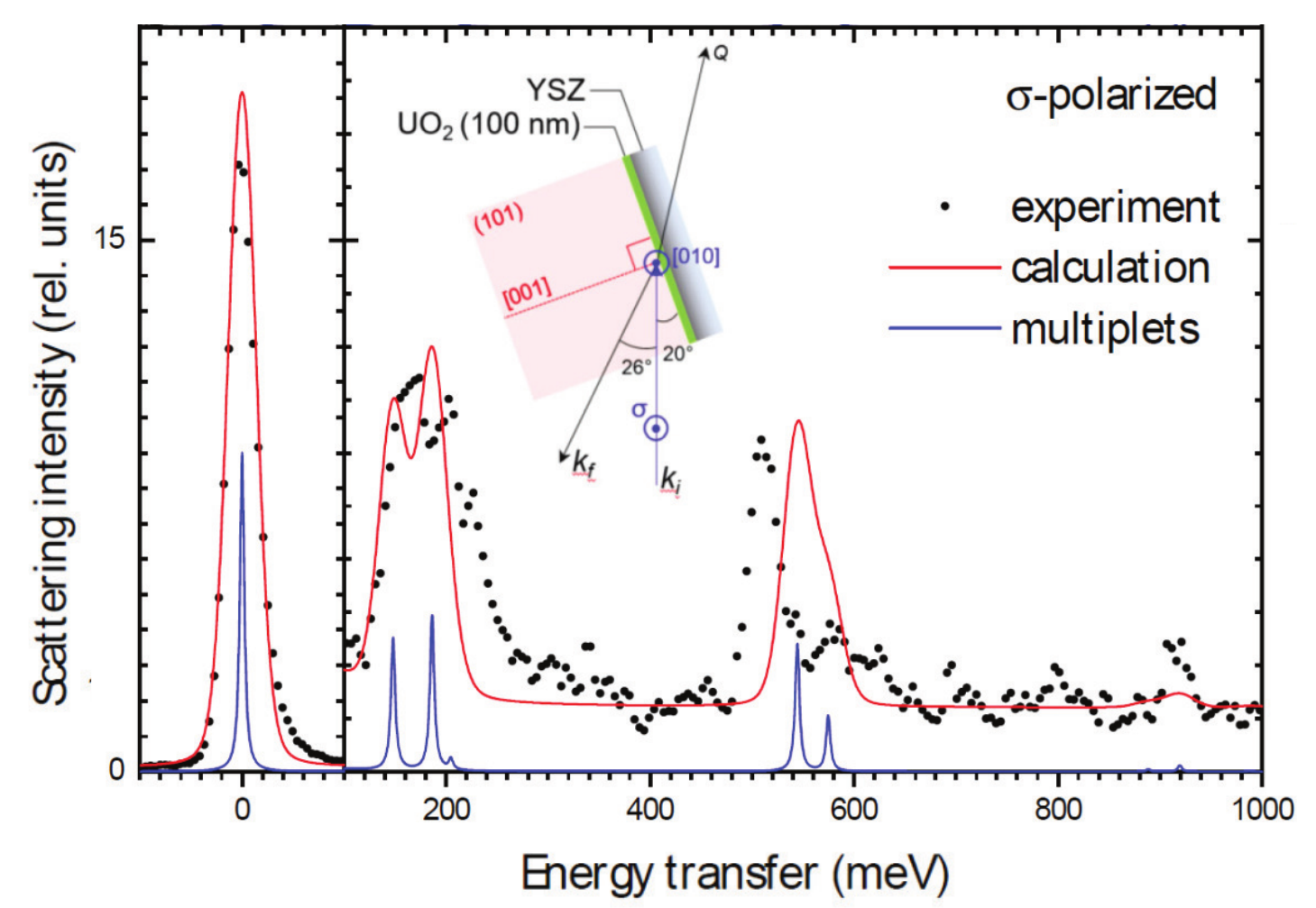}}
\caption{ 
U $N_4$-edge RIXS spectra of UO$_2$ at $T$ = 15 K. Data were taken with photon energy of 778 eV for $\sigma$ linear polarization, as shown in the inset. The red lines show the RIXS spectrum calculated for the U$^{4+}$ 5$f^{2}$ configuration and crystal-field parameters consistent with inelastic neutron scattering. The blue lines show the underlying multiplet peaks (without line broadening) of the crystal-field excitations around 180 meV and $^{3}F_{2}$ multiplet around 550 meV.
Adapted from \cite{lander21}.
 \label{softRIXSUO2}}
\end{figure}

In \cite{kvashnina17}, the authors report full core-to-core RXES maps for UPd$_3$, USb, USn$_3$, and URu$_2$Si$_2$. The experiment was performed at the ID26 beamline of ESRF \cite{glatzel21} by scanning the incident energy across the U $M_{4,5}$ edges at different emission energies around the $M\alpha$ and $M\beta$ emission lines. For UPd$_3$, a localized system with 5$f^{2}$ configuration, incident energy scans at the maximum of the emission lines show a shift of the white line by $\sim$0.2 eV compared to UO$_2$ (also a localized system with  5$f^{2}$ electron configuration but with an empty conduction band), and that the white line shift for 5$f^{2}$ (UPd$_3$) and 5$f^{3}$ (USn$_3$) in intermetallic compounds is smaller than in ionic compounds. For URu$_2$Si$_2$, where an itinerant character of the 5$f$ electrons is expected, the spectral features observed at the $M$ edges suggest almost tetravalent U atoms, with a 5$f^{2}$ configuration. This is in contrast with the results of high-energy-resolution core-level and valence-band photoelectron spectroscopy studies \cite{fujimorishin12} and with the analysis of $L_3$ RXES data \cite{booth16} that give a value $n_f$ close to 3 for the 5$f$ occupation number, but in agreement with the conclusions of polarized neutron scattering \cite{santini00,ressouche12}, NIXS experiments \cite{sundermann16}, and high-resolution RIXS measurements at the U $O$ edges \cite{wray15}. The discrepancy with the $L_3$ RXES conclusions about $n_f$ is probably due to 5$f$-6$d$ hybridization effects that are more relevant at the $L_3$ edge, where the 5$f$ shell is interrogated only indirectly. The degree of 5$f$ electron localization in 
URu$_2$Si$_2$ is examined by \cite{jeffries10} using a spin-orbit sum rule analysis of EELS spectra at the U $N_{4,5}$ edge.

As mentioned above, experiments at the $M$ edges require clean surfaces, because of the small penetration depth of few hundreds nm, see Table \ref{attlengths}. This requirement is much more stringent at the $N_{4,5}$ edges (4$d \rightarrow $ 5$f$), where the penetration depth is of a few nm only. The reward for the efforts required is an energy resolution that can be as good as 30-35 meV, in an energy range up to 1 eV.

Figure \ref{softRIXSUO2} shows the  RIXS spectrum at the uranium $N_4$ absorption edge (778 eV), measured  at 15 K on  beamline   I21 of the Diamond Light Source  \cite{zhou22} on atomically flat, epitaxial UO$_2$ films ($\sim$100 nm thickness) \cite{lander21}. With a resolution of 35 meV, the crystal field excitations 
within the $^{3}H_{4}$ ground state multiplet are clearly resolved between $\sim$140 and 180 meV, confirming earlier inelastic neutron scattering (INS) studies \cite{amoretti89}. INS measurements on UO$_2$ failed to detect non-dipolar, higher-energy intermultiplet excitations. RIXS, instead, shows excitations at 520-580 meV due to transitions towards components of the $^{3}F_{2}$ excited multiplet. Measurements with $\pi$-polarization at the $N_5$ edge (not shown in Fig. \ref{softRIXSUO2}) also establish a peak at
$\sim$ 920 meV due to a transition to the $^{3}H_{5}$ multiplet.
RIXS experiments around the $O$-edges (5$d \rightarrow$ 5$f$) have been attempted to determine the oxidation state of curium in oxide forms \cite{kvashnina07}.  At this edge, however, it is very complicated to get access to bulk properties by RIXS measurements due to the small penetration length (see Table \ref{attlengths}).

These problems are also illustrated in the work  on URu$_2$Si$_2$ using the $O$ resonance by \cite{wray15}. They do suggest the data are only compatible with the uranium having a 5$f^{2}$-like ground state (in agreement with the NIXS work of \cite{sundermann16}), but they propose that the wave functions have a dominant $J_z$ =
 $| \pm$$3 \rangle$ component rather than  $| \pm$$4 \rangle$ as proposed in the NIXS work \cite{sundermann16}. Because of the very low penetration of the  $\sim$100 eV incident photon beams at this resonance energy, bulk properties may not be sampled, and, in addition, the modeling is complicated due to extreme shallowness of the created core-hole interaction.
 
 RIXS experiments clearly have a bright future, especially at the $M$ and $N$ edges. New work at the $M$ edges (at the DESY synchrotron, Hamburg) has shown signs of multiplets in a number of uranium systems with the present resolution of $\sim$150 meV, while new experiments at Diamond (UK), with a resolution of $\sim$100 meV at the uranium $N$ edges, have seen a signal in the 5$f^{1}$ U system U$_{3}$O$_{8}$, and almost no measurable signal in UN, so this field is presently very active. It is limited by the number of instruments available and their intensity and resolution.

\section{NIXS experiments}\label{secNIXS}

NIXS at the $O_{4,5}$ (5$d_{3/2,5/2}\rightarrow$ 5$f$) absorption edges, is a bulk-sensitive technique exploiting multipole transitions  from core 5$d$ to valence 5$f$ states. For small values of the scattering vector, $\mathbf{Q}$, the NIXS spectra are dominated by the dipole-allowed transitions encapsulated within the giant Fano resonance. For high $\mathbf{Q}$ values, the intensity of the dipole transitions becomes negligible and the spectral response is dominated by dipole-forbidden transitions appearing as well-resolved multiplet structure in the pre-edge region \cite{macrander96,gurtubay05,larson07,haverkort07,veenendaal08,vanderlaan12}. These features provide information on states with symmetries different than those interrogated by electric-dipole  transitions in XAS, and are sensitive to atomic-environment, valence, and hybridization effects \cite{gordon08}. 

Compared to most other x-ray spectroscopy techniques, NIXS is intrinsically intensity-limited. On the other hand,  the absence of an intermediate state make easy to model the experimental data in a quantitative way. Another advantage of NIXS is that shallow energy edges are probed by hard x-rays. Bulk-sensitive information is, therefore, obtained. Moreover, the high penetration depth of hard x-rays ($\sim$5 $\mu$m in UO$_2$ for 10 keV incident photons) makes feasible the study of  encapsulated samples, which is a prerequisite in the case of transuranium materials.

Far away from resonance conditions, the radiation-matter interaction is dominated by a term proportional to the square of the vector potential $\textbf{A}$.  Within the first Born approximation, the NIXS double differential cross section is obtained by expanding the transition operator $\exp(\mathrm{i} \mathbf{Q} \cdot \textbf{r})$ in terms of spherical harmonics, and is given by Eq.~(\ref{eq:DDCS}).




The potential of NIXS for characterizing the dynamical electron-density response in actinide materials has been demonstrated in a few studies of oxides and intermetallic compounds at the $O_{4,5}$ absorption edges \cite{caciuffo10,bradley10,sengupta11,sundermann16,sundermann18,sundermann20}. Attempts to carry out NIXS experiments at the $N$ edges of actinides have been frustrated by the weakness of the signal.

\begin{figure}
\centerline{\includegraphics[width=1.0\columnwidth]{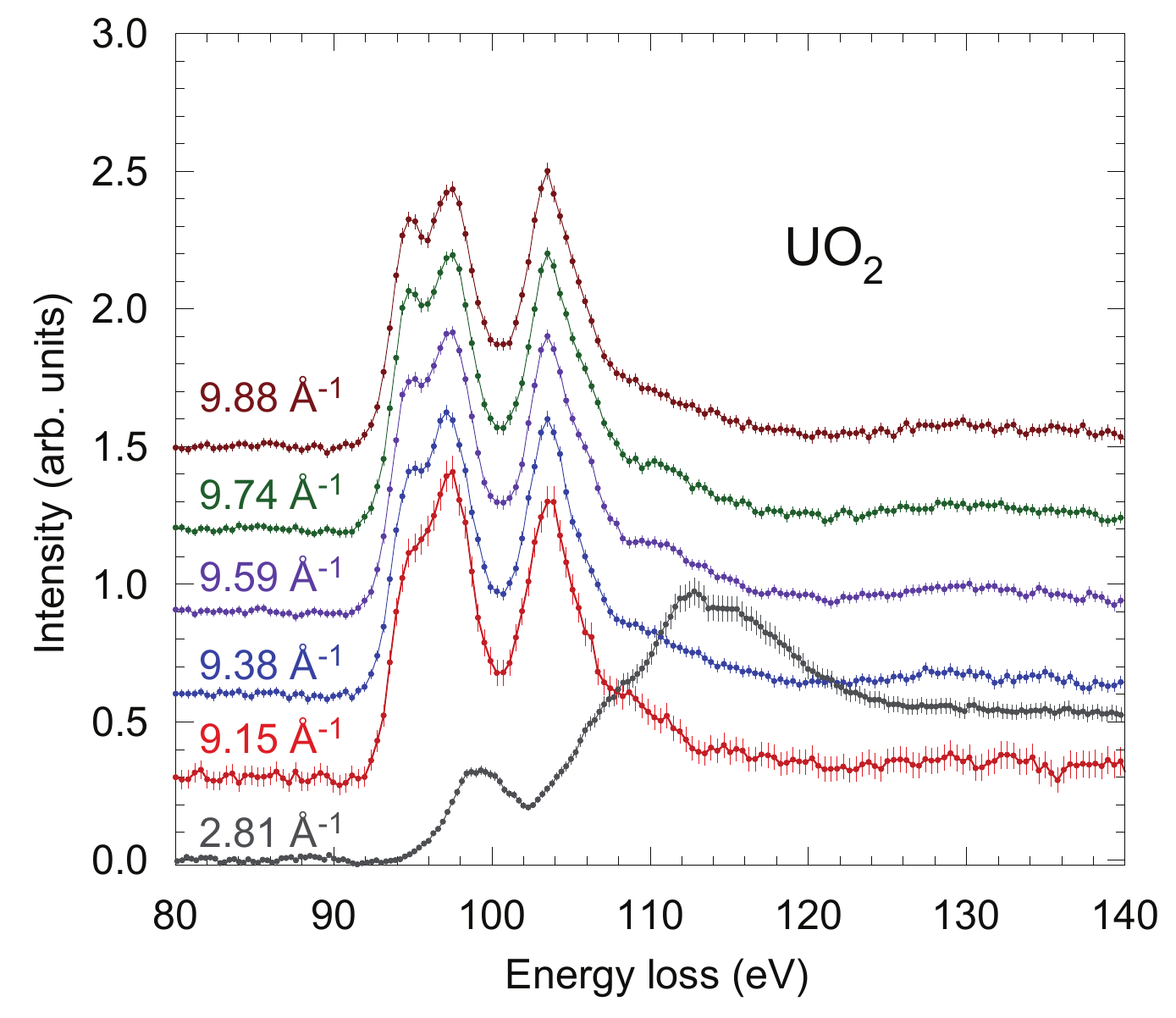}}
\caption{ NIXS spectra  for UO$_2$ at the uranium $O_{4,5}$ edges measured with a fixed final energy of 9.689 keV, at different values of the scattering vector $Q$. Data at different $Q$ values are normalized to the peak intensity of the feature centered at about 104 eV. Adapted from \cite{caciuffo10}.
\label{NIXSUO2}}
\end{figure}

Figure~\ref{NIXSUO2} shows, as an example, the NIXS room-temperature spectra obtained for UO$_{2}$ in an energy loss ($\hbar\omega  =  E_{g}  - E_{f}$) range encompassing the uranium $O_{4,5}$  absorption edges \cite{caciuffo10}. Data were collected at the ID16 inverse-geometry, multiple-analyzer-crystal spectrometer \cite{verbeni09} of ESRF, with a resolution of
$\sim$1.3 eV at a final photon energy $E_f$ of 9.689 keV. The sample was a UO$_{2}$ single crystal with the exposed external surface perpendicular to $\left [111 \right]$.

At $Q$ = 2.81~\AA$^{-1}$, the response is dominated by the dipole transition and shows the broad, asymmetric Fano profile due to resonant decay into continuum states, as observed by XAS measurements \cite{kalkowski87}.
Above $Q \sim$ 9~\AA$^{-1}$, the higher multipole transitions appear at lower energies, with three resolved features at 94.9, 97.3, and 103.6~eV at $Q$ = 9.88~\AA$^{-1}$, providing a fingerprint for the uranium ground state (the radial and angular dependence of the NIXS cross section for the $O_{4.5}$ edge of a U$^{4+}$\,5$f^2$ configuration is discussed in \cite{sundermann16,caciuffo10}).

An excellent agreement was found between the experimental data shown in Fig.~\ref{NIXSUO2} and many-electron atomic spectral calculations in intermediate coupling that allows one to identify the origin of the observed multipole transitions \cite{caciuffo10,vanderlaan12}. Similar experiments were also extended to transuranium materials, investigating 5$f^{3}$, 5$f^{4}$, and 5$f^{5}$ configurations in NpO$_2$, PuO$_{2}$, and $\beta$-Pu$_{2}$O$_{3}$, respectively \cite{sundermann20}.

The sum rule for the branching ratio of the electric multipole transitions from a core hole to a spin-orbit split valence state probed by NIXS has been derived by \cite{vanderlaan12b}, and is given in Eq.~(\ref{eq:BRk}). It shows that the effect of the valence spin-orbit interaction on the branching ratio strongly depends on $k$, the rank of the transition. These effects are very large at the end of the lanthanide series, as in the case of the Er 4$f^{11}$ $M_{4,5}$ spectra, but are observable also for light actinides, as for the U 5$f^{2}$ $O_{4,5}$ transition, providing additional information about the electronic structure of the investigated material \cite{vanderlaan12b,vanderlaan12}.

\begin{figure}
\centerline{\includegraphics[width=1.0\columnwidth]{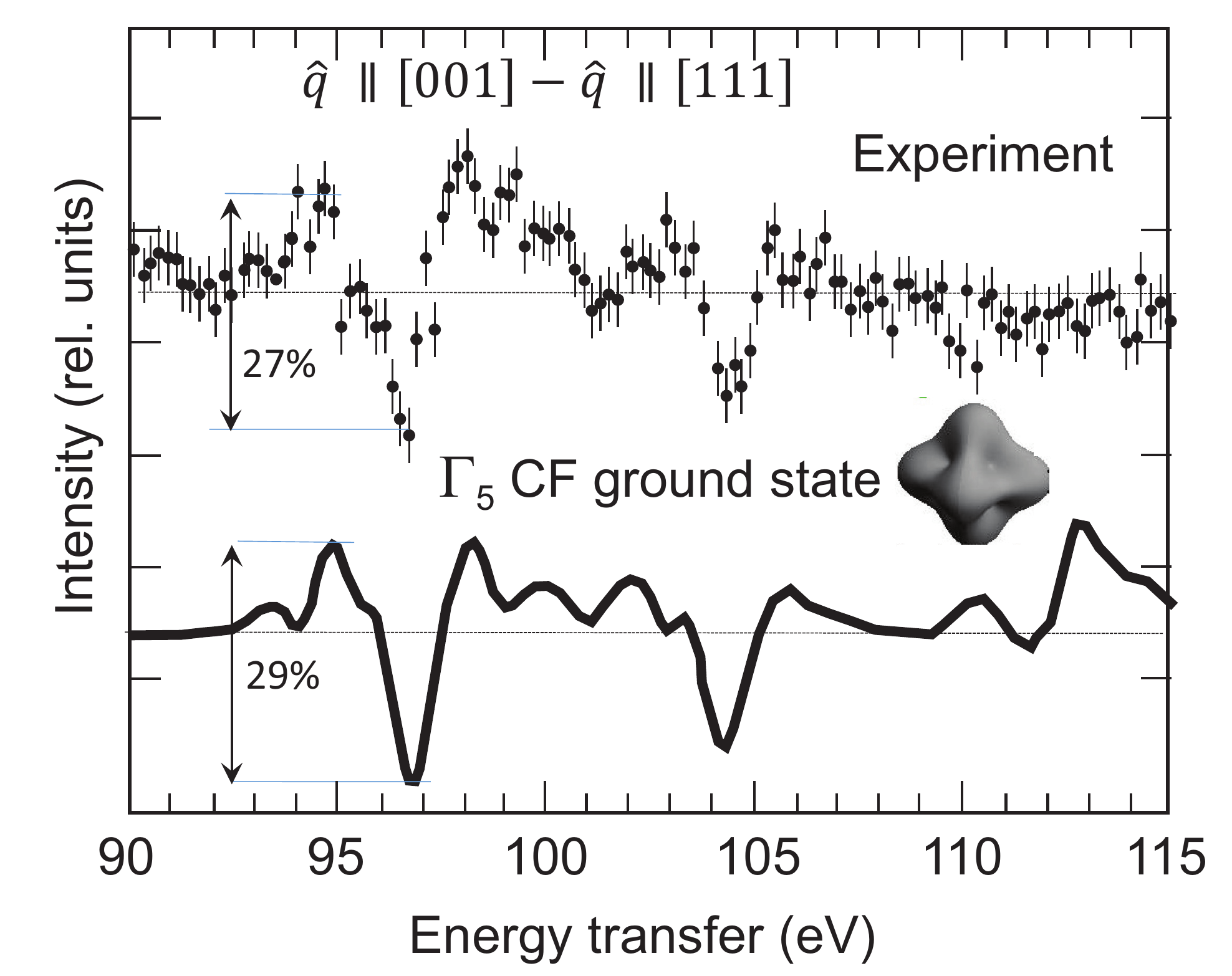}}
\caption{Difference between the experimental NIXS spectra measured at the uranium $O_{4,5}$ edges in single-crystal UO$_2$ with momentum transfer direction $\bm{\mathbf{Q}} = \mathbf{Q}/Q$ along
$\left[001\right]$ and $\left[111\right]$. Calculations assuming a $\Gamma_5$ triplet crystal field ground state and a crystal field potential strength compatible with inelastic neutron scattering experiments are shown by the solid line. 
Adapted from \cite{sundermann18}. 
 \label{NIXSdichroism}}
\end{figure}

Contrary to the dipole transition, which in cubic systems cannot exhibit any directional anisotropy, the multipole transitions in the NIXS process  branch to representations with distinct angular dependence.  This fact introduces an anisotropy of the NIXS signal, with measurable differences of the spectral intensity along different directions of the momentum transfer. Such a directional dichroism can be exploited for gaining information about symmetry and strength of the crystal field acting on the scattering atoms \cite{willers12}. The potential of the method has been demonstrated for the tetragonal  URu$_{2}$Si$_{2}$ intermetallic \cite{sundermann16} and for cubic UO$_2$ \cite{sundermann18}, for which the crystal field is firmly established by inelastic neutron scattering experiments \cite{amoretti89,magnani05}. Fig. \ref{NIXSdichroism} shows the comparison between experimental data and simulations in the case of UO$_2$.

The ability to detect these anisotropies in NIXS experiments depending on the direction of the scattering vector have been important in studies of the UM$_2$Si$_2$ series of compounds, which includes the heavy-fermion superconductor URu$_2$Si$_2$, for which
neither the electron count (i.e., either $f^{2}$ or $f^{3}$) nor the associated ground-state crystal-field wave functions have been firmly established prior to this NIXS investigation \cite{sundermann16}.

\begin{figure}
\includegraphics[width=1.0\columnwidth]{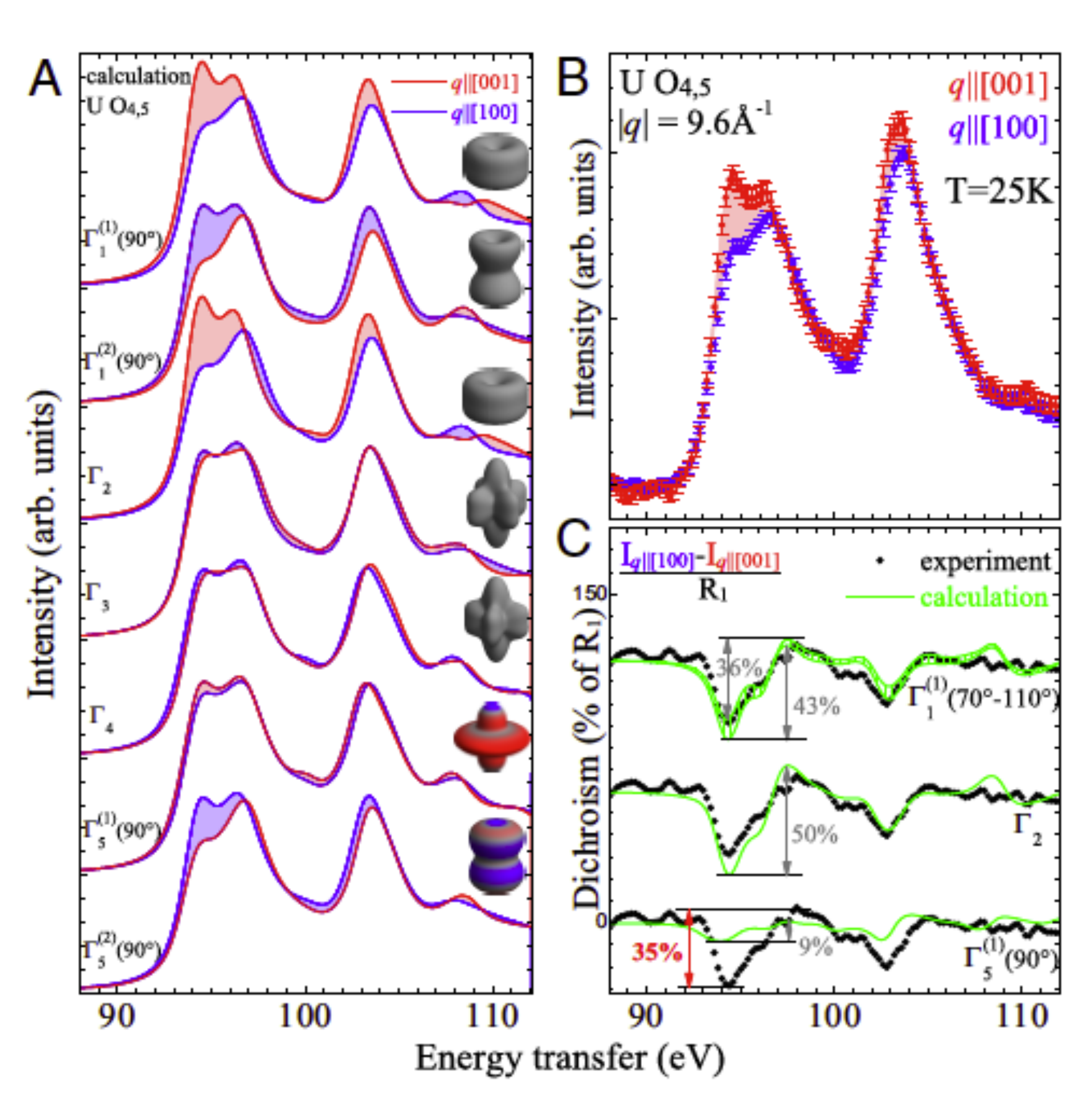}
\caption{(A–C) NIXS measurements of the U $O_{4,5}$-edge of URu$_2$Si$_2$ for $Q$ = 9.6 \AA$^{-1}$ and corresponding calculations for 5$d^{10}$5$f^{2}$ $\rightarrow$ 5$d^{9}$5$f^{3}$. (A) Simulation of S($\mathbf{Q}$,
$\omega$) of U crystal-field states for $J$ = 4 in D$_{4h}$ symmetry for the two directions $\mathbf{Q} \parallel [001]$ (red) and $\mathbf{Q} \parallel [100]$ (blue).
Insets show the corresponding electron densities. (B) NIXS data for momentum transfers
$\mathbf{Q} \parallel  [100]$ (blue) and $\mathbf{Q} \parallel  [001]$ (red) at $T$ = 25 K.  (C) Dichroism at 25 K in percentage defined as difference
$I( \mathbf{Q} \parallel [100]) - I( \mathbf{Q} \parallel [001])$ 
relative to peak height R$_{1}$ as defined in the isotropic spectrum; data (black dots) and calculations (green lines) for the crystal-field states with the correct sign of the directional dichroism. Here the data points have been convoluted with a Gaussian of 0.5 eV FWHM.
Taken from \cite{sundermann16}. \label{sundermann}}
\end{figure}

The crystal field in the case of the lower (tetragonal) symmetry of URu$_2$Si$_2$ \cite{sundermann16},  where the $J$ = 4 multiplet splits into five singlets and two doublets, is more complicated as compared to UO$_2$ \cite{sundermann18}. The full wave functions are given in  \cite{sundermann16}. An examination of the calculated anisotropies in Fig.~\ref{sundermann}(A) shows that only the $\Gamma_1$ and $\Gamma_2$ states agree with the experimental data in Fig.~\ref{sundermann}(B). These differences in the directional signal are shown explicitly in Fig.~\ref{sundermann}(C), where a more quantitative evaluation of the ground-state wave function is given. Crucially, the wave function must contain a large proportion of the $J_z$ = $| \pm$$4 \rangle$ state. The ground states of a 5$f^{3}$ configuration cannot explain the experimental results. This is the first time that the ground state wave function was identified in URu$_2$Si$_2$, so it is an important achievement. Many theories have been proposed to explain the properties of this material, see \cite{mydosh11,mydosh20} for reviews, and there has been confusion as to whether to start any theory from a 5$f^{2}$ or 5$f^{3}$ configuration. Neutron inelastic scattering has not been able to observe any well-defined crystal-field levels, certainly because of the mixing of the 5$f$ states with the conduction states \cite{mydosh11,mydosh20}. Interestingly, a polarized-neutron study \cite{ressouche12} of the magnetization induced with a magnetic field found the results consistent with only a 5$f^{2}$ configuration,
also with the $\Gamma_1$ or $\Gamma_2$ states the likely ground states. The NIXS experiment \cite{sundermann16} found no temperature dependence of the signal, so these results are not able to make any statements about the apparent hidden order that appears at 17 K. This is a rare example of an intermetallic material exhibiting atomic-like multiplets, although the $5f$ states are mostly itinerant.

A second important NIXS paper was published by \cite{amorese20} reporting an investigation of a series of single crystals of UM$_2$Si$_2$ compounds, with M = Fe, Ni, Ru, and Pd. For the NIXS experiments the authors found that all four compounds had very similar anisotropies as discussed previously for URu$_2$Si$_2$ \cite{sundermann16}, showing that as a basis for discussing these materials one needs to start from a 5$f^{2}$ multiplet structure. In addition to NIXS, they also examined the 4$f$ core states of these same materials using a high-energy (5.945 keV) photon beam with an overall instrumental resolution of 0.3 eV. These core levels are between 370 and 400 eV and their position (and width) depend on the valence state. UPd$_3$ was used for the nominally localized 5$f^{2}$ state, and UCd$_{11}$ for the 5$f^{3}$ state. The authors argue that the transition metal $d$-states play an important role as they mix with the U 5$f_{5/2}$ band strongly in the Fe compound, and such mixing becomes smaller for Ru and Ni, and is almost negligible for Pd. Finally, the authors show how the properties can be placed on a Doniach-like phase diagram \cite{amorese20}. Both experiments were performed at the PETRA beamlines in Hamburg, showing that this third generation machine is now also performing experiments on actinides, at least with uranium.

\section{High-resolution IXS experiments}\label{secIXS}

High-resolution inelastic x-ray scattering (IXS) at third-generation SR sources is a well-established technique for mapping phonon branches with meV energy resolution. High-performance spectrometers using crystals in near-backscattering geometry and efficient focusing optics \cite{burkel00} allow one to measure phonon dispersion curves in crystals with volumes as small as 10$^{-4}$ mm$^{3}$ and in epitaxial thin films less than 500 nm thick \cite{dastuto02,rennie18}. Compared with neutrons, this is a crucial advantage for studying radioactive materials or, in general, systems for which large single crystals are not available, or high pressure limits the crystal size. The intrinsic background in IXS experiments is very low, the energy resolution is decoupled from energy transfer, and the momentum transfer is energy-independent. A drawback of the technique is that the scattering cross-section is proportional to the square of the atomic number, making it challenging to observe the contributions of light atoms to the vibrational spectra. 

The ID28 beamline at ESRF is an example of high-performance IXS spectrometer. In the incident energy range of interest for actinide systems ($\sim$17-24 keV) a flat Si perfect crystal monochromator in back-scattering geometry, temperature controlled in the mK region, affords an energy resolution of about 1.5-3 meV when the analyzer is thermally stabilized to 6$\times$10$^{-4}$ K. Properly oriented single crystal diamond slabs provide ideal windows if sample encapsulation is mandated by safety reasons.

The IXS cross-section for single- phonon scattering is proportional to the scattering function $S(\mathbf{Q}, \omega)$ \cite{krisch07}
\begin{equation}
\begin{aligned}
S(\mathbf{Q}, \omega) & = { }  \sum_{d, \mathbf{q},j} \left | \frac{f_{d}(Q)e^{i\mathbf{Q} \cdot \mathbf{r}_{d}}e^{-W_{d}}(\bm{\varepsilon}_{d}^{\,\mathbf{q},j} \cdot \mathbf{Q})}{\sqrt{M_{d}}}\right |^{2} \\
& \times \frac{\delta (\hbar\omega \pm E_{j}(\mathbf{q}))}{E_{j}(\mathbf{q})} \left \langle n(E_{j}, T)+
\frac{1}{2}\pm\frac{1}{2} \right \rangle ,
\end{aligned}
\label{phononsCS}
\end{equation}
\noindent 
where $\mathbf{Q}=\mathbf{G}  +  \mathbf{q}$ with   reciprocal lattice vector $\mathbf{G}$  and   phonon quasi-momentum $\mathbf{q}$. $f_{d}$ is the atomic form factor of the $d$-th atom located at position
$\mathbf{r}_{d}$ of the unit cell, with Debye-Waller factor $W_{d}$. Phonons with polarization vector
$\mathbf{\bm{\varepsilon}}_{d}^{\,\mathbf{q},j}$
and polarization index $j$ (taking values 1 to 3$n$, where $n$ is the number of atoms per unit cell)
have energy $E_j(\mathbf{q})$. The plus (minus) sign in the Bose factor  $\left \langle n(E_{j}, T) + \frac{1}{2}\pm\frac{1}{2} \right \rangle$ at temperature $T$ indicates phonon creation (annihilation). The sum in Eq.~(\ref{phononsCS}) is over the $n$ values of $d$, the values of $\mathbf{q}$ in the first Brillouin zone (given by the number of unit cells in the crystal), and over the 3$n$ values of $j$.
Note that here the scattering factor is necessary as all the electrons contribute and they have their spatial distribution. This is not the case in resonant scattering experiments.

The measurements of the dispersion curves in face-centered cubic (\textit{fcc}) $\delta$-plutonium performed on large-grain polycrystalline samples are exemplary for illustrating the potential of IXS in the study of actinide materials \cite{wong03}. The results provide a qualitative validation of the predictions of DMFT calculations \cite{dai03}, showing that this theoretical approach is appropriate to describe not only the structure but also (at least qualitatively) the dynamics of strongly correlated electron systems. In particular, the experiment confirms the softening of the T$\left[111\right]$ modes and the low-shear elastic modulus $C'$, reflecting the strong coupling between the lattice structure and the 5$f$ valence instabilities. This work, which was carried out  at room temperature on the ESRF ID28 beamline, was described in a longer article \cite{wong05}.  In another attempt to determine whether the soft mode in the $\left[111\right]$ direction further softened with reduced temperature, the experimental team resorted to measuring at the APS the diffuse scattering \cite{wong04}, which contains information on the phonons, eventually concluding that there were no effects as a function of lowering the temperature \cite{wong04a}. There is still, of course, the open question about what happens above room temperature, but that raises further safety concerns.

\begin{figure}
\centerline{\includegraphics[width=1.0\columnwidth]{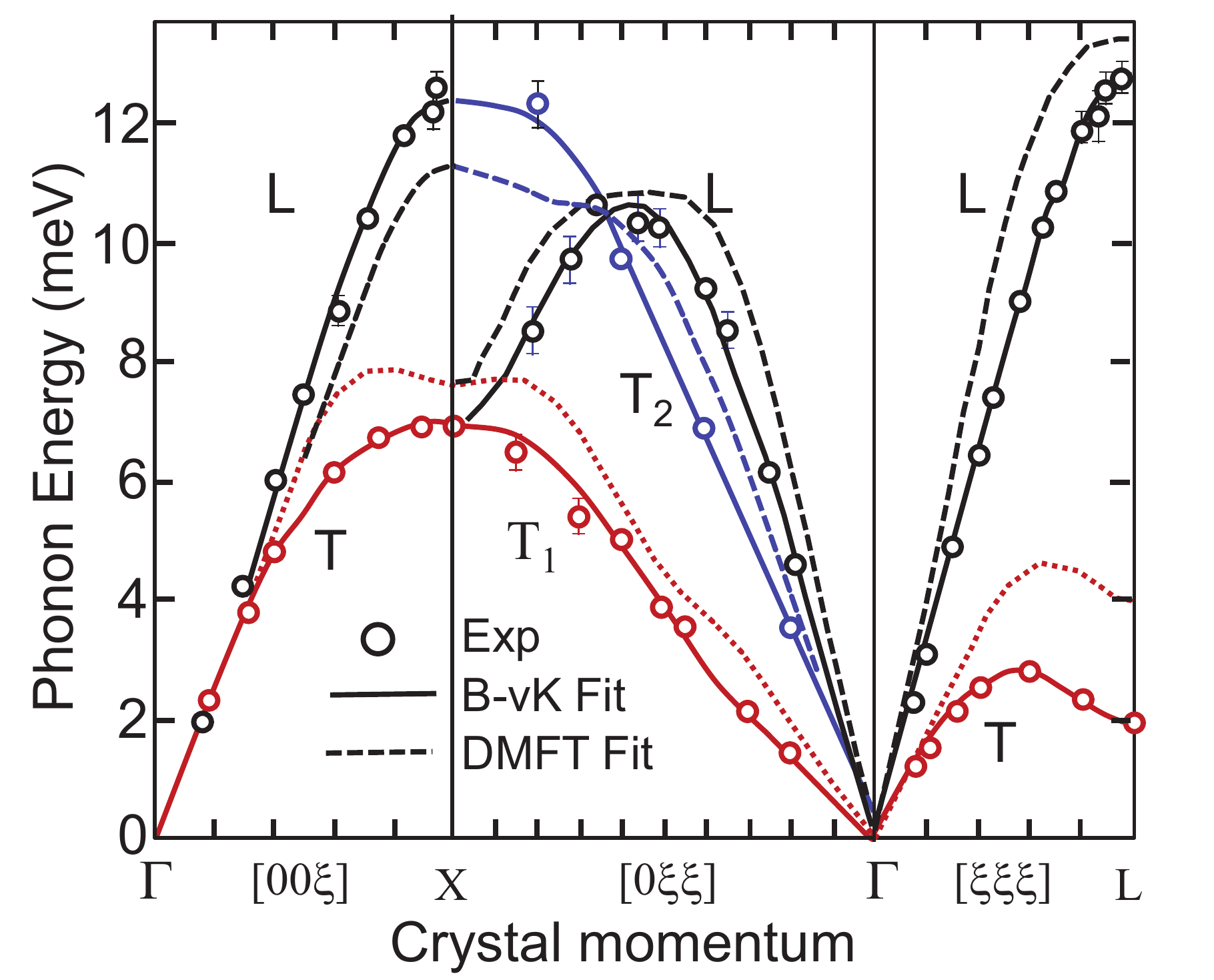}}
\caption{ Phonon dispersion curves at $T$ = 300 K along high-symmetry directions in  $\delta$-Pu stabilized by 0.6 weight \% Ga alloying (\textit{a} = 0.4621 nm). The experimental data are shown as circles (black longitudinal (L) modes; red and blue transverse (T) modes). The branches T$_{1}$ and T$_{2}$ propagating along the $\left[0 \xi \xi\right]$ direction are polarized along $\left< 011 \right>$ and $\left<100\right>$, respectively. The solid curves are the fourth-nearest neighbor Born–von Kármán (B-vK) model fit. The dashed curves are calculated dispersions for pure $\delta$-Pu based on dynamical mean field theory (DMFT) \cite{dai03}.
Adapted from \cite{wong03}.
 \label{DispDeltaPu}
}
\end{figure}

The low-temperature properties of uranium metal were a mystery since elastic constant measurements in the 1960s found a most unusual behavior. The phonons were measured \cite{crummett79} with neutron inelastic scattering in 1979, showing unusual softening along the
$\left[ 100 \right]$
direction of the orthorhombic structure of $\alpha$-uranium. Following these measurements, a charge-density wave (CDW) was found to develop at 43 K \cite{lander94}. $\alpha$-U is the only element to exhibit such a CDW at ambient pressure. However, it was not until 2008 that a theory was presented for the phonons \cite{bouchet08}, and this immediately suggested that the phonon anomaly should be suppressed by pressure --- a prediction confirmed by experiments \cite{raymond11} with ID28 at the ESRF using pressures up to 20 GPa. Later, experiments with thin films were successful in placing the
$\left[ 100 \right]$ axis of $\alpha$-U under tensile stress \cite{springell14}, caused by interaction with the substrate, and the results were a CDW developing at 120 K, much higher temperature than in the bulk. The combination of theory and experiment showed the importance of the electron-phonon couplings in the metal, and their momentum dependence. Properties such as the equation of state are affected by these considerations \cite{dewaele13}.

The phonons in $\alpha$-U were also measured at APS in 2003 when the instrument to measure dispersion relations in single crystals there was quite new \cite{manley03}. These measurements, combined with others using neutrons, led to the discovery of a new effect in $\alpha$-U at room temperature, that actually disappears above 650 K, which was described as an intrinsic localized mode (ILM) \cite{manley06,manley08}. These results fit well into the earlier reports of a large phonon softening in the phonon-density of states of $\alpha$-U, as measured with neutrons \cite{manley01}, over the same temperature range. However, the overall significance of these observations within the phase diagram for uranium is less clear, and needs further elucidation.
 
Recently, the IXS spectrometer has been pushed to new limits by measuring both diffuse scattering and the phonon dispersion curves from a thin (300 nm) epitaxial film of U-Mo alloys. These materials have been of interest for many years, and might find applications as advanced nuclear fuels, but single crystals are not available and there has been controversy over whether the structures are body-centered cubic ($bcc$) or something more complicated. Growing epitaxial films turned out to be relatively simple, and the subsequent diffuse scattering patterns \cite{chaney21} showed that the structure is essentially $bcc$ but superposed on that symmetry is correlated disorder, where the local symmetry is lower, as if the uranium atoms prefer to have neighbors reminiscent of the low symmetry found in the element at room temperature, and not the high symmetry demanded of $bcc$. The correlation length of such disorder depends on the composition, but is of the order of 5-10 nm.

The phonon dispersion curves, shown in Fig.~\ref{DispDeltaPu}, are close to those calculated by theory for this composition, except that they show large linewidths. The latter are much broader than expected for an alloy, and are due to the correlated disorder in the material. Thus, we see new effects in these materials that have been studied for many years. An earlier study of the phonons at the APS of a U-Mo alloy used very small single crystals of dimensions $<$ 200 $\mu$m cut out by laser techniques from the melt \cite{brubaker19}. They also found very wide phonons. However, the study assumed that because no extra phases were found with a standard Rietveld analysis, there is no diffuse scattering in this system. Unfortunately, the diffuse scattering, as first observed by \cite{yakel69}, will not amount to even 1\% of the strong \textit{bcc} peaks, but its presence is crucial to understanding the formation and properties of the alloys. Given the intensity available at such synchrotron beams, it may be wise to search routinely in alloy systems for possible diffuse scattering.

In a similar vein, new experiments \cite{paolasini21} at ID28 on the phonon linewidths of UO$_2$ below room temperature (this time using a single crystal rather than a thin epitaxial film) have shown that the linewidth broadening in UO$_2$ is only along the
$\left[100\right]$ direction, and not in the other directions. Such anisotropic broadening has consequences for the thermal conductivity, which for a cubic material should not be anisotropic.

These experiments were performed with IXS since with neutrons the magnetic cross-section is appreciable and one wants a technique that is sensitive to only the vibrational spectra. The resolution was increased to 1.4 meV. In an exactly similar vein, the acoustic phonons of URu$_2$Si$_2$ were examined at the APS  and anomalous widths were found for some of the vibrational modes \cite{gardner16}.

IXS was used to measure the phonon dispersion in  NpO$_{2}$ \cite{maldonado16}, for which available single crystals are far too small for inelastic neutron scattering. The results observed along the three high-symmetry directions have been used to validate first-principles density functional theory calculations, taking into account third-order anharmonicity effects in the quasiharmonic approximation.

The phonons in the heavy fermion superconductor PuCoGa$_5$ were also reported by \onlinecite{raymond06}. The experiment was performed at room temperature and the theory required a Hubbard U = 3 eV to fit the phonons, suggesting rather localized $5f$ states, which was also the conclusion from the XMCD experiments reported in Fig. \ref{XMCDPu115}.

In the case of polycrystalline or powder samples, if multiphonon effects are negligible, it is possible to acquire the phonon density of states by summing $S(\mathbf{Q}, \omega)$ in all directions for each value of momentum transfer $\mathbf{Q}$. In practice, this is obtained by summing IXS datasets measured at many different values of $\mathbf{Q}$ \cite{bosak05}. This technique was used to determine the phonon density of states (PDOS) for both $\alpha$- and $\delta$-Pu, using the instruments at the APS with an incident energy of 23.8 keV \cite{manley09}. In an elemental system the PDOS may be related directly to the heat capacity, and this relationship seems to work well for the $\alpha$-Pu phase. However, this is not the case for the $\delta$-Pu. It appears that most of the entropy stabilizing $\delta$-Pu at high temperature comes from unconventional sources, including electron correlations, and possibly intrinsic phonon softening, although the latter has not been found up to room temperature.

On polycrystalline compounds, IXS has also been applied to study the PDOS of Ga-doped PuO$_2$ \cite{manley12} and of fluorine-doped and undoped NpFeAsO, an analogue of iron arsenide high-temperature superconductors \cite{walters15}. In the case of the PuO$_2$ work the PDOS was deduced from various theories and compared with experiment. Perhaps surprisingly, the theoretical results differed by a considerable amount. The best fit was found for DFT+U and for DMFT. For much more on theory for thermal conductivity in the actinide oxides, see  \cite{hurley22}.

\begin{figure}
\centerline{\includegraphics[width=1.0\columnwidth]{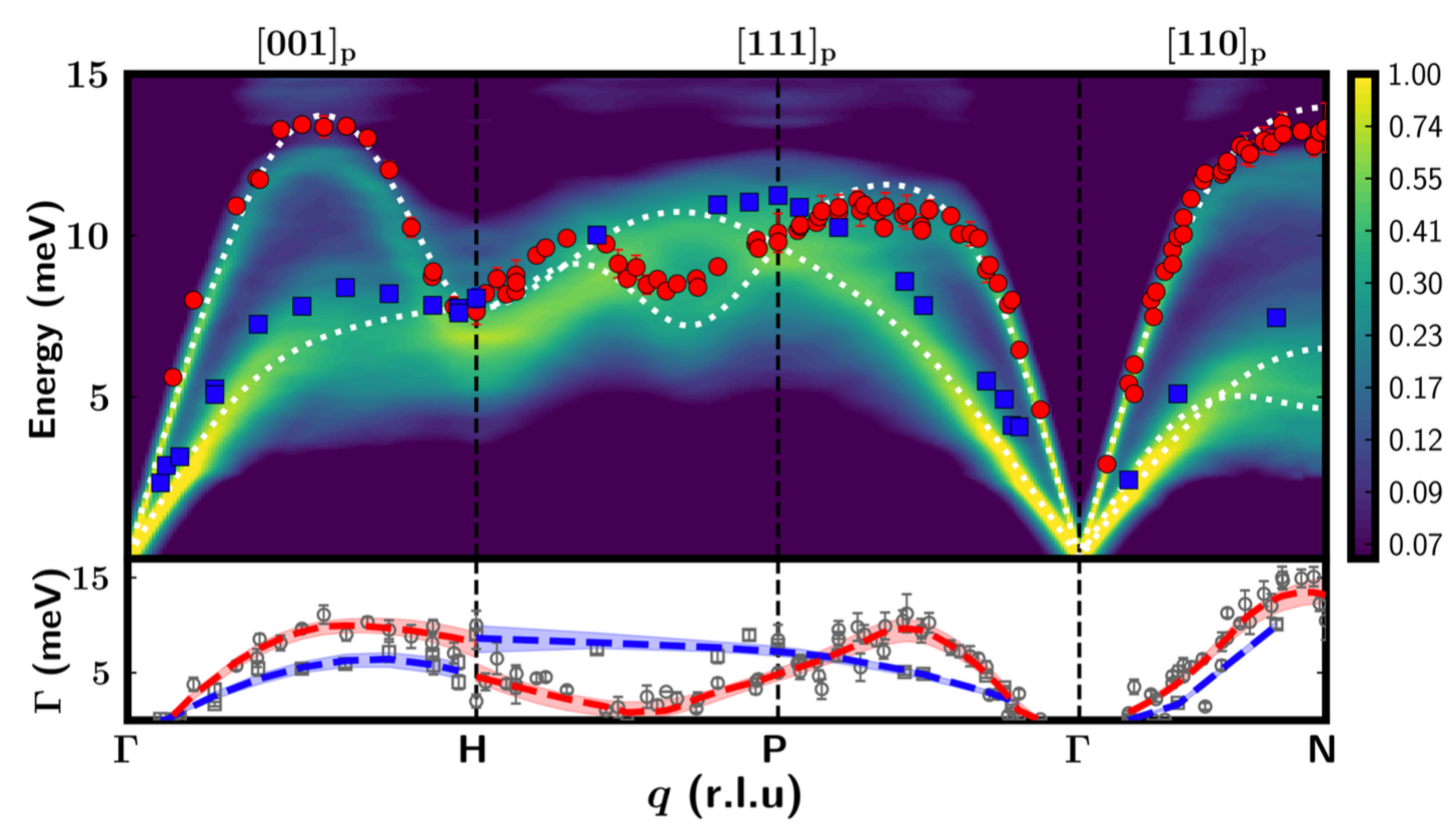}}
\caption{ Phonon energy (top panel) and linewidth (bottom panel) dispersions on epitaxial U$_{1-x}$Mo$_{x}$ thin film alloys. Transverse (longitudinal) acoustic modes are shown as blue squares (red circles) for the alloy with 23 at.\% Mo. Theoretical results from a virtual crystal approximation for an alloy with 25 at.\% Mo are shown as dashed white lines. The full spectral function is plotted as a  
log$_{0.6}$ color map to rescale the intensity divergence at gamma. All directions are within the parent Brillouin zone. The bottom panel shows raw linewidths,  $\Gamma_0$, as gray squares (TA) and circles (LA). Deconvoluted linewidths  $\Gamma_d$ are shown by dashed blue (TA) and red (LA) trend lines. 
Adapted from \cite{chaney21}.
\label{PhononsUMo}
}
\end{figure}

\section{PES and ARPES experiments}\label{secPES}

Photoemission techniques, which involve an incident photon and an emitted electron, are, of course, fundamentally different to all the experiments discussed in the present article. Photoemission spectroscopy originates in the photoelectric effect, first explained by Einstein in 1905 (and for which he won the Nobel Prize in 1921), and was further exploited in the 1960s by Siegbahn and collaborators in Sweden (Nobel Prize in 1981). An excellent review was published in \cite{reinart05}. As explained in this review, the majority of photoemission experiments are performed with specialized laboratory equipment, but clearly since the incident beam consists of photons, there are advantages to using synchrotron beams for the incident photons. Two obvious ones come to mind (a) investigations that can profit from tuning the incident photons to a known resonance of one of the elements in the sample and (b) investigations that are intensity limited, e.g., those that measure the signal as a function of the specific directions in a solid, e.g., performing angle-resolved photoemission (ARPES), which requires an intense beam to map much data. As we shall show below, both of these techniques have been used for actinide samples at synchrotrons.

Early PES work on actinides was reported by \cite{veal74} on Th, U, and their oxides, and by  \cite{naegele84} on Am. This technique has made a considerable contribution to the research on actinides, and we note a large number of studies performed on transuranium samples at both Los Alamos National Laboratory and in those of the European Commission’s Joint Research Centre in Karlsruhe, Germany. Examples are studies of Pu metal at both places \cite{arko00,gouder01,havela02}, of the superconductor PuCoGa$_5$ \cite{joyce03,eloirdi09}, of PuSb and PuTe \cite{gouder00}, of PuN \cite{havela03}, and of Am metal and compounds \cite{gouder05}. 

These are only some of the experimental results reported, and none have used synchrotron radiation. 

On the other hand, early synchrotron work \cite{tobin03} focused on using the technique of \textit{resonant} photoemission, with the incident energy tuned to the $O_{4,5}$ edges of $\sim$ 110 eV, and involved both absorption and photoemission experiments. The differences they reported between $\alpha$- and $\delta$-Pu metal are actually quite small, and there is always at such low incident photon energies a question of whether the true bulk properties are being examined. Both phases show that the states at the Fermi energy are dominated by the 5$f$ electrons \cite{vanderlaan10,gouder01}.

\subsection{Core-level photoemission}\label{clpes}

\begin{figure}
\includegraphics[width = 0.4\textwidth]{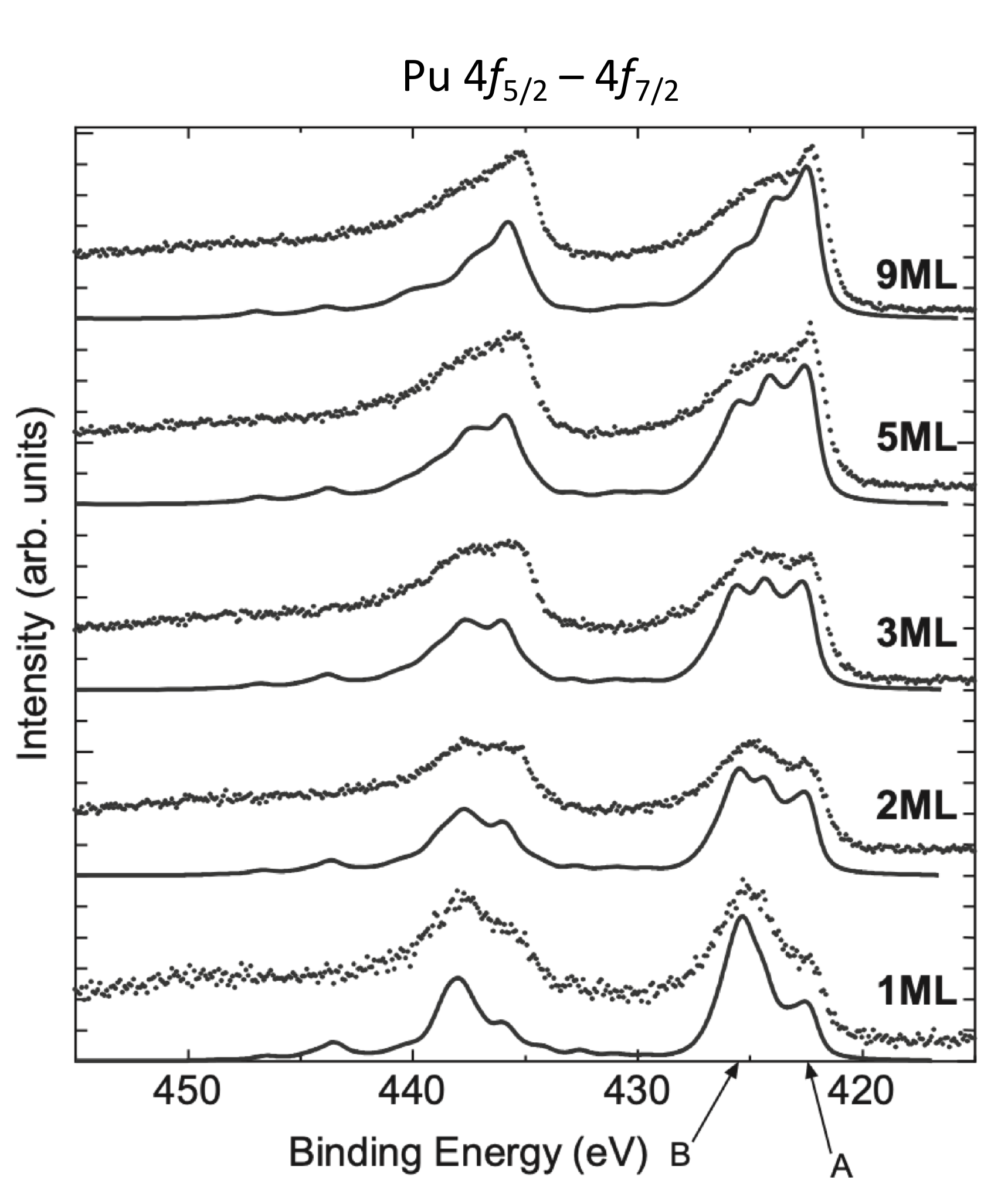}
\caption{The 4$f$ PES spectra for  1, 2, 3, 5, and 9 monolayers (ML) Pu thin films. Experimental results (dots) show a gradual increase in the main to satellite intensity ratio with increasing layer thickness (experimental data from \onlinecite{gouder01}).
The arrows A and B  indicate the energy positions of the main and satellite peaks, respectively, which are only well separated in the case of the most localized case of 1ML.
The experimental spectra are compared to Anderson impurity calculations (drawn lines)
 with hybridization parameter $V$ = 0.25, 0.35, 0.4, 0.45, and 0.55 eV for increasing number of MLs  \cite{vanderlaan10}. The spectrum for 9 ML Pu resembles that of $\delta$-Pu metal, whereas 1 ML Pu resembles Am. Reproduced from \onlinecite{vanderlaan10}.}
\label{fig:Pu4f}
 \end{figure}

The average number of $5f$ electrons making up the valence state in plutonium metal together with the
electronic fluctuations on each metal site has been a subject of debate. For the $\delta$ phase of Pu, where
compared to the $\alpha$ phase, increased localization (more atomic-like character) leads to decreased overlap and
volume increase, an $f$ count close to either 5 or 6 has been proposed depending on the type of electronic
structure calculation. To resolve the controversy,  the Pu $4f$ photoemission spectrum has been analyzed, which
displays well screened and poorly screened peaks that can be used as a measure for the degree of localization.
Detailed Anderson impurity model calculations, including the full multiplet
structure for Pu $4f$ photoemission, have been compared to experimental results obtained from 1 to 9
monolayers thin films of Pu on Mg (Fig.~\ref{fig:Pu4f}) and from Pu metal in the $\alpha$ and $\delta$ phases \cite{vanderlaan10}. The trend in the satellite to main
peak intensity ratio as a function of the Pu layer thickness gives a clear indication that Pu metal has a mainly $5f^5$-like
ground state. For the Pu allotropes and thicker films an $f$ count of 5.22 is obtained with a Coulomb interaction
$U$ = 4 eV. 
The calculated results for the $f$ counts are given in Table \ref{tab:Pu-weights}.
As seen, the standard deviation $\sigma$ corresponding to the spread over the $f$ counts strongly increases with layer thickness, which  indicates an increased delocalization due to stronger $5f$ fluctuations.
A strong increase in delocalization is also found   going from $\delta$ to $\alpha$ Pu, whereas the average $f$ count $\langle n \rangle$ is barely changing.

\begin{table}
\caption{Ground state $5f$ weights in \% for Pu thin films and $\alpha$ and $\delta$ Pu obtained from calculated PES spectra in Fig.~\ref{fig:Pu4f} with hybridization parameter $V$. Derived from this are  the average value $\langle n \rangle$ and the standard deviation $\sigma$ over the $f$ count \cite{vanderlaan10}. }
\begin{tabular}{l|r|r|r|r|r|r}
\hline \hline
  & $V$ (eV) & $5f^4$ & $5f^5$  & $5f^6$  & $ \langle n \rangle$ & $\sigma$ \\ 
  \hline
1 ML             & 0.25 & 1.3 & 81.4 & 17.3 & 5.16  & 0.16 \\
2 ML            & 0.35 & 2.6 & 74.7 & 22.7 & 5.20  & 0.21 \\
3 ML            & 0.40 & 3.4 & 72.1 & 24.5 & 5.21  & 0.23 \\
5 ML            & 0.45 & 4.2 & 69.9 & 25.9 & 5.22  & 0.25 \\
9 ML           & 0.55 & 5.7 & 66.4 & 27.8 & 5.22   & 0.29 \\
$\alpha$ Pu & 0.55 & 5.7 & 66.4 & 27.8 & 5.22   & 0.29 \\
$\delta$ Pu & 1.10 & 9.6 & 58.8 & 31.6 & 5.22   & 0.36 \\ 
\hline \hline
\end{tabular}
\label{tab:Pu-weights}
\end{table}

Further insight in the screening process of the photoexcited core electron can be gained from a simple two-level model \cite{vanderlaan81,vanderlaan10}. Consider an initial state $\psi_g$ composed of two basis states $\psi_a$  and $\psi_b$   (e.g., $f^5$ and $f^6$) with  energy difference $\Delta = E_b - E_a$ and mixed by hybridization with  matrix element  $V = \langle  \psi_a |H| \psi_b \rangle$. Introducing the mixing parameter $\theta$, defined by $\tan 2 \theta = 2V/\Delta$, the ground state can be written as 
\begin{equation}
\psi_g = \psi_a \sin \theta - \psi_b \cos \theta .
\end{equation}
After electron emission, the final-state basis functions are $\psi'_a  = \epsilon^\dagger c \psi_a$  and $\psi'_b  = \epsilon^\dagger c \psi_b$,
where $c$ is the annihilation operator of a core electron and  $ \epsilon^\dagger $ the creation operator of a continuum electron.
 The states have an energy difference $\Delta' = E'_b - E'_a$  and mixing  $V' = \langle  \psi'_a |H| \psi'_b \rangle$, with a mixing parameter  $\theta'$ defined by  $\tan 2 \theta' = 2V'/\Delta'$. This gives the ‘bonding’ and ‘antibonding’ final states as
 \begin{align}
& \psi_M = \psi'_a \sin \theta' - \psi'_b \cos \theta' . \nonumber \\
& \psi_S = \psi'_a \cos \theta' + \psi'_b \sin \theta'  ,
\label{eq:psi-MS}
\end{align}
with an  energy separation of 
\begin{equation}
E_S - E_M = \sqrt{ \Delta'^2 +4V'^2}.
\label{eq:EnergySeparation}
\end{equation} 
Substituting Eq.~(\ref{eq:psi-MS}) into the transition probability
$ I_n = | \langle f_n | \epsilon^\dagger c | g \rangle |^2 $, valid in the sudden approximation,
gives the relative intensity ratio of the satellite to main peak as
\begin{align}
\frac{I_S}{I_M} = \frac{ | \langle \psi_S | \epsilon^\dagger c | \psi_g \rangle |^2}
{ | \langle \psi_M | \epsilon^\dagger c | \psi_g \rangle |^2}
& = \left( \frac{ \sin \theta' \cos \theta - \cos \theta' \sin \theta}{ \cos \theta' \cos \theta + \sin \theta' \sin \theta} \right) ^2
\nonumber \\
&= \tan^2 (\theta'  - \theta)  ,
\end{align}
where the labels $M$ and $S$ refer to the main and satellite peak, respectively. This equation demonstrates the important fact that the satellite intensity depends only on the difference in hybridization between the initial and final state. Thus, if the PE process induces no change in the hybridization ($\theta' = \theta$), then all intensity goes into the main peak and the satellite intensity vanishes. Since PES creates a hole in the core level, there will usually be a change in the energy difference between the two basis states due to screening, so that $\Delta' \ne \Delta$  and a satellite peak will be present.

As an example, consider the ground state as a mixture of  $|5f^n \rangle$ and  $|5f^{n+1} \underline{k}\rangle$  and the final state a mixture of $|\underline{c} 5f^n \epsilon \rangle$  and  $|\underline{c} 5f^{n+1} \underline{k} \epsilon \rangle$, where $\underline{k}$ denotes a reservoir of hole states near the Fermi level, and $\underline{c}$ denotes a core hole. The underlying physical picture is one in which the $f$ electrons fluctuate among the two different atomic configurations by exchanging electrons with a reservoir. Quantum mechanically, the electrons can, for short periods of time, preserve their atomic character in a superposition of two atomic valence states with different number of $5f$ electrons, while   maintaining their metallic, delocalized hopping between neighboring sites. Correlations are strongest when the electrons are on the same atom. If the energy difference between the initial states is taken as $\Delta$, then the energy difference between the final states is $\Delta' = \Delta - Q_{cf}$ , where $Q_{cf}$ is the $c$-$5f$ Coulomb interaction. For $Q_{cf} = 0$ we obtain only the main peak. When the Coulomb interaction is switched on, the satellite peak appears. If   $Q_{cf} > \Delta$  then the satellite peak is at a lower intensity than the main peak. This gives a well-screened peak that can have a higher intensity than the main peak, depending on the precise values of $Q_{cf}$,   $\Delta$, and $V$. 
In localized systems the core hole potential gives rise to a poorly screened photoemission peak. In metallic systems, on the other hand, the core hole can be screened by valence electrons from surrounding atoms, giving rise to a well screened peak, which is at lower binding energy compared to the unscreened peak. 

\begin{figure}
 \includegraphics[width = 0.45\textwidth]{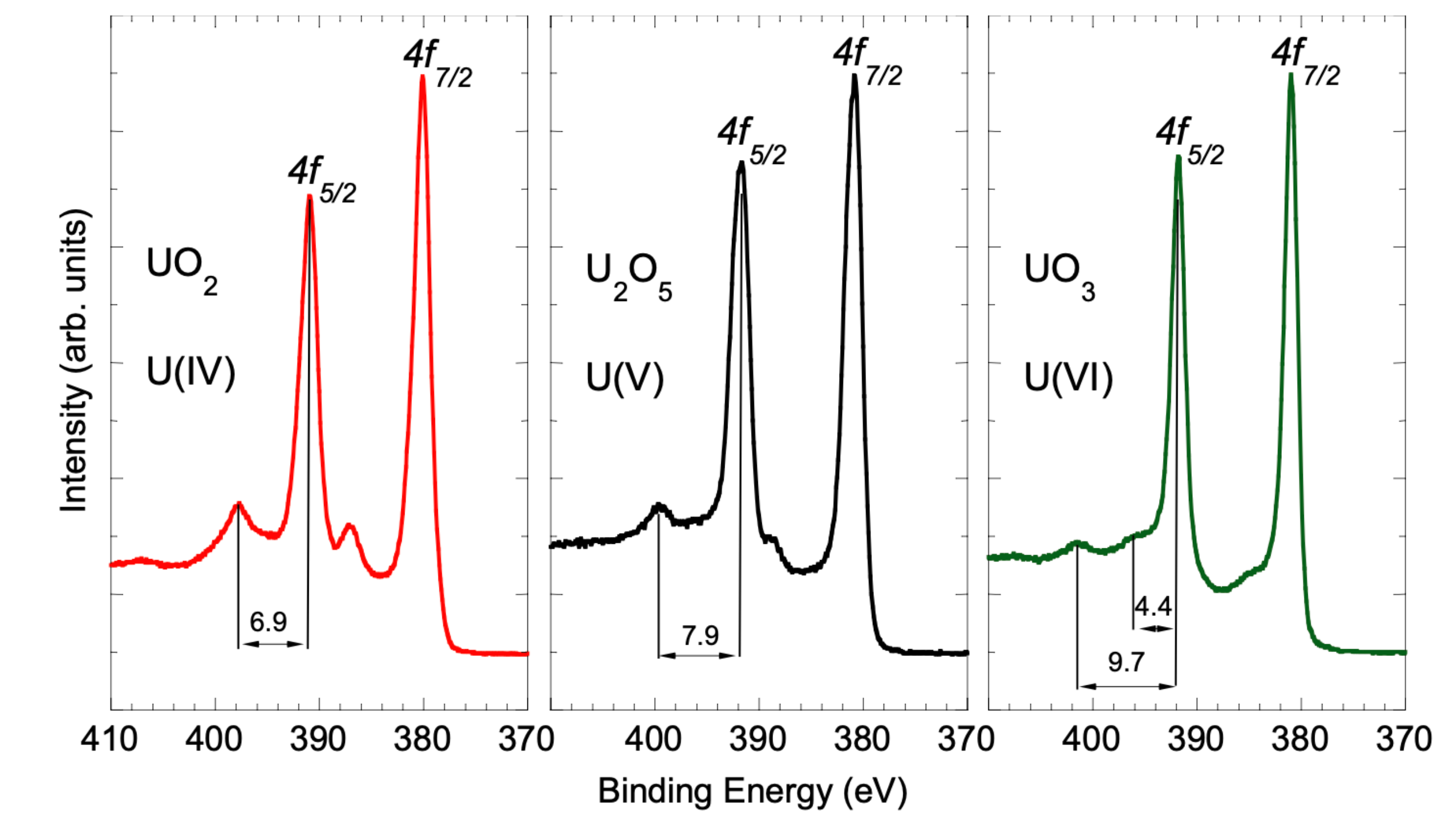}
\caption{Uranium 4$f$ core level x-ray Photoemission Spectra recorded for U(IV) in UO$_2$ (left panel), U(V) in U$_2$O$_5$ (central panel), and U(VI) in UO$_3$ (right panel). The relative energy between the satellite peak and the 4$f_{5/2}$ (4$f_{7/2}$) emission line is used as a marker for the oxidation state of the uranium atoms. Figure reproduced from \cite{gouder18}.
}
\label{U-4f}
 \end{figure}          

Apart from the satellite to main peak intensity ratio, one can also look at their energy separation (cf., Eq.~(\ref{eq:EnergySeparation}).
\cite{ilton11} calculated peak separations to determine uranium oxidation states.
Figure \ref{U-4f} shows the binding energy separation between primary U $4f$
peaks and associated satellites for different oxidation states of uranium \cite{gouder18}.
Both satellite-primary peak binding energy separations, as well as intensities to a lesser extent, are relatively insensitive to composition/structure and can be used to both identify
and help quantify U oxidation states in mixed valence phases.

\subsection{ARPES}\label{arpes}

Angle-resolved photoemission spectroscopy (ARPES) using a laboratory  source was already reported from Japan on the compound USb in \cite{kumigashira00}. Surprisingly, these results and calculations suggested that the 5$f$ electrons in USb have a ‘dual character’, partly localized and partly itinerant. Earlier work had suggested that the 5$f$ electrons in USb were almost fully localized, as are the 4$f$ electrons in CeSb. Of course, the answer to this question may depend on the choice of the probe used for the investigation.

A first ARPES measurement at SPring-8 was reported in 2006 on the compound UFeGa$_5$ \cite{fujimori06}, in which the 5$f$ states are essentially itinerant, and the spectra are quite well reproduced by LDA theory. The following year, they published a rather complete ARPES study of the superconducting (and antiferromagnetic) compound UPd$_2$Al$_3$ \cite{fujimori07}.

In this material the photoemission experiments showed an increase of the weight of the 5$f$ contribution, as the temperature was lowered, offering an explanation of why the high-temperature properties appear to show localized 5$f$ states, whereas at the lowest temperatures, when the material becomes first an antiferromagnet and then (in addition) a superconductor, the 5$f$ states are better described as itinerant (see Fig. 5 of \onlinecite{fujimori07}).
These experiments used incident energies of 400 and 800 eV, where the difference in these two spectra can be related to that originating from the 5$f$ states.
They also demonstrated strongly dispersive 5$f$ bands in the compound UB$_2$, in which the 5$f$ bands are fully itinerant \cite{ohkochi08}.

Given the intense interest in URu$_2$Si$_2$ it is not surprising that a major effort with ARPES has been made on this material. Fortunately, there is an excellent review that considers all the work, done with both synchrotron and laboratory instruments \cite{durakiewicz14}. Many features of URu$_2$Si$_2$ were established by these techniques. One interesting aspect is that Durakiewicz concludes that the 5$f$ count in this material is close 2.6, whereas the NIXS and RIXS experiments discussed in earlier sections of this review come out with a number closer to 2.0. Again, it should be emphasized that this number may be dependent on the technique used. A more recent study of ThRu$_2$Si$_2$ at SPring-8 \cite{fujimori17a} shows the difference between this material, with no 5$f$ states, and URu$_2$Si$_2$. The authors conclude that there is a strong hybridization between the 5$f$ states and the ligand states in URu$_2$Si$_2$.

\begin{figure}
\centerline{\includegraphics[width=0.9\columnwidth]{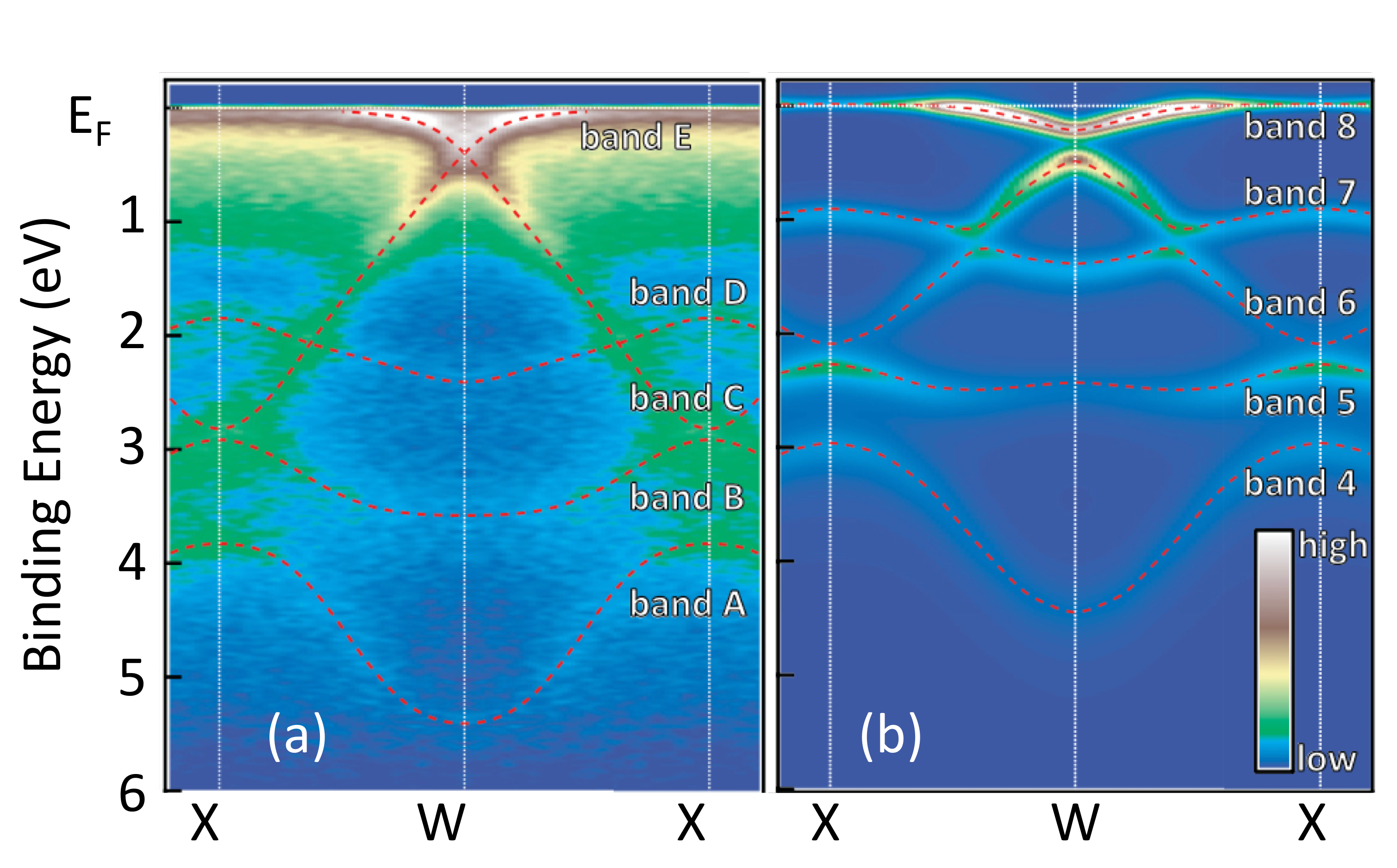}}
\caption{ARPES spectra and comparison with the results of band-structure calculations for single crystals of UN. (a) Symmetrized ARPES spectra measured along the X-W-X line. Dashed curves are guides for the eye. (b) Simulation of ARPES spectra based on the band-structure calculation treating all U 5$f$ electrons as  itinerant. Taken from \cite{fujimori12}.
\label{fujimori12}
}
\end{figure}

An interesting study at SPring-8 was performed on UN  \cite{fujimori12}. This simple (NaCl structure type) compound has been controversial for many years, especially since first investigations in the 1960s. The results from ARPES experiments are shown in Fig.~\ref{fujimori12}.
Whereas not all the bands in the figure involve 5$f$ states, there is strong evidence that all such 5$f$ states are itinerant. These conclusions strongly support earlier neutron work \cite{holden84} as well as more recent RIXS experiments \cite{lawrencebright22}. Neither neutron or RIXS work have observed any sign of the crystal-field states which are a major component of any 5$f$ localization.

A good summary of the work published at SPring-8 has been published by \cite{fujimori16} and by the Los Alamos group \cite{beaux11}, who have performed experiments at both the ESRF and the Wisconsin synchrotron on uranium compounds, as well as with laboratory systems. Many of these experiments were performed with hard x-ray photoelectron spectroscopy (HAXPES), where the incident energies were as high as 7.6 keV, eliminating questions about whether the bulk properties were examined.  In reviewing the experiments done at synchrotrons, \cite{fujimori16}  make the point that they use photon energies of between 400 and 500 eV, whereas laboratory experiments, using, e.g., the He-I and He-II radiations use energies of normally less than 100 eV. As we have discussed often in this review, such low photon energies must be regarded as extremely surface sensitive (see Table \ref{attlengths}), so the conclusions may depend on the preparation of the surface and the ultra-vacuum conditions in the chamber.

More recent studies at SPring-8 have been made of the UX$_3$ (X = Al, Ga, and In) systems and a relatively complete picture of their Fermi surfaces elucidated \cite{fujimori17}. There are strong electron correlations in these systems, so the magnetic ordering is of the weakly itinerant 5$f$ states. The mixing between the 5$f$ and anion $p$ states shows why a strong signal was obtained at the Ga $K$-edge in UGa$_3$ in resonant diffraction experiments \cite{mannix01}.

A recent study has also been reported on UTe$_2$, a material of much current interest \cite{ran19}, and the authors \cite{shick21} conclude  that there are both localized and itinerant features seen in the angle-integrated spectra, and they believe the ground state is predominantly 5$f^3$.

\begin{figure}
\centerline{\includegraphics[width=0.9\columnwidth]{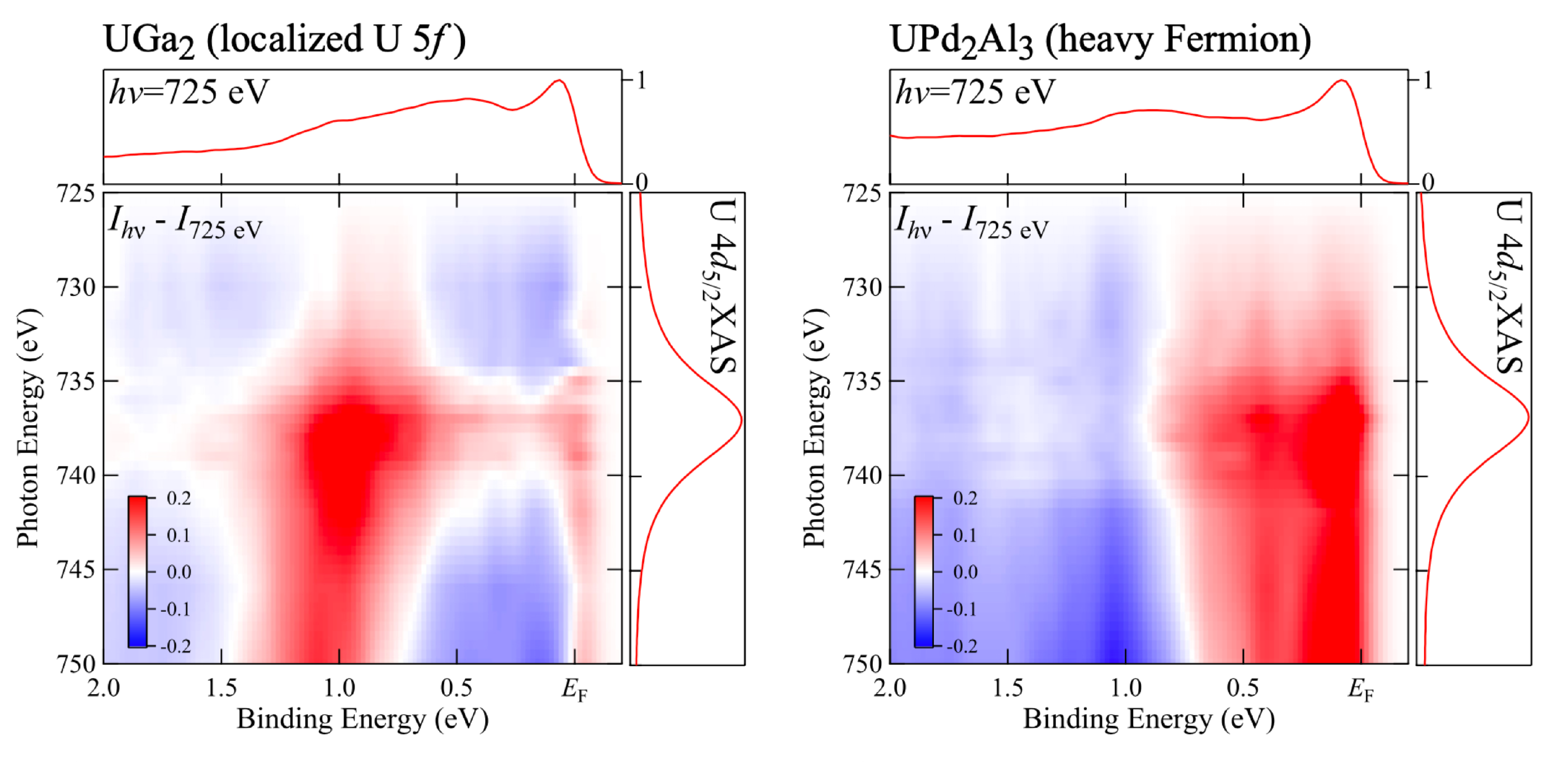}}
\caption{Angle-integrated spectra showing the difference between photoemission spectroscopy taken with the energy tuned near to the $N_{5}$ edge at 737 eV and the off-resonance condition of 725 eV. The 5$f$ enhancement shows the weight of the signal in UGa$_2$, where the 5$f$ electrons are essentially localized, is some $\sim$1 eV below $E_{\mathrm{F}}$, whereas in the heavy-fermion system UPd$_2$Al$_3$ they are basically at $E_{\mathrm{F}}$.
Taken from \cite{fujimori19}. 
\label{fujimori19}
}
\end{figure}

Finally, resonant photoemission was reported from the SPring-8 group for the first time on a series of U compounds \cite{fujimori19}. This was performed by tuning the incident photons to the $N_{4,5}$ resonance, where the core 4$d$ states are promoted to the empty 5$f$ states and there is an enhancement of the signal arising from the 5$f$ states. This is the same transition energy as used for successful RIXS work reported in Sec.~\ref{secRIXS}. The enhancement in the signal is less than that observed at the Ce $M_{4,5}$ edges (3$d$ $\rightarrow$ 4$f$) but  can still be useful. They have found the enhancement at the two edges roughly equivalent so report scans on-resonance (737 eV for the $N_{5}$ 4$d_{5/2}$ $\rightarrow$ 5$f$) and off-resonance at 725 eV, and then plot the difference spectra. This is shown for UGa$_2$ and UPd$_2$Al$_3$ in Fig.~\ref{fujimori19}.

Although UGa$_2$ (which is a ferromagnet with a large magnetic moment of $\sim$3 $\mu_{\mathrm{B}}$)  
appears to have a localized $5f$ component \cite{kolomiets21} it is probably better to consider it as a band system. One can see some weight at $E_{\mathrm{F}}$ attributed to the U 5$f$ states, even if the major weight is around $-$1 eV.

The series of experiments performed with ARPES at SPring-8 have been important in a number of ways in actinide research. Perhaps the most crucial aspect is that they have produced high-quality data that can be tested against the latest theoretical predictions. In this respect the resonance experiments described in \cite{fujimori19} are particularly valuable and have already shown that the behavior of the 5$f$ electrons is often more complicated than expected; reproducing these results from first-principles theories is a considerable challenge. On the negative side, it is most unfortunate that we have so little such data on transuranium samples. Of course, given that the detected particles are electrons, it is clear that the chambers would become contaminated if transuranium samples were used, as no encapsulation of the sample is possible with the technique. Whether this could be overcome with a determined effort is still unclear. However, imagine the information that, e.g., could be obtained if resonant ARPES could be determined on the series AnCoGa$_5$ with An = U, Np, and Pu. We could follow the 5$f$ states as they progress from a probably itinerant UCoGa$_5$, through to partial localization of the 5$f$’s in NpCoGa$_5$ (which orders antiferromagnetically at 47 K with a magnetic moment of 0.8 $\mu_{\mathrm{B}}$), and then to the heavy-fermion superconductor PuCoGa$_5$, which has the highest $T_c$ of any heavy-fermion material with $T_c$  = 18 K. Many other such examples could be proposed, but this one is especially interesting, and it turns out that single crystals of all of these three compounds have been made in Japan. 

\section{Conclusions}

After the first synchrotrons began operations in the late 1970s and 1980s actinides were soon examined, notably with XAS \cite{kalkowski87a,bertram89}, XMCD \cite{collins95a}, and EXAFS measurements \cite{silva95,conradson98} at the actinide $L$ edges, high-pressure experiments using energy dispersive detectors \cite{benedict93}, and high-resolution XRD to investigate fine details of the diffraction patterns \cite{grubel91}. The success of these early generation synchrotrons utilized across all parts of the periodic table led, of course, to more powerful  light sources being built, and the advantages for the study of actinides increased. They now represent some of the most powerful tools available for such research.

This article has focused on experiments performed in the field of condensed-matter physics in the last 20 years at the ESRF in Grenoble, France, with references to work at other synchrotrons, notably Diamond in the UK, the APS at ANL (USA), and SPring-8 in Japan. Of course, many experiments have been performed in the domain of chemistry and environmental sciences; efforts to study the dissolution of UO$_{2}$ in water \cite{springell15,rennie18a} span these various fields. They demand synchrotron radiation, as diffraction signals are required from epitaxial films of $<$10 nm thickness. 

In concluding this article, we would like to highlight six examples that we believe to have been “game changers” for actinide research. 

The first (Sec.~\ref{secXRDhp}) is the capability to observe samples at the $\mu$g level, allowing pressures up to 100 GPa to be applied in diamond-anvil cells, and the development of the angular-dependent data collection so sophisticated data analysis can be used to determine the crystallographic structures, see Fig.~\ref{pdepAmCmU}. These experiments, extended to transuranium samples \cite{lindbaum01,heathman2005}, illustrated the failure of the density-functional theory (DFT) for actinides, and spurred the subsequent work with the dynamical mean-field theory (DMFT) that is frequently mentioned in the text. These experiments could only have been done on third generation synchrotrons. 

The second example (Sec.~\ref{materialsscience}) concerns the field of materials science, and involves very narrow beams of usually high-energy photons (often E $>$ 40 keV). Combined with tomography, this gives an extremely detailed view of defects and pore structure, and unlike the electron microscope, the technique is non-destructive. Fig. \ref{thomas2020} gives one example, \cite{thomas20}. This field may also benefit from the time structure of future FEL’s, as one can imagine materials subjected to various dynamical stresses and being able to follow the microscopic changes, again in a non-destructive way.

The third example (Sec.~\ref{secREXSEQ}) is the discovery in the actinide oxides and UPd$_3$ of ordering of the electric quadrupolar moments \cite{santini09} below room temperature. In this case, intensity is not always the main problem, but complex instrumentation is needed to measure the polarization dependence of the scattered photons, and their azimuthal dependence, see Fig. \ref{NpO2azimuth}. In the case of NpO$_2$, e.g., the nature of the phase transition at 25 K \cite{paixao02} had been the source of speculation for more than 50 years. How common this phenomenon is in the actinides is still an open question, but the tools to find such effects are now available.

The fourth example is that the development of the technique of XMCD (Sec.~\ref{secXMCD}) led to a direct method of determining the absolute values of the spin and orbital moments in actinide materials. Prior to this, the only technique available was using polarized neutrons, but this requires sizable single crystals. Now these quantities can be measured with $\mu$g of polycrystalline samples. The values, and especially their ratio, are of major importance for actinide research, as with itinerant 5$f$ electrons there is a tendency for a partial quenching of the orbital moment \cite{lander91}. Now that the $\left< T_z \right>$ values calculated with intermediate coupling \cite{vanderlaan96} can be used, the orbital to spin ratio is readily determined. Figure \ref{XMCDToverS} represents perhaps the most convincing evidence given so far that intermediate coupling is a crucial requirement of the physics and chemistry of actinide materials in general.

The fifth example concerns the inelastic scattering of x-rays (Sec.~\ref{secIXS}). Here again, the capability to determine the phonon spectra from $\mu$g’s of material as at the ID28 beamline has been of great importance. The phonons of $\delta$-Pu (Fig.~\ref{DispDeltaPu} and \onlinecite{wong03}) was a major achievement, especially since the theoretical work was actually published in advance \cite{dai03}, and gave the first major “credibility test” to the DMFT theory at that time. More recently, the development of grazing incidence IXS has led to important new information on the radiation damage in UO$_2$ \cite{rennie18} and on the complexity of the U-Mo alloy system, see Fig. \ref{PhononsUMo}. The sensitivity of these experiments is outstanding; especially considering the sample mass is less than 100 $\mu$g. 

The measurement of ARPES from a large number of uranium compounds at the Japanese SPring-8 facility (our sixth example, Sec. \ref{secPES}) represents a tour-de-force for actinide research. The data, especially that using resonance at the uranium $N$-edge offer an unprecedented snap-shot of the electronic structure and the role of the $5f$ states in these materials, see Figs. \ref{fujimori12}-\ref{fujimori19}. Although considerable efforts have already been made to match these to present-day theory, they remain a challenge for future, no doubt more sophisticated, theories. As we describe in the text, it is vital that ways be found to extend this work to transuranium materials so that a better overview can be found of the changes in the $5f$-electron behavior as their number is increased.

The future of actinide research at synchrotrons is clearly a bright one. A new generation of machines, and associated complex instrumentation are coming online that will undoubted benefit new initiatives. Free-electron lasers (FEL’s) with extremely high peak brightness and pulse durations from a few to hundreds of femtoseconds are now starting to operate, pushing the timescale limits of spectroscopy and structure studies \cite{liermann21}. Diffraction-limited storage rings (DLSR) such as MAX-IV in Sweden \cite{tavares18}, Sirius in Brazil \cite{liu14}, and ESRF-EBS in France \cite{raimondi16,chenevier20}, with emittance approaching the diffraction limit and delivering ultra-small x-ray beam sizes, start demonstrating their potential as an extraordinarily powerful tool for the investigation of complex systems and emerging phenomena. These machines will provide an increase in average brightness and coherent flux of about two orders of magnitudes, compared to third generation x-ray sources, with obvious applications for high pressure studies, microscopy, coherent diffraction, high-resolution imaging, and spectroscopy \cite{eriksson14,frenkel14}. The possibility to achieve
sub-$\mu$m focal-spot diameters at the sample position will extend the use of x-ray absorption and emission spectroscopy methods to single actinide nanoparticles and colloids, providing powerful characterization tools for the study of surface defects and particle-size-dependent structural details, important for a better understanding of environmental effects and catalytic properties. 
High-pressure diffraction experiments will benefit both from the higher brightness and from the smaller beam size, giving access to higher pressures in smaller diamond anvil cells \cite{mcmahon14}. For actinides, this will give the opportunity to extend phase diagrams and observe the destabilization of localized 5$f$ states in Curium and beyond.
All such instrumentation will greatly benefit our understanding of this complex row of elements in the periodic table.
X-ray photon correlation spectroscopy (XPCS) \cite{grubel04,lim14} will benefit from the increased coherence of the incident beam enabling studies of phase transitions and critical phenomena in strongly correlated electron systems at picosecond time scales \cite{shpyrko14}. A source at the diffraction limit in the hard x-ray range, with high coherent flux and an almost round beam footprint (thanks to the strongly reduced horizontal emittance), will make scanning coherent diffractive imaging techniques, such as x-ray ptychography \cite{thibault14}, available for coherent imaging experiments on nuclear waste forms and irradiated fuels with mesoscopic spatial resolution, between the nanometer and the micrometer length scales.
A strong reduction of the focus size will be reflected in an improved energy resolution of spectroscopy techniques, like IXS, RIXS, NIXS, and ARPES, enabling a precise observation of magnetovibrational and multipole excitations, splitting of electronic multiplets by spin-orbit and crystal field interactions, and dispersion of band structure states \cite{schmitt14,rotenberg14}. This is particularly important for understanding the subtle interplay between competing interaction mechanisms in actinide materials poised at the brink of electronic, lattice, and magnetic instabilities. 

\begin{acknowledgments}

A number of colleagues from Joint Research Centre Karlsruhe, Universities of Parma (Italy), K\"oln (Germany), Oxford and Bristol (UK), European Synchrotron Radiation Facility, Diamond Light Source, DESY synchrotron, Brookhaven and Los Alamos National Laboratories, and several other institutions, have been involved in the studies presented in this review. The inspiration shown by the late Marty Blume at BNL in the early days of this research is gratefully acknowledged.

We thankfully acknowledge helpful discussions with M. Altarelli, L. Braicovich, S. P. Collins, A. Fujimori, G. Ghiringhelli, K. Kvashnina, and A. Severing. We would like to thank the Radioprotection Services, and especially Patrick Colomp, at the ESRF for their collaboration in the safe running of transuranium experiments on (so far) at least ten different beamlines.
\end{acknowledgments}

 \bibliography{RMPSRactinides}

\end{document}